\documentclass[11pt]{article}
\usepackage[letterpaper,includefoot,top=1in,bottom=1in,left=1in,right=1in]{geometry}
\usepackage[centertags,sumlimits]{amsmath}
\usepackage{amsfonts}
\usepackage{amssymb,graphics,psfrag}
\usepackage{array,epsfig,multirow,stmaryrd,graphicx}
\usepackage[all,cmtip]{xy}
\usepackage{mathrsfs}
\usepackage[bookmarks,breaklinks]{hyperref}
\usepackage{multirow}

\numberwithin{equation}{section}
\numberwithin{table}{section}

\newcommand{\Div}{D}

\newcommand{\Z}{\mathbbm{Z}}

\renewcommand{\Im}{\operatorname{Im}}
\renewcommand{\Re}{\operatorname{Re}}

\newcommand{\beq}{\begin{equation}}
\newcommand{\eeq}{\end{equation}}
\newcommand{\be}{\begin{equation}}
\newcommand{\ee}{\end{equation}}
\newcommand{\bea}{\begin{eqnarray}}
\newcommand{\eea}{\end{eqnarray}}   
\newcommand{\ben}{\begin{eqnarray*}}
\newcommand{\een}{\end{eqnarray*}}                  
\newcommand{\ba}{\begin{aligned}}
\newcommand{\ea}{\end{aligned}}
\newcommand{\bt}{\begin{tabular}}
\newcommand{\et}{\end{tabular}}
\newcommand{\bc}{\begin{center}}
\newcommand{\ec}{\end{center}}

\newcommand{\cO}{\mathcal{O}}

\newcommand{\cC}{\mathcal{C}}
\newcommand{\cD}{\mathcal{D}}
\newcommand{\cL}{\mathcal{L}}

\newcommand{\cK}{\mathcal{K}}
\newcommand{\cN}{\mathcal{N}}

\newcommand{\cG}{\mathcal{G}}
\newcommand{\cA}{\mathcal{A}}

\newcommand{\cB}{\mathcal{B}}
\newcommand{\cF}{\mathcal{F}}

\newcommand{\cV}{\mathcal{V}}

\newcommand{\cM}{\mathcal M}

\newcommand{\I}{\text{Im}}
\newcommand{\R}{\text{Re}}

\newcommand{\bbZ}{\mathbb{Z}}
\newcommand{\bbR}{\mathbb{R}}

\newcommand{\bbP}{\mathbb{P}}

\newcommand{\nn}{\nonumber}

\newcommand{\cref}{{\bf [check ref]}}

\newcommand{\tr}{\mathrm{Tr}\:}

\newcommand{\CN}{{\cal N}}

\def\Z{\mathbb{Z}}

\def\C{\mathbb{C}}

\def\P{\mathbb{P}}

\def\Kern{\operatorname{Ker}}
\def\Image{\operatorname{Im}}

\def\mc{\mathcal}

\newcommand\T{\rule{0pt}{2.6ex}}
\newcommand\B{\rule[-1.2ex]{0pt}{0pt}}

\entrymodifiers={+!!<0pt,\fontdimen22\textfont2>}

\begin{document}

\baselineskip=16pt
\setlength{\parskip}{6pt}

\begin{titlepage}
\begin{flushright}
\parbox[t]{1.8in}{
BONN-TH-2008-16\\
0811.2996\ [hep-th]}
\end{flushright}

\begin{center}

\vspace*{1cm}

{\LARGE \bf The D5-brane  effective action and superpotential \\[.5cm] in $\mc N=1$
compactifications}

\vskip 1cm

\begin{center}
        \normalsize \bf{Thomas W.~Grimm, Tae-Won Ha, Albrecht Klemm and Denis Klevers} \footnote{\texttt{grimm, tha, aklemm, klevers@th.physik.uni-bonn.de}}
\end{center}
\vskip 0.75cm

{\em Bethe Center for Theoretical Physics\\[.1cm]
and\\[.1cm]
Physikalisches Institut der Universit\"at Bonn \\[.1cm]
Nussallee 12, 53115 Bonn, Germany}

 \vspace*{0.75cm}

\end{center}

\vskip 0.0cm

\begin{center} {\bf ABSTRACT } \end{center}

The four-dimensional effective action for D5-branes in generic 
compact Calabi-Yau orientifolds is computed by performing a Kaluza-Klein reduction.
The $\cN=1$ K\"ahler potential, the superpotential, the gauge-kinetic coupling
function and the D-terms are derived  in terms of the geometric data 
of the internal space and of the two-cycle wrapped by the D5-brane.
In particular, we obtain the D5-brane and flux superpotential by 
integrating out four-dimensional three-forms which couple 
via the Chern-Simons action. Also the infinitesimal complex 
structure deformations of the two-cycle induced by the 
deformations of the ambient space contribute to the F-terms. 
The superpotential can be expressed in terms of relative 
periods depending on both the open and closed moduli.
To analyze this dependence we blow up along the two-cycle and 
obtain a rigid divisor in an auxiliary compact threefold with 
negative first Chern class. The variation of the mixed Hodge 
structure on this blown-up geometry is equivalent to the original 
deformation problem and can be analyzed by Picard-Fuchs equations.  
We exemplify the blow-up procedure for a non-compact Calabi-Yau 
threefold given by the canonical bundle over del Pezzo surfaces.

\bigskip

\hfill November, 2008
\end{titlepage}

\section{Introduction}

In recent years much progress has been made in the study of supersymmetric 
four-dimensional effective actions arising from Type II compactifications with 
D-branes and background fluxes \cite{Douglas:2006es,Blumenhagen:2006ci,Denef:2008wq, Lust:2004ks}. In these set-ups gauge theories are localized on
space-time filling D-branes while chiral matter arises along their intersections \cite{Blumenhagen:2005mu,Blumenhagen:2006ci,Lust:2004ks}. 
For consistent compactifications one needs to include an orientifold projection by
dividing out the reversal of the world-sheet parity as well as some geometric involutions. 
The orientifold planes are located on the fixpoint set of the geometric involution and 
have to cancel the positive tension of the space-time filling D-branes. In order that the four-dimensional 
effective theory admits $\cN=1$ supersymmetry, all D-branes have to preserve the 
same supersymmetry as the orientifold planes. The implementation of all the consistency 
requirements in phenomenologically appealing set-ups has been successfully carried out 
for compactifications with O3 and O7-planes and the corresponding 
D-branes \cite{Blumenhagen:2005mu,Blumenhagen:2006ci,Lust:2004ks,Denef:2008wq}. 
 
In this work we will focus on the class of orientifold compactifications
admitting O5-planes and space-time filling D5-branes. Our aim is 
to determine the four-dimensional $\cN=1$ effective action for the 
D5-brane moduli coupling to the closed string zero modes from 
the internal Calabi-Yau orientifold geometry. This will be done by performing
a Kaluza-Klein reduction of the Dirac-Born-Infeld and Chern-Simons action for
the D5-brane. Similar analysis for D3- and D7-branes on generic Calabi-Yau
orientifolds has been carried out in 
refs.~\cite{Camara:2003ku,Grana:2003ek,Camara:2004jj,Jockers:2004yj,Jockers:2005zy}. For D5-branes the
effective action including the bulk couplings has been evaluated for orbifold
compactifications \cite{Kors:2003wf,Lust:2004cx}. A review of these results can be found in 
ref.~\cite{Blumenhagen:2006ci,Lust:2004ks}.

In the open string sector one finds $\cN=1$ vector multiplets
for the gauge theory on the D-brane. In addition, there are chiral multiplets 
parametrizing the Wilson line moduli as well as the deformations of the D-brane.
For D5-branes the Wilson lines arise if the wrapped Riemann surface is of
genus one or higher. In this work we will not consider intersecting branes,
such that there are no additional charged matter fields. In the $\cN=1$ effective
four-dimensional theory the kinetic terms for all chiral multiplets must arise
from a K\"ahler potential. We will be able to derive its explicit form,
generalizing the expressions for the closed string moduli found in
refs.~\cite{Grimm:2004ua}. It will be shown that the deformation moduli of the 
D5-brane correct the $\cN=1$ complex coordinates on the K\"ahler moduli space
of the closed sector, while the Wilson line moduli correct the dilaton complex 
coordinate. The situation is thus similar to the one encountered in
compactifications of the Type I string \cite{Blumenhagen:2006ci,Lust:2004ks}.

In order that the D5-branes preserve the $\cN=1$ supersymmetry of the background
they have to wrap holomorphic cycles in the internal space \cite{Becker:1995kb}. In
addition, also the combination of the NS--NS B-field and the gauge flux on the
D5-brane have to vanish. However, this will no longer be the case if one
considers fluctuations around the background configuration. One expects that
in this case there will be a scalar potential induced for these variations. We will
explicitly derive this potential by reducing the D-brane effective actions and
show that it splits into F- and D-term contributions. In
order to do that, we find that it is crucial to also include non-dynamical
three-forms arising in the reduction of the bulk R--R fields. These couple
via the Chern-Simons action to the D5-brane moduli and induce additional
contributions to the scalar potential.  Moreover, we also have to account for
contributions in the Dirac-Born-Infeld action which are induced by
the variations of the complex structure of the ambient Calabi-Yau
orientifold. These new insights together with explicit knowledge of the $\cN=1$
K\"ahler metric on the field space allow us to compute the $\cN=1$
superpotential and D-terms by direct dimensional reduction of the bosonic
D-brane actions. 

In our study of $\CN=1$ theories with D5-branes and background fluxes
the superpotential $W$ is of particular interest. Its holomorphicity protects it from 
perturbative corrections and allows it to be computable using topological models 
associated to the physical string
\cite{Witten:1997ep,Brunner:1999jq,Kachru:2000ih,Kachru:2000an,Aganagic:2000gs,
Aganagic:2001nx,Marino:2004uf}.\footnote{The second key quantity of the 
open string is the gauge kinetic function, which encodes the annulus contributions.}
It thus plays a crucial role in the extension of mirror symmetry between Type IIA and Type IIB compactifications on mirror dual Calabi-Yau manifolds $X$ and $Y$
from the closed to the open string topological sector. 
In fact, the mirror dual of the D5-brane superpotential in Type IIA is,
in the mirror large radius expansion, the generating function for suitably counted 
holomorphic disk world-sheet instantons ending on a D6-brane 
wrapped on a special Lagrangian submanifold of $X$.
This is analogous to the closed string case where the prepotential 
is the generating function for suitably counted genus zero world-sheet instantons on $X$ in Type IIA.
Recall that the prepotential in Type IIB topological model 
is computed by considering the dependence of the holomorphic three-form $\Omega$
on the complex structure of the Calabi-Yau manifold $Y$. More precisely, 
one notes that the variation of the Hodge structure of $H^3(Y)$ with respect to 
the complex structure deformations has a flat Gau\ss -Manin connection 
and leads to a system of differential equations
for the holomorphic three-form called Picard-Fuchs 
equations.
The periods of $\Omega$ contain the information 
about the Type IIB prepotential, together with preferred coordinates defining the mirror map to Type IIA. This 
provides a much simpler calculation for this quantity in the Type IIB than in the Type IIA theory due to the use of 
classical geometry. 
Note that mapping the Type IIB to the mirror Type IIA configurations also provides 
an extension of the latter to stringy length scales. 

Including also the open string sector and deriving the superpotential 
is more involved\footnote{An idea alternative to the follwing discussion has been developed and exploited for concrete examples in \cite{Baumgartl:2007an} where the superpotential has been computed using methods from conformal field theory and Landau-Ginzburg techniques.}. 
It has been realized in \cite{Lerche:2002ck,Lerche:2002yw,Aganagic:2000gs,Aganagic:2001nx} 
that in the generalization to the open string sector 
the variation of the Hodge structure has to be replaced by the variation of 
the mixed Hodge structure. This replacement is due to the D5-brane contribution to the 
superpotential $W$ that is calculated by an integral of the holomorphic 
three-form $\Omega$ over a three-chain $\Gamma$, whose boundary includes the 
curve $\Sigma$ on which the D5-brane is supported \cite{Witten:1997ep} 
\beq
W_{\text{open}} = \int_\Gamma \Omega\ .
\eeq 
Then, the variation of the mixed Hodge structure leads again to 
Picard-Fuchs systems for the periods and the chain integrals
as well as to preferred coordinates. 
This has been extensively studied for non-compact 
Calabi-Yau manifolds with special brane configurations
\cite{Aganagic:2000gs,Aganagic:2001nx,Marino:2004uf}. 
In particular, the open string superpotential has been analyzed in some generality and depth for a 
non-compact toric Calabi-Yau manifold with Harvey-Lawson type branes
\cite{Aganagic:2000gs,Aganagic:2001nx}. 
In these cases the chain integral reduces to an integral of a meromorphic 
one-form over a one-chain. Since all basic ideas are realized here in an 
elementary fashion and all quantities can be explicitly calculated, we will 
recall this construction in section \ref{sec:n1-special-geometry}. 
In the non-compact case the results have been
obtained for all world-sheet topologies and checked successfully against
calculations in the Type IIA models using the topological vertex, localization 
and large $N$ methods at various points in the moduli space \cite{Aganagic:2003db,Graber:2001dw,Marino:2004uf}.

Significant  progress in the extension of these ideas to compact Calabi-Yau spaces has been made in \cite{Walcher:2006rs}, where the open string Picard-Fuchs system for the rigid  
special Lagrangian, defined as the fixpoint locus of the anti-holomorphic 
involution in the quintic in $\mathbb{P}^4$, has been suggested and 
the resulting predictions for the disk instantons have been checked. Also the order of the 
obstruction of the open string moduli by $W$ in  certain situations 
has been analyzed earlier in \cite{Kachru:2000ih}. So far, however, the compact examples are
restricted to very special cases and the general dependence of $W$ on 
the open string deformations has not passed independent checks.
Recently, extending the works \cite{Lerche:2002ck,Lerche:2002yw} to compact examples, a suggestion for a related problem with open string deformations 
has been made in \cite{Jockers:2008pe} together with some predictions for disk orbifold 
instantons, whose status is unclear. The crucial idea in the works \cite{Lerche:2002ck,Lerche:2002yw} is to
introduce, following Griffiths \cite{Griffiths:1979rt}, an auxiliary divisor containing the curve whose
deformations model deformations of the chain integrals. For example, in 
\cite{Jockers:2008pe}, it is claimed that the Picard-Fuchs equations for a meromorphic 
differential defined via the auxiliary divisor on a Calabi-Yau space are solved 
by the chain integrals which define the superpotential and the preferred open 
coordinates.  

In this work we propose an alternative route and map the calculation of the 
superpotential to the ordinary deformation space of pure Hodge theory  on a manifold $\tilde
Y$, which is obtained by blowing up along the curve $\Sigma$ which supports the D5-brane. This blow-up procedure 
replaces $\Sigma$ by the projectivization of its normal bundle which is a divisor $D$ in $\tilde Y$, so that
the mixed Hodge structures of $H^3(Y,\Sigma)$ is equivalent to 
$H^3(\tilde Y,D)$. The manifold $\tilde Y$  has negative Chern class. It has 
a single holomorphic three-form, which vanishes however on $D$. 
In this way we can argue that the original deformation problem in $H^3(Y,\Sigma)$ is mapped 
to the complex structure deformations of $\tilde Y$. As we start with an arbitrary 
$\Sigma$ in $Y$ the construction is very general, however concrete 
calculations are relegated to forthcoming work.

This paper is organized into two parts. The first part is dedicated to the 
derivation of the four-dimensional effective action governing the low-energy 
dynamics of the D5-brane system. 
In section \ref{D5reduction} we perform the Kaluza-Klein reduction of the 
Dirac-Born-Infeld and Chern-Simons action of the D5-brane. 
We summarize the four-dimensional spectrum of the closed and open string 
sector in section \ref{4Dspectrum}.
Additionally, in section \ref{relations}, we discuss the interdependency of 
the bulk and brane moduli focusing on the complex structure deformations of the 
Calabi-Yau $Y$ and the fluctuations of the position $\Sigma$ of the wrapped 
D5-brane. We obtain relations useful throughout our whole first order analysis 
of the effective action.
In \ref{reduction_D5} we turn to the detailed calculation 
of the effective action. 
Particular emphasis is put on the computation of the 
scalar potential discussed in section \ref{subsec:fluxpotential}. 
We show that crucial F-type potential terms are contributed by the interaction 
and kinetic terms of non-dynamical three-form fields coupling via the 
Chern-Simons action to the D5-brane. 
Finally, we also identify D-terms due to a non-vanishing combination 
of the NS--NS B-field and the gauge flux on the D5-brane as well as 
background NS--NS fluxes.  

In section \ref{subsec:N=1data} we cast the results of the dimensional reduction into the 
standard $\cN=1$ supergravity form by determining the $\mathcal{N}=1$
characteristic data. In section \ref{subsec:Kpot} we summarize the $\mathcal{N}=1$ complex coordinates 
that are corrected due to scalar fields arising from the D5-brane and derive
the K\"ahler potential by bringing the kinetic terms of chiral multiplets
into the standard $\cN=1$ form.
We encounter a no-scale like property of the K\"ahler potential 
which enables us to derive the effective $\mathcal{N}=1$ superpotential in
section \ref{subsec:suppot}. 
We complete the characteristic data in section \ref{subsec:gauging}
by giving the gauge kinetic functions for the brane and bulk vectors and 
analyzing the gauging of shift symmetries. We conclude 
by evaluating the D-term potential due to the gauged shift symmetries and 
show that this perfectly matches the result of the dimensional reduction.

In the second part of the paper we turn to 
a more mathematical treatment of the $\cN=1$ superpotential. In 
section \ref{sec:n1-special-geometry} we complete the D5-brane superpotential 
into the chain integral expression of \cite{Witten:1997ep,Aganagic:2000gs}. Following 
\cite{Lerche:2002ck,Lerche:2002yw}, we unify it with the flux superpotential to a 
pairing in relative cohomology. 
This allows us to study the dependence of the superpotential on the 
complex structure and D5-brane moduli in more detail.
After reviewing the calculations yielding the superpotential for non-compact Calabi-Yau threefolds in section \ref{non-compactCY}, we introduce the necessary mathematical tools for the general analysis in sections \ref{closedHodge} and \ref{rel_cohom}.
First, we 
review the situation with complex structure moduli only in section \ref{closedHodge}. 
After this motivation and a brief repetition 
of relative cohomology in section \ref{rel_cohom}, we present the main idea of our analysis of the 
open-closed moduli dependence of the superpotential. In section \ref{sssec:curves-to-divisors} 
we construct an auxiliary divisor in an auxiliary K\"ahler threefold by 
blowing up the curve $\Sigma$ wrapped by the D5-brane.
In section \ref{subsec:PF} we  discuss in detail how these auxiliary geometrical 
objects are helpful to analyze the moduli dependence of the superpotential. 
The presented mathematical machinery can be applied to potentially derive open-closed Picard-Fuchs equations obeyed by the effective 
superpotential. 
We conclude in section \ref{blow-up-example} with an example of the described blow-up procedure by considering non-compact curves in the total space of the canonical bundle over the del Pezzo surface $\mc B_3$.

Our paper has four appendices which provide more detailed computations and definitions
omitted in the main text. The appendix \ref{app:derivationofpotential} contains standard expressions for the 
$\cN=2$ gauge-coupling functions. In appendix \ref{kinmix} we 
determine the kinetic mixing between the bulk and brane vectors. 
In appendix \ref{app:F-termscalarpot} we present the detailed calculation of the $\mathcal{N}=1$ 
F-term scalar potential and list the explicit form of the $\cN=1$ K\"ahler metric and its inverse. 
Finally, in appendix \ref{app:mixed-hodge} we describe the mixed Hodge structure in more detail.


\section{The D5-brane action \label{D5reduction}}

In this section we derive the four-dimensional effective action of 
Type IIB string theory on a generic Calabi-Yau orientifold with O5-planes and
D5-branes extended along Minkowski space and wrapped on an internal two-cycle.
We begin with the discussion of the four-dimensional field content in section \ref{4Dspectrum}.
On the one hand, it arises from the Kaluza-Klein zero modes for the fields in
the ten-dimensional Type IIB bulk supergravity action.
On the other hand, the D5-brane dynamics are encoded by the Dirac-Born-Infeld 
and Chern-Simons action. They describe the dynamics and couplings of the open string modes that 
are localized on the D5-brane world-volume. We discuss some special relations
between the open and closed string modes in section \ref{relations}.
In section \ref{reduction_D5} we proceed with the discussion of the 
calibration conditions for supersymmetric D5-branes in orientifolds with 
O5-planes and work out the complete effective action of the D5-brane by
performing the dimensional reduction of both the Dirac-Born-Infeld and
Chern-Simons action. This also includes a discussion of the global consistency 
conditions imposed to cancel R--R tadpoles. Finally, in section \ref{subsec:fluxpotential} 
we conclude with the derivation and discussion of the complete scalar
potential due to the presence of the D5-brane and possible R--R and NS--NS background 
fluxes.

\subsection{The four-dimensional spectrum \label{4Dspectrum}}

Here we discuss the four-dimensional spectrum emerging from compactification of the Type IIB theory.
We start our discussion by fixing the background geometry of our setup.
In the following, we consider the direct product of a compact Calabi-Yau orientifold $Y/\mathcal{O}$ and 
flat Minkowski space $\mathbb{R}^{1,3}$ with metric in the string frame given by 
\begin{equation} \label{eqn:metric}
        ds^2_{10}=\eta^{\text{SF}}_{\mu\nu}dx^\mu dx^\nu+2g_{i\bar \jmath}dy^id\bar{y}^{\bar \jmath}\,.
\end{equation}
We are interested in compactifications which allow the inclusion of
space-time filling D5-branes and O5-planes which preserve $\cN=1$ supersymmetry in
four space-time dimensions. This fixes the orientifold projection to be 
of the form \cite{Brunner:2003zm}
\begin{equation} \label{OactionO5}
        \mathcal{O} = \Omega_p \sigma^*\ ,\qquad \quad \sigma^* J = J \ ,\qquad \qquad \sigma^* \Omega = \Omega\ . 
\end{equation}
Here $\Omega_p$ is the world-sheet parity reversal and $\sigma$ is a 
holomorphic and isometric involution of the compact Calabi-Yau manifold $Y$.
The spectrum consists of two classes of fields. Firstly, there are zero modes
arising from the expansion of
the ten-dimensional closed string fields into harmonics of the internal 
space. Secondly, one finds zero modes arising from open strings 
ending on the D5-branes. In the following we will discuss both sets of fields
in turn.  

\subsubsection{The closed string spectrum}\label{sec:closedStringSpectrum}

In order to determine the zero modes from the closed string 
sector, we first recall the massless bosonic  spectrum of the Type IIB theory.
It consists of the ten-dimensional metric  $g_{10}$, the anti-symmetric two-form 
$B_2$ as well as the dilaton $\phi$ in the NS--NS sector. The R--R sector
comprises the form fields $C_{0}$, $C_{2}$, $C_{4}$ $C_{6}$ and $C_{8}$
with field strengths \cite{Bergshoeff:2001pv}
\begin{equation} \label{10dfieldstrength}
        G^{(p)}=\begin{cases}
                                dC_{0}&         p=1,\\
                                dC_{p-1}-dB_2\wedge C_{p-2}& \text{else} \ .
                       \end{cases}
\end{equation}
Note that not all degrees of freedom in the $C_p$ are physical 
and we have to additionally impose the duality constraints  \cite{Bergshoeff:2001pv}
\begin{equation}\label{selfduality}
        G^{(1)}=\ast_{10}G^{(9)},\qquad G^{(3)}=(-1)\ast_{10}G^{(7)},\qquad
        G^{(5)}=\ast_{10}G^{(5)}\ .
\end{equation}

As in $\cN=2$ Calabi-Yau compactifications the four-dimensional fields arise
in the expansions of the ten-dimensional fields into harmonic forms of $Y$. 
However, in the orientifold setup only fields survive which are invariant
under the projection $\cO$ given in \eqref{OactionO5}. One first recalls that
$g_{10}$, $\phi$ as well as $C_2,\ C_6$ are even, while $B_2$ and $C_0,\ C_4\
, C_8$ are odd under the world-sheet parity operation $\Omega_p$. This implies 
that 
\beq \label{transf_sigma}
  \sigma^*g_{10}=g_{10}\ , \qquad \sigma^* B_2 = -B_2\ , \qquad \sigma^* \phi
  = \phi\ , \qquad \sigma^* C_p = (-1)^{(p+2)/2} C_p\ .
\eeq

The expansions of the ten-dimensional fields as well as of $J$ and $\Omega$ 
into harmonics of $Y$ have to be in accord with \eqref{transf_sigma} and \eqref{OactionO5}.
One thus splits the Dolbeault cohomology groups into the two eigenspaces $H^{(p,q)}_\pm(Y)$ under 
$\sigma^*$ with eigenvalues $\pm 1$, respectively.
We introduce a basis $(\omega_\alpha ,\omega_a)$ of $H_+^{(1,1)},\, H_-^{(1,1)}$    
with dual basis $( \tilde \omega^\alpha,\tilde \omega^a)$ of $H_+^{(2,2)},\,
H_-^{(2,2)}$ such that 
\beq
  \int_Y \omega_\alpha\wedge \tilde{\omega}^\beta=\delta^\beta_\alpha\ ,\qquad
  \int_Y\omega_a\wedge \tilde{\omega}^b=\delta^b_a\ ,
\eeq
where $\alpha,\beta = 1,\ldots,h^{(1,1)}_+$ and $a,b = 1,\ldots,h^{(1,1)}_-$.
Moreover, we denote by  $(\alpha_K,\beta^K)$, $(\alpha_{\tilde K},\beta^{\tilde K})$ a real symplectic basis of
$H^3_+$ and $H^3_-$, respectively. 
This basis is chosen such that the intersection pairings take the  form
\beq \label{def-alpha_beta}
        \int_Y \alpha_K\wedge \beta^L=\delta^L_K\ ,\qquad\int_Y \alpha_{\tilde
          K}\wedge \beta^{\tilde L}=\delta^{\tilde L}_{\tilde K}\ ,
\eeq
and vanish otherwise. 
Note that the holomorphic three-form $\Omega$ is contained in $H^3_+(Y)$, 
hence, $K=0,\dots, h^{(2,1)}_+$, but $\tilde K=1,\dots,h^{(2,1)}_{-}$.
Our conventions are summarized in Table~\ref{tab:cohomology-basis}.

\begin{table}[h!]
        \centering
        \begin{tabular}{|c|c||c|c|}
                \hline
                \rule[-0.2cm]{0cm}{0.6cm}  $(+1)$-Eigenspace& Basis & $(-1)$-Eigenspace& Basis\\
                \hline 
                \rule[-0.2cm]{0cm}{0.7cm}  $H^3_+(Y)$ & $\alpha_K,\beta^K$ & $H^3_{-}(Y)$ & $\alpha_{\tilde K},\beta^{\tilde K}$\\
                \hline
                 \rule[-0.2cm]{0cm}{0.7cm}   $H^{(1,1)}_{+}(Y)$ & $\omega_\alpha$ & $H^{(1,1)}_{-}(Y)$ & $\omega_a$\\
                \hline 
                 \rule[-0.2cm]{0cm}{0.7cm}    $H^{(2,2)}_{+}(Y)$ & $\tilde{\omega}^\alpha$ & $H^{(2,2)}_{-}(Y)$ & $\tilde{\omega}^a$\\
                \hline
        \end{tabular}
        \caption{Basis of the cohomology groups.}
        \label{tab:cohomology-basis}
\end{table}

To determine the four-dimensional bulk spectrum we use the 
cohomology basis of Table \ref{tab:cohomology-basis} and expand the NS--NS as well as the R--R 
fields. Let us start with the holomorphic three-form $\Omega$.
In accord with \eqref{OactionO5} we expand 
\beq
   \Omega = X^K(z) \alpha_K - \cF_{K}(z) \beta^K\ .
\eeq
The $2h^{(2,1)}_+ + 2$ coefficient functions $X^K,\cF_K$ are the periods of
$\Omega$. They can be expressed as period integrals for a symplectic homology 
basis $(A_K,B^K)$ dual to $(\alpha_K,\beta^K)$ as
\beq \label{def-AB}
  X^K = \int_{A_K} \Omega \ , \qquad \cF_K = \int_{B^K} \Omega\ .
\eeq
where $\int_{A_L} \alpha_K=\delta^L_K=-\int_{B^K} \beta^L$.
The periods depend on the 
complex structure deformations $z^\kappa$, $\kappa=1,\ldots,h^{(2,1)}_+$ of
$Y$. We denote the complex $h^{(2,1)}_+$-dimensional field space spanned by
$z^\kappa,\bar z^{\bar{\kappa}}$ by $\cM^{\rm cs}$. Infinitesimally the 
$z^\kappa$ parameterize the variations of the internal Calabi-Yau
metric with purely holomorphic or anti-holomorphic indices 
\beq \label{def-chi}
   \delta g_{ij} = \frac{i\mathcal{V}}{\int \Omega \wedge \bar{\Omega}}\
   \Omega_{j}^{\ \bar \imath \bar \jmath} \, (\bar
   \chi_{\bar \kappa})^{\phantom{i}}_{\bar \imath \bar \jmath i}\,  \delta
   \bar z^{\bar \kappa} \ ,
\eeq
where we introduced the string-frame volume $\mathcal{V} = \int_Y
d^6 y \sqrt{g}$ of the Calabi-Yau manifold $Y$. We also denoted by
$\chi_\kappa$ the basis of $H^{(2,1)}_+(Y)$. 
This cohomology thus also determines the change of $\Omega$ under complex
structure deformations as 
\begin{equation} 
        \partial_{z^\kappa}\Omega=\chi_{\kappa}-K_{\kappa}\Omega\ ,
\label{variationofomega}
\end{equation}
where $K_\kappa$ will later be identified with the first derivative of the
K\"ahler potential on $\cM^{\rm cs}$.
We note that there are only $h^{(2,1)}_+$ complex structure deformations
$z^\kappa$ which preserve \eqref{OactionO5}. In special coordinates they are expressed through the $h^{(2,1)}_+ +1$ periods 
$X^K$ as $z^\kappa = X^\kappa / X^0$. Here one uses the fact that $\Omega$ is only defined up to 
holomorphic rescalings. In the effective four-dimensional theory the $z^\kappa(x)$ will be complex scalar fields 
and correspond to bosonic components of $h^{(2,1)}_+$ chiral multiplets. 

Similarly, we proceed with the remaining NS--NS fields and expand
\begin{equation} \label{exp_JB}
        J=v^{\alpha}(x)\omega_{\alpha}\ ,\qquad B_2=b^a(x)\omega_a\ , \qquad
        \phi=\phi(x)\ ,
\end{equation}
where $(v^\alpha,b^a,\phi)$ are scalars in four space-time dimensions.
The R--R fields are expanded as
\begin{eqnarray}
        \label{eqn:kk-expansion-form-fields}
        C_6 &=& A_{(3)}^K\wedge\alpha_K+\tilde{A}^{(3)}_K\wedge\beta^K +
        \tilde{c}^{(2)}_\alpha \wedge \tilde\omega^\alpha+h\, m_6\ ,  \nn \\
        C_4 &=& V^{\tilde K}\wedge\alpha_{\tilde K}+U_{\tilde
        K}\wedge\beta^{\tilde K}+\tilde{\rho}^a_{(2)} \wedge
        \omega_a+\rho_a\tilde\omega^a\ ,\\
        \nn C_2 &=& \mathcal{C}_{(2)}+c^\alpha\omega_\alpha\ ,
\end{eqnarray}
where $m_6 = \Omega \wedge \bar \Omega/\int_Y \Omega \wedge \bar \Omega$ 
is a top form on $Y$ normalized such that $\int_Y m_6 =1$.
In \eqref{eqn:kk-expansion-form-fields} the $(A_{(3)}^K, \tilde{A}^{(3)}_K)$ are
three-forms, $( \tilde{c}^{(2)}_\alpha, \tilde{\rho}^a_{(2)},\mathcal{C}_{(2)})$ are two-forms, $(V^{\tilde K},
U_{\tilde K})$ are vectors and $(h,\rho_a,c^\alpha)$ are scalars in the
four non-compact dimensions of $\mathbb R^{1,3}$.

Let us comment on the general expansion 
\eqref{eqn:kk-expansion-form-fields} before turning to the D5-brane sector.
Note that due to the duality constraints \eqref{selfduality} not all degrees of freedom in
\eqref{eqn:kk-expansion-form-fields} are physical. On the level of the
four-dimensional effective action one can
eliminate half of the degrees of freedom in the R--R fields by introducing 
Lagrange multiplier terms. However, in order to couple the bulk fields to the
brane sector, it turns out to be useful to work with the democratic formulation
\eqref{eqn:kk-expansion-form-fields}. Only at the very end of our analysis we
will choose a set of physical degrees of freedom and eliminate the remaining
fields using~\eqref{selfduality}. This will leave us with $h^{(1,1)}_+$ chiral multiplets 
with bosonic components $(v^\alpha,c^\alpha)$, $h^{(1,1)}_-$ chiral multiplets with 
components $(b^a,\rho_a)$ and the chiral dilaton multiplet $(\phi,h)$. In addition, there 
are $h^{(2,1)}_-$ vector multiplets with vectors $V^{\tilde K}$, cf.\ Table \ref{tab:fieldcontent}.

A second point to note is that the expansion
\eqref{eqn:kk-expansion-form-fields} also contains three-form fields 
$(A_{(3)}^K, \tilde{A}^{(3)}_K)$. Clearly, in four space-time dimensions a
massless three-form does not carry dynamical degrees of freedom. However,
we will show that the inclusion of the three-forms is crucial to determine the
scalar potential of a compactification with background fluxes and
D-branes from a purely bosonic reduction. In case these terms are omitted a
fermionic reduction must be invoked to derive the induced brane and flux
superpotential as done, for example, for D7-branes and D5-branes in refs.~\cite{Jockers:2005zy,Martucci:2006ij}.

\subsubsection{The open string spectrum}
\label{openspectrum}

Let us now include space-time filling D5-branes into our setup. In general,
they can be arranged in a complicated way as long as the consistency constraints 
for the compactification are met. We consider a stack of $N$ D5-branes on a two-cycle $ \Sigma$ in 
$Y$. If $ \Sigma$ is in the fix-point set of the involution $\sigma$,
the D5-branes lie on top of an orientifold five-plane and $ \Sigma$ is its own
$\sigma$-image. More generally $\Sigma$ can be mapped to a two-cycle $ \Sigma'=\sigma( \Sigma)$ which is
not pointwise identical to $ \Sigma$.

In this work we will mostly focus on the simplest situation, for which $N=1$, $ \Sigma
\cap  \Sigma' =0$ and $ \Sigma,  \Sigma'$ are in different homology classes. Hence, we
consider one D5-brane on $ \Sigma$ and its image brane on $ \Sigma'$. For this
situation the pair of the D5-brane and its image D5-brane is merely an auxiliary description of a single
smooth D5-brane wrapping a cycle in the orientifold $Y/\mathcal{O}$. 
On $Y$ it is natural to define
\begin{equation}
    \Sigma_+ =  \Sigma+ \Sigma'\ , \qquad  \Sigma_- =  \Sigma -  \Sigma'\ ,
        \label{sigma-plus}
\end{equation}
where $ \Sigma_+$ is the union of $ \Sigma$ and $ \Sigma'$ while $ \Sigma_-$ contains the
orientation reversed cycle $\Sigma'$. Clearly, one finds that $\sigma(\Sigma_\pm)=
\pm \Sigma_\pm$.

Let us first discuss the degrees of freedom due to 
$U(1)$ Wilson lines arising from non-trivial one-cycles on the six-dimensional D5-brane
world-volume. These enter the expansion
of the $U(1)$ gauge boson $A(\xi)$ on the D5-brane as 
\begin{equation} \label{gauge-expansion}
        A(x,u^a)=A_{\mu}(x)dx^{\mu}P_{-}(u^a)+a_I(x)A^I(u^a)+\bar{a}_{\bar
        I}(x)\bar{A}^{\bar I}(u^a)\ .
\end{equation}
Here we introduce real coordinates $\xi$ on the world-volume of the D5 where 
we distinguish $\xi=(x,u^a)$, $a=1,2$, for the Minkowski space and the 
two-cycle $\Sigma_+$, respectively. We denote complex coordinates for $\Sigma_+$ 
by $u$, $\bar{u}$ in the complex structure induced by the ambient space $Y$, 
i.e.~by the complex coordinates $y^i$, $\bar{y}^{\bar \imath}$. The one-forms 
$A^I=A_{\bar u}^I d\bar{u},\,\bar{A}^{\bar I}=\bar{A}_u^{\bar I}du$ denote 
a basis of the Dolbeault cohomology
$H_{-}^{0,1} (\Sigma_+ )$ and $H_{-}^{1,0} (\Sigma_+ )$, respectively, and $P_{-}$
is the step function equaling $1$ on $\Sigma$ and $-1$ on $\Sigma'$. 
Note that generally the $U(1)$ field strength $F=dA$ can admit a background flux
$\langle F\rangle=f$. Since $F$ is negative under $\sigma$, this flux enjoys the expansion 
\beq \label{D5-flux}
  f=f^a\iota^{\ast}\omega_a=f^a(\iota^{\ast}\omega_a)_{u\bar u}du\wedge d\bar{u}\ ,
\eeq
where $\iota^{\ast}\omega_a$ are the pullbacks of the basis $\omega_a$ of
$H_-^{(1,1)}(Y)$ introduced in Table \ref{tab:cohomology-basis}. As we will recall
later on, $F$ naturally combines with the NS--NS B-field into the combination
$ \ell F- \iota^{\ast} B_2$ with $\ell =2\pi \alpha'$.

The dynamics of the D5-brane is more complicated and is encoded by fluctuations of the embedding map
$\iota:\Sigma_+ \hookrightarrow Y$. These fluctuations are described by
sections $\zeta$ of the normal bundle $N_Y\Sigma_+$ of $\Sigma_+$ and its conjugates $\bar\zeta$. 
In other words, they give rise to real sections $\hat\zeta$ in $H^0_+\left(\Sigma_+,N_Y^\bbR \Sigma_+\right)$
which enjoy the expansion
\begin{equation} \label{zeta_expansion}
        \hat\zeta=\hat \zeta^{\cA} \hat s_{\cA}= \zeta + \bar \zeta=\zeta^A \,s_A+\bar{\zeta}^{\bar
        A}\, \bar{s}_{\bar A}\ .
\end{equation}
Here the split into $\zeta$ and $\bar \zeta$ arises from the choice of complex structure on the real normal bundle $N_Y^{\mathbb{R}}\Sigma_+$ which decomposes into the holomorphic normal 
bundle $N_Y\Sigma_+$ and anti-holomorphic normal bundle
$\overline{N_Y\Sigma_+}$. In particular, we will mostly work with $\zeta \in
H^0_+\left(\Sigma_+,N_Y\Sigma_+\right)$ instead of its real counterpart $\hat \zeta$.
In  \eqref{zeta_expansion} we also introduced the real basis $\hat s_\cA$ and 
the complex basis $s_A$ and $\bar{s}_{\bar A}$ of the respective cohomology
groups. The coefficients $\zeta^A$ in this expansion become fields $\zeta^A(x)$ in 
the four-dimensional effective theory.

We conclude by summarizing the $\mathcal N=1$ field content in four dimensions emerging from the bulk and the brane sector in Table \ref{tab:fieldcontent}. The precise organization of these fields into $\mc N=1$ complex coordinates is postponed to section \ref{subsec:N=1data}. 

\begin{table}
\begin{center}
 \begin{tabular}{|c||c|c|c|c|c|}
 \cline{1-3} \cline{5-6}
      & \multicolumn{2}{c|}{{closed}   \rule[-0.2cm]{0cm}{0.6cm}  } & 
      &\multicolumn{2}{c|}{{open}}\\ \cline{1-3} \cline{5-6}
 \rule[-0.2cm]{0cm}{0.6cm}  Type & \quad Number \quad \ & Fields & & Number & Fields\\ \cline{1-3} \cline{5-6}
\multirow{4}*{chiral multiplet} &\T\B $h^{(1,1)}_+$ & $
     \rule[-0.2cm]{0cm}{0.6cm} \quad  t^\alpha=(v^\alpha,c^\alpha)$ \quad \ &&
        \multirow{2}*{\quad $h^0_+(\Sigma_+,N\Sigma_+)$\quad} & \multirow{2}*{$\zeta^A$}\\ \cline{2-3}
        & \T\B $h^{(1,1)}_- $ & 
         \rule[-0.2cm]{0cm}{0.6cm}  $P_a=(b^a,\rho_a)$ & & &\\ \cline{2-3} \cline{5-6}
        & \T\B $1$ &
         \rule[-0.2cm]{0cm}{0.6cm}  $S=(\phi,h)$ 
        &&\multirow{2}*{$h^{(1,0)}(\Sigma_+)$}&\multirow{2}*{$a_I$}\\
        \cline{2-3} 
        &\T\B  $h^{(2,1)}_+$ & $z^\kappa$ & & &   \\ \cline{2-3}
\cline{1-3} \cline{5-6}
vector multiplet & \T\B $h^{(2,1)}_-$& $V^{\tilde K}$ && $1$ & $A$\\
\cline{1-3} \cline{5-6}

 \end{tabular}
\caption{The  spectrum cast into multiplets of the four-dimensional $\mc N=1$ supersymmetry.}
\label{tab:fieldcontent}
\end{center}
\end{table}

\subsection{Special relations on the $\mathcal{N}=1$ moduli space} \label{relations}

In this section we discuss a subtlety in the decomposition \eqref{zeta_expansion}.
The notion of  $\zeta^A$ being a complex
scalar field depends on the background complex structure chosen on the
ambient Calabi-Yau $Y$, i.e.\ on the split \eqref{zeta_expansion}, ${N^\mathbb{R}}_Y\Sigma_+\otimes\mathbb{C}=N_Y \Sigma_+\oplus \overline{N_Y \Sigma_+}$, into holomorphic and anti-holomorphic parts.
To explore this dependence further it is natural to consider the contractions 
of the $s_A$ with the holomorphic $(3,0)$-form $\Omega$, the $(2,1)$-forms
$\chi_\kappa$ introduced in \eqref{def-chi} and their complex conjugates. 
In the background complex structure defined at $z_0$ we find, in the cohomology of $Y$ as well as in the cohomology of $\Sigma$, that
\beq
  s_A \lrcorner\Omega(z_0) = 0 \ , 
   \qquad \quad   s_A
  \lrcorner\bar \chi_\kappa(z_0) =0 \ ,\qquad \quad s_A \lrcorner \bar \Omega(z_0) = 0 \ .
\label{backgroundcs}
\eeq
These contractions vanish on $Y$ since there are no non-trivial $(2,0)$-forms
in $H^2(Y)$. Moreover, they also vanish on $\Sigma$ for a supersymmetrically
embedded D5-brane. As we will recall in section \ref{reduction_D5}, every two-form pulled back 
to $\Sigma$ has to be proportional to the $(1,1)$-K\"ahler form $J$.
Therefore, only $s_A \lrcorner\chi_\kappa$ can be a non-trivial $(1,1)$-form on $\Sigma$. Note that also $s_A \lrcorner\chi_\kappa$  is trivial in the
cohomology of $Y$ due to the primitivity of $H^{(2,1)}(Y)$.

However, in the four-dimensional effective theory we also have to allow for
possible fluctuations around the supersymmetric background configuration, including
those corresponding to complex structure deformations of $Y$.
The holomorphic three-form $\Omega$ as well as the complex scalars $\zeta$ are
then functions of the complex structure parameters $z^\kappa$. Now, the notion
of holomorphic and anti-holomorphic coordinates for $Y$ expressed by
$\Omega(z)$ has not to be aligned with the splitting into complex 
scalars (\ref{zeta_expansion}) in general.
To exemplify this, we consider the pullback
$\iota^\ast(s_A\lrcorner\Omega(z))$  on $\Sigma$. For $z=z_0+\delta z$ near a 
background complex structure $z_0$ we expand $\Omega(z)$ to linear order in 
$\delta z$ to obtain
\begin{equation} 
        \iota^\ast(s_A\lrcorner\Omega(z))= (1-K_{\kappa}\delta z^ {\kappa})
         \iota^\ast(s_A\lrcorner\Omega(z_0))+
         \iota^\ast(s_A\lrcorner\chi_\kappa(z_0))\delta z^\kappa
          =\iota^\ast(s_A\lrcorner\chi_\kappa(z_0))\delta z^\kappa,
        \label{eqn:expansion-omega-a}
\end{equation}
where we used (\ref{variationofomega}) and \eqref{backgroundcs}.
In other words, the form $s_A\lrcorner\Omega$ is a $(2,0)$-form on $\Sigma$ in the complex
structure $z$ but a $(1,1)$-form on $\Sigma$ in the complex structure $z_0$ to
linear order in the complex structure variation $\delta z$. Here we used the 
fact (\ref{backgroundcs}) that $s_A\lrcorner\Omega$ vanishes in the background
complex structure $z_0$ when the complex structure of $Y$ and $\Sigma$ are
aligned. 
However, a similar argument shows that
\begin{equation}
(s_A\lrcorner\bar\Omega)(z)=(s_A\lrcorner\bar\chi) (z^\kappa)= 0\ ,
\label{eqn:zeta-omega-bar-zero}
\end{equation}
even to linear order in $\delta z^\kappa$. These forms only appear at higher
order in the complex structure variations as we will discuss in section
\ref{sec:n1-special-geometry}.

The above considerations allow us to describe the metric deformations of the
induced metric $\iota^\ast g$ on the two-cycle $\Sigma_+$. In general, both the
complex structure deformations of $Y$ and the fluctuations of the embedding
map $\iota$ contribute. Here, we will discuss those variations
$\delta(\iota^\ast g)$ originating from complex structure deformations and
postpone the analysis of all possible metric variations to section
\ref{DBIreduction}. Analogously to (\ref{def-chi}) the complex structure
deformations on $\Sigma_+$ are encoded in the purely holomorphic metric variation
\begin{equation} \label{internalmetricvariation}
 \iota^\ast(\delta g)_{uu}= \frac{2i v^{\Sigma}}{\int\Omega\wedge\bar{\Omega}}
 \iota^\ast( s_A\lrcorner\Omega)_{uu}(\iota^\ast g)^{u\bar u}
  \iota^\ast(\bar{s}_{\bar B}\lrcorner \bar{\chi}_{\bar{\kappa}})_{\bar{u}u}
  \mathcal{G}^{A\bar B}\ \delta \bar{z}^{\bar \kappa}\ .
\end{equation}
Here we have introduced the volume of the holomorphic two-cycle $\Sigma_+$ as
\begin{equation}
 v^{\Sigma}=\int_{\Sigma_+}d^2u\sqrt{g}=\int_{\Sigma_ +}\iota^\ast J
\end{equation}
and a natural hermitian metric $\mathcal{G}_{A\bar B}$ given by
\begin{equation} \label{openmetric}
 \mathcal{G}_{A\bar B}=-\frac{i}{\mathcal V}\int_{\Sigma_+}s_A\lrcorner \bar{s}_{\bar B}\lrcorner (J)\iota^\ast J.
\end{equation}
We will show later on that it can be obtained by dimensional reduction, cf.~section \ref{DBIreduction}. Thus, it can be identified with the metric for the moduli $\zeta$ on the open string moduli space and is independent of the coordinates $u,\bar u$.

The metric variation \eqref{internalmetricvariation} can be explained by application of some useful formulas for the open string moduli space. 
First, we use the fact that $H^{(1,1)}(\Sigma_+)$ is spanned by the pullback $\iota^\ast J$. This can be exploited to rewrite the pullback of any closed $(1,1)$-form $\omega$ to $\Sigma_+$ in cohomology, cf.~\eqref{pullbackformula}. Especially for $\iota^\ast(s_A\lrcorner \chi_{\kappa})$ we obtain
\begin{equation}
        \iota^\ast(s_A\lrcorner \chi_{\kappa})=\frac{\iota^\ast J}{v^{\Sigma}}\int_{\Sigma_+}\iota^\ast(s_A\lrcorner \chi_{\kappa})\ ,
\end{equation}
which can be written after multiplication with $\mc V^{-1}\mc G^{A\bar B}g(s_C,\bar s_{\bar B})$ and using \eqref{openmetric} as 
\begin{equation} \label{metricidentity}
 \int_{\Sigma_+} \iota^\ast(s_A \lrcorner \chi_{\kappa})=-\frac{v^{\Sigma}}{\mathcal V}\int_{\Sigma_+}g(s_A,\bar{s}_{\bar B})\mathcal{G}^{\bar{B}C} \iota^\ast(s_C\lrcorner\chi_{\kappa}).
\end{equation}
We evaluate this for every choice of $s_A$ and compare the coefficients on both sides to relate the metric on the normal bundle $N_Y\Sigma$ and the metric $\mc G^{A\bar B}$.  

Thus, the identity \eqref{metricidentity} allows us to infer the metric variations \eqref{internalmetricvariation} from the complex structure deformations on $Y$. First, we consider the pullback to $\Sigma_+$ of the metric variations $\delta g_{ij}$, cf.~\eqref{def-chi}, of the ambient Calabi-Yau $Y$
\begin{equation}
        \iota^\ast(\delta g)_{u u}=\frac{i\mathcal{V}}{\int \Omega \wedge \bar{\Omega}}\
   \Omega_{u}^{\ \bar \imath \bar \jmath} \, (\bar
   \chi_{\bar \kappa})^{\phantom{i}}_{\bar \imath \bar \jmath u}\,  \delta
   \bar z^{\bar \kappa}.
\end{equation}
Then we replace, motivated by \eqref{metricidentity}, the inverse metric $g^{i\bar \jmath}$ occurring in the contraction of $\bar{\chi}_{\bar \kappa}$ and $\Omega$ by $s^i_As^{\bar \jmath}_{\bar B}\mc G^{A\bar B}$ to obtain our ansatz for the induced metric deformation on $\Sigma_+$ given in \eqref{internalmetricvariation}.

However, there are some remarks in order. Since there are no $(2,0)$-forms 
on $\Sigma_+$ in the background complex structure $z_0$, the form 
$\iota^\ast(s_A\lrcorner\Omega)$ should vanish identically. Thus, in 
order to make sense of the metric variation (\ref{internalmetricvariation}) 
we have to consider it, following the logic of (\ref{eqn:expansion-omega-a}), 
in the complex structure $z=z_0+\delta z$. 
Applying this to (\ref{internalmetricvariation}) we expand $\delta(\iota^\ast g)$ to linear
order in $\delta z$, i.e.~$\iota^\ast(\delta g)_{uu}(z)=
\iota^\ast(\delta g)_{uu}(z_0)+\iota^\ast(\delta g)_{u\bar{u}}(z_0)\cdot\delta z$, to obtain
\begin{equation} \label{internalmetricvariation2}
 \iota^\ast(\delta g)_{u\bar{u}}(z_0)=
 \frac{2i v^{\Sigma}}{\int\Omega\wedge\bar{\Omega}} 
 \iota^\ast( s_A\lrcorner\chi_{\kappa})_{\bar{u}u}(\iota^\ast g)^{u\bar u}
  \iota^\ast(\bar{s}_{\bar B}\lrcorner \bar{\chi}_{\bar{\kappa}})_{\bar{u}u}
  \mathcal{G}^{A\bar B}\ \delta z^{\kappa}\delta \bar{z}^{\bar \kappa}\ .
\end{equation}
Here we emphasize the change in type from purely holomorphic indices $\delta
g_{uu}$ at $z$ to mixed type $\delta g_{u\bar u}$ at $z_0$. It is important to
note that there are no metric deformations linear in the complex structure
parameter $\delta z$ nor any of pure type.

We have just stressed that the analysis of the open string moduli space
depends on the chosen background complex structure encoded by the moduli $z^\kappa$.
It is hence natural that the complex structure parameters $z^\kappa$ of $Y$
and the open string moduli $\zeta^A$ should be treated on an equal footing to
characterize the structure of the $\cN=1$ field space. This led the authors of
refs.~\cite{Lerche:2002ck,Lerche:2002yw} to introduce $\cN=1$ special geometry for open-closed fields 
and we will explore this in our context further in section 
\ref{sec:n1-special-geometry}. In the next sections we derive 
the four-dimensional effective D5-brane action and 
show that the superpotential is naturally encoded by the 
forms $s_A\lrcorner\Omega$ and $s_A\lrcorner\chi_\kappa$.

\subsection{Reduction of the D5-brane action \label{reduction_D5}}

Now we are prepared to derive the four-dimensional effective action of the
D5-brane in a Calabi-Yau orientifold. It is obtained by reducing the 
bulk supergravity action $S_{\text{IIB}}$ as well as the effective D-brane
actions using a Kaluza-Klein reduction. 
The string-frame Type IIB action is used in its democratic form 
\begin{eqnarray} \label{bulkaction}
        S_{\text{IIB}}^{\text{SF}}=\int\,\tfrac12 e^{-2\phi}R *_{10}
        1-\tfrac14
        \,e^{-2\phi}\left(8d\phi\wedge\ast_{10}d\phi-H\wedge\ast_{10}
        H\right)+\tfrac18\,\sum_{p\ \text{odd}}G^{(p)}\wedge\ast_{10}G^{(p)},
\end{eqnarray}
where $H = dB_2$ and the R--R field strengths have been introduced in \eqref{10dfieldstrength}
and obey the duality constraints \eqref{selfduality} imposed on the level of the equations
of motion. In addition, one includes the string-frame D5-brane action 
\beq \label{eqn:effectiveaction}
        S^{\text{SF}}_{\text{D5}}=-\mu_5\int_{\mathcal{W}}d^6\xi
        e^{-\phi}\sqrt{-\text{det}\left(\iota^{\ast}\left(g_{10}+B_2\right)-\ell F\right)}+\mu_5\int_{\mathcal{W}}\sum_{q\ \text{even}} \iota^\ast(C_q) \wedge e^{\ell F-\iota^\ast(B_2)} \ .
\eeq
The two parts of $S^{\text{SF}}_{\text{D5}}$ are the Dirac-Born-Infeld and
Chern-Simons action, respectively. The Kaluza-Klein reduction of the bulk
action \eqref{bulkaction} on the orientifold background introduced in
section \ref{4Dspectrum} has been carried out in ref.~\cite{Grimm:2004uq} and
we refer to this work for further details. Here we will mainly
concentrate on the reduction of the D5-brane action
\eqref{eqn:effectiveaction} and later include the contributions entirely due
to bulk fields in the determination of the $\cN=1$ characteristic functions.

It is important to note that there are conditions on the D5-branes in a 
supersymmetric orientifold background. These calibration conditions have 
been determined in \cite{Becker:1995kb,Marino:1999af}. For vanishing background fields 
these conditions imply $\Sigma$ to be a holomorphic curve, i.e.~the embedding 
$\iota$ to be a holomorphic map obeying ${\partial \bar{y}^{\bar \imath}(u^a)}/{\partial u}=
{\partial y^{ i}(u^a)}/{\partial \bar u}=0$. In particular, this implies a
natural choice of complex structure on $\Sigma$ by aligning it with the
ambient complex structure using the holomorphic embedding. As a consequence
the volume form on $\Sigma$ is just proportional to the pullback of $J$, a
well-known fact for complex submanifolds of K\"ahler manifolds. 
Moreover, the D5-branes have to obey the same calibration conditions as the 
O5-planes arising as fix-point set of the holomorphic involution $\sigma$.
This fixes the supersymmetric calibration condition also in the presence of
a non-vanishing NS--NS B-field and a background gauge field configuration completely. 
Explicitly,  the calibration conditions on the D5-brane background reduce to 
\begin{equation}
        du^1\wedge du^2\sqrt{-\text{det}\left(\iota^{\ast}\left(g_{10}+B_2\right)-\ell
        F_{ab} 
        \right)}=\iota^{\ast} J+i \langle \ell F-\iota^{\ast} B_2 \rangle\ ,
\end{equation}
where we restrict the consideration to the internal coordinates of the D5-brane.
This implies by separating into imaginary and real part the two conditions 
\begin{equation} \label{calibration1}
        \langle \iota^{\ast} B_2-\ell F\rangle =0\,,\qquad
         du^1\wedge du^2\sqrt{-\text{det}\left(\iota^{\ast} g_{10}\right)}=\iota^{\ast} J\,.
\end{equation}
Once again, these formulas are given in the string frame and can be
translated to the Einstein frame by multiplying the second equation of 
(\ref{calibration1}) by $e^{\phi}$. Note that the first condition in \eqref{calibration1}
implies that a non-vanishing flux $f$ on the D5-brane as in \eqref{D5-flux} has to be
cancelled in the background by a non-vanishing B-field. Clearly, we still need
to include the variations of $B_2$ around such a vacuum configuration. We will denote 
the variations of the two-form part of $\iota^{\ast} B_2-\ell F$ on $\Sigma_-$ by 
\beq \label{def-Ba}
   \cB^\Sigma =  \cB^a \, \int_{\Sigma_-}\iota^{\ast} \omega_a = \int_{\Sigma_-}
   \iota^{\ast} B_2-\ell F\ , \qquad \qquad \cB^a(x) = b^a(x)-\ell f^a  \ ,
\eeq  
where $f^a$ is the background flux \eqref{D5-flux} of the D5-brane field
strength.

\subsubsection{Dirac-Born-Infeld action and tadpole cancellation \label{DBIreduction}}

In the following we will perform the Kaluza-Klein reduction of the
Dirac-Born-Infeld action given in \eqref{eqn:effectiveaction}.
Firstly, we expand the determinant using  
\begin{equation} \label{taylor}
        \sqrt{\text{det}\left(\mathfrak{A}+\mathfrak{B}\right)}=
         \sqrt{\text{det}\mathfrak{A}}\cdot\left[1+\tfrac{1}{2}\tr
        \mathfrak{A}^{-1}\mathfrak{B}+\tfrac{1}{8}\left(\left(\tr\mathfrak{A}^{-1}\mathfrak{B}\right)^2
         -2\tr\left(\mathfrak{A}^{-1}\mathfrak{B}\right)^2\right)+\ldots\right]\ .
\end{equation}
Here, the matrix $\mathfrak{A}$ encodes the background configuration of the
Minkowski spacetime and the six-dimensional Calabi-Yau for which we can 
use \eqref{calibration1}. Additionally, $\mathfrak{B}$ contains the fluctuations around this
background. These are precisely the fluctuations of the embedding $\iota$ of
the two-cycle $\Sigma_+$ parametrized by the fields $\zeta^A$ of
\eqref{zeta_expansion}, the Wilson lines $a_I$ introduced in
\eqref{gauge-expansion} as well as the perturbations about the calibrated NS--NS B-field defined in (\ref{calibration1}) and about the background complex structure.
We use the normal coordinate expansions of the metric (\ref{eqn:metric}) and the NS--NS B-field (\ref{exp_JB}) on the D5-brane world-volume as well as the metric variation $\delta (\iota^\ast g)_{u\bar u}$ of (\ref{internalmetricvariation2}) to obtain 
\begin{eqnarray}
        \iota^{\ast}g_{10}&=& \mathcal{V}^{-1} e^{2\phi}\eta_{\mu
          \nu}dx^{\mu}\cdot dx^{\nu} +(\iota^\ast g+\delta (\iota^\ast g))_{u\bar u} du\cdot d\bar{u}+
          g(\partial_{\mu}\zeta,\partial_{\nu}\bar{\zeta})dx^{\mu}\cdot dx^{\nu}\;,\\
\iota^{\ast}B_2-\ell \cF &=&\cB^a\iota^{\ast}\omega_a-\ell F +
      \cB^a\, \iota^{\ast}\omega_a(\partial_{\mu}\zeta,\partial_{\nu}\zeta)dx^{\mu}\wedge dx^{\nu}\;\label{B-fieldexp},
\label{pullbackvariation}
\end{eqnarray}
where $\ \cdot \ $ is the symmetric product and $\cV$, $g_{u\bar u}$ are the string frame volume and the induced hermitian metric on $\Sigma_+$. Note that the Minkowski metric $\eta$ is rescaled to the four-dimensional
Einstein frame\footnote{Recall
  that the four-dimensional metric in the Einstein frame $\eta$ is related to
  the string frame metric $\eta^{\rm SF}$ via $\eta = e^{-2\phi} \cV\, \eta^{\rm SF}$.}. The combination $\cB^a$ containing the fluctuations of the internal B-field and
the D5-brane background flux was introduced in \eqref{def-Ba}.
Using this we obtain 
\begin{eqnarray}
        \mathfrak{A}&=&
                    \begin{pmatrix}
                                \mathcal{V}^{-1}e^{2\phi}\eta_{\mu\nu} & 0&0\\
                        0 & 0& g_{u\bar u}\\
                        0& g_{u\bar u}& 0
                     \end{pmatrix},\\
        \mathfrak{B}&=&\begin{pmatrix}
                (2g+\mathcal{B}^a\omega_a)(\partial_{\mu}\zeta,\partial_{\nu}\bar{\zeta})-\ell F_{\mu\nu}& -\ell\partial_{\mu} \bar{a}_{\bar J}\bar{A}_u^{\bar J} &-\ell\partial_{\mu}a_IA_{\bar u}^I\\
                        -\ell\partial_{\nu} \bar{a}_{\bar J}\bar{A}_u^{\bar J}&0&(\delta g+\tfrac12\cB^a\omega_a)_{u\bar u}\\
                        -\ell\partial_{\nu}a_IA_{\bar u}^I& (\delta g-\tfrac12\cB^a\omega_a)_{u\bar u}& 0
                       \end{pmatrix}\;,
\end{eqnarray}
where we omitted the pullback $\iota^\ast$ for notational convenience.
Only the terms
\begin{equation}
        \tfrac12\tr\mathfrak{A}^{-1}\mathfrak{B}-\tfrac14\tr\big((\mathfrak{A}^{-1}\mathfrak{B} )^2\big)
\end{equation}
of the Taylor expansion (\ref{taylor}) contribute to the effective action up to quadratic order in the fields.
We insert the result into the first part of (\ref{eqn:effectiveaction})
and use \eqref{calibration1} to obtain the four-dimensional action 
\begin{equation} \label{eqn:DBI}
         S_{\text{DBI}}
         =\text{-}\mu_5\int\Big[\tfrac{\ell^2e^{-\phi}}{4}  v^{\Sigma}F\wedge \ast
         F+\tfrac{\ell^2e^{\phi}}{\mathcal{V}}\mathcal{C}^{I\bar J}
         da_I\wedge \ast d\bar{a}_{\bar
         J}+\tfrac12e^{\phi}\mathcal{G}_{A\bar B}d\zeta^A\wedge\ast d\bar{\zeta}^{\bar
         B}+V_{\rm DBI}\ast 1\Big]
\end{equation}
in the four-dimensional Einstein frame.
The potential term in \eqref{eqn:DBI} is of the form
\beq
  V_{\rm DBI} = \frac{e^{3\phi}}{2\mathcal{V}^2}\Big(v^{\Sigma}+\frac{2i\mathcal{G}^{A\bar B}}{\int\Omega\wedge\bar{\Omega}}\int_{\Sigma_+}s_A\lrcorner \chi_{\kappa}\int_{\Sigma_+}\bar s_{\bar B}\lrcorner \bar \chi_{\bar\kappa}\delta z^{\kappa}\delta \bar{z}^{\bar \kappa}+\frac{(\mathcal{B}^{\Sigma})^2}{8v^{\Sigma}} \Big)\ .\label{DBIpotential}
\eeq  
   In the following we
   will discuss the separate terms appearing in the action $S_{\text{DBI}}$ in turn.

The first term in (\ref{eqn:DBI}) is the kinetic term for the $U(1)$ gauge boson
$A$. The gauge coupling is thus given by $1/g_{\rm D5}^2 = \tfrac 12\mu_5 \ell^2 e^{-\phi}
v^{\Sigma}$, where $v^{\Sigma}$ is
the volume of the two-cycle $\Sigma_ +$ using the calibration \eqref{calibration1}.
The second term is the kinetic term for the Wilson line moduli $a^I$. The
appearing metric takes the form 
\beq
   \mathcal{C}^{I\bar J}=\frac12\int_{\Sigma_ +}A^I\wedge\ast_2\bar{A}^{\bar J}=\frac i2\int_{\Sigma_ +}A^I\wedge\bar{A}^{\bar J}\,,
\eeq
where we have used $\ast_2 \bar{A}^{\bar J}=i\bar{A}^{\bar J}$ on the
$(1,0)$-form basis introduced in \eqref{gauge-expansion}.
The third term in \eqref{eqn:DBI} contains the field space metric for the
deformations $\zeta^A$ and is of the form
\beq  \label{eqn:metrics}
        \displaystyle\mathcal{G}_{A\bar
 B}=-\frac{i}{2\mathcal{V}}\int_{\Sigma_+}s_A\lrcorner \bar s_{\bar
 B}\lrcorner\left(J\wedge
 J\right)=\frac{\mathcal{K}_\alpha}{2\mathcal{V}}\mathcal{L}_{A\bar
 B}^{\alpha}\ , \qquad \quad
        \displaystyle\mathcal{L}_{A\bar
 B}^\alpha=-i\int_{\Sigma_+}s_A\lrcorner\bar{s}_{\bar
 B}\lrcorner \tilde\omega^\alpha \ ,
\eeq
where $\mathcal{K}_\alpha=\int \omega_\alpha\wedge J\wedge J$ and we have used 
$J\wedge J=\mathcal{K}_\alpha\tilde{\omega}^\alpha$.

Finally, let us comment on the potential terms $V_{\rm DBI}$. In fact, the first of the three terms represents an NS--NS tadpole and takes the form of a
D-term. To guarantee a consistent compactification with D-branes, we have to
ensure R--R as well as  NS--NS tadpole cancellation.
Hence the two-cycle $\Sigma_+$ wrapped by the D5-brane has to lie in the 
same homology class as an O5-plane arising from the fix-points of $\sigma$.
Consequently, we have to add the contribution of the orientifold plane
\begin{equation} \label{eqn:orieaction}
        S_{\text{ori}}^{\text{SF}}=\mu_5\int_{\mathcal{W}_{\text{orie}}}d^6\xi
        e^{-\phi}
        \sqrt{-\text{det}\left(\tilde\varphi^{\ast}\left(g_{10}+B\right)\right)}\
        \quad  \rightarrow
        \quad S_{\text{ori}}^{\text{EF}}  = \mu_5 \int\frac{e^{3\phi}}{2\mathcal{V}^2}v^{\Sigma}\ast 1\ ,
\end{equation}
to the action (\ref{eqn:DBI}). Here we again applied a calibration
condition of the form \eqref{calibration1} to obtain the two-cycle volume $v^{\Sigma}$.
Having rescaled $S_{\text{ori}}^{\text{SF}}$ into the Einstein frame 
one  compares it with
(\ref{eqn:DBI}) and notes that the O5-plane contribution precisely cancels the
D-term potential of the D5-brane. 

The last two terms in $V_{\rm DBI}$ describe deviations of the calibration conditions
(\ref{calibration1}). The first potential term accounts for the metric
deformations (\ref{internalmetricvariation2}) induced by the change of the
ambient complex structure and the second term describes the field fluctuation
$\mathcal{B}^a$ of the NS--NS B-field of (\ref{def-Ba}). Later on, we will
show that this term is actually a D-term consistent with the 
analysis of \cite{Douglas:2000ah}. Clearly, both terms vanish at the supersymmetric ground
state with the calibration conditions (\ref{calibration1}).
Let us comment on the dimensional reduction yielding these two terms.
The evaluation of $\text{Tr}\mathfrak{A}^{-1}\mathfrak{B}$ in the expansion (\ref{taylor}) of the DBI-action yields a term given by 
\begin{equation}
        \delta \cL_{\delta g}=\frac{ie^{3\phi}v^{\Sigma}\mathcal{G}^{A\bar B}}{\cV^2\int \Omega\wedge\bar{\Omega} }\delta z^\kappa\delta \bar{z}^{\bar \kappa}\int_{\Sigma_+}\iota^\ast( s_A\lrcorner\chi_{\kappa})_{\bar{u}u}(\iota^\ast g)^{u\bar u}(\iota^\ast g)^{u\bar u}
  \iota^\ast(\bar{s}_{\bar B}\lrcorner \bar{\chi}_{\bar{\kappa}})_{\bar{u}u}\ \iota^\ast J\ . 
\end{equation}
This is the only contribution to the four-dimensional effective action originating from the metric variation $\delta (\iota^\ast g)$ of (\ref{internalmetricvariation2}) that is relevant at our lowest order analysis.
As discussed before, cf.~section \ref{relations}, the $(1,1)$-form $\iota^\ast J$ is essentially the only non-trivial element in the cohomology $H^{(1,1)}(\Sigma_+)$. Thus, we can rewrite the pullback of any closed $(1,1)$-form $\omega$ to the two-cycle $\Sigma_+$ as
\begin{equation}
        \iota^\ast \omega=\frac{\int_{\Sigma_+} \omega}{v^{\Sigma}}\iota^\ast J
\label{pullbackformula}
\end{equation}
in cohomology, where we used again $\int_{\Sigma_+} \iota^\ast J=v^{\Sigma}$. In particular, we can apply this to the closed $(1,1)$-forms $s_A\lrcorner \chi_{\kappa}$ to obtain the second term of $V_{\rm DBI}$ given in (\ref{DBIpotential}).
Considering the fluctuation $\mathcal{B}^a$, the only contribution arises from $\text{Tr}(\mathfrak{A}^{-1}\mathfrak{B})^2$ in (\ref{taylor}). Then, we obtain the
four-dimensional effective term  
\begin{equation}
\delta \cL_{\cB}=  \frac{e^{3\phi}}{16 \cV^2 }\, \cB^a \cB^b
\int_{\Sigma_+}(\iota^\ast \omega_{a})_{u\bar u}\ (\iota^\ast\omega_{b})_{u\bar u}\ 
g^{u\bar u}g^{u\bar u}\iota^\ast J\ .
\end{equation}
Again we use (\ref{pullbackformula}) to expand 
$P_-\ \cB^a \iota^\ast \omega_a=\cB^{\Sigma}\, {\iota^\ast J}/{v^{\Sigma}}$
in the cohomology $H^{(1,1)}(\Sigma_+)$ and obtain the geometrical dependence of the volume $v^{\Sigma}$ of the cycle
as given in \eqref{DBIpotential}. Here, $P_-$ again denotes the step function introduced in (\ref{gauge-expansion}).
Later on in section \ref{subsec:N=1data}, we show explicitly that the above
results of the dimensional reduction are necessary to match the F- and D-term potential
arising from a superpotential $W$ and a gauging of a shift symmetry by the $U(1)$ vector $A$ on the D5-brane, respectively.

\subsubsection{Chern-Simons action}

Let us now turn to the dimensional reduction of the Chern-Simons part of the D5-brane action.
For this purpose we need the normal coordinate expansion of the R--R fields
(\ref{eqn:kk-expansion-form-fields}) pulled back to the world-volume of the
D5-brane. Here we will only display the relevant terms for the reduction of
the Chern-Simons action which read
\begin{eqnarray}
   (\iota^* C_p)_{i_1\ldots i_p} &=& \tfrac{1}{p!}C_{i_1 \ldots i_p} +\tfrac{1}{p!}\zeta^n\partial_nC_{i_1 \ldots i_p}-\tfrac{1}{(p-1)!}\nabla_{i_1}\zeta^nC_{ni_2 \ldots i_p}+\tfrac{1}{2p!}\zeta^n\partial_n(\zeta^m\partial_mC_{i_1 \ldots i_p}) \\
 &-&\tfrac{1}{(p-1)!}\nabla_{i_1}\zeta^n\zeta^m\partial_m C_{ni_2 \ldots i_p}+\tfrac{1}{2(p-2)!}\nabla_{i_1}\zeta^n\nabla_{i_2}\zeta^m C_{nmi_3 \ldots i_p}+\tfrac{p-2}{2p!}R_{ni_1 m}^{j}\zeta^n\zeta^m C_{ji_2 \ldots i_p} \ ,\nonumber
\end{eqnarray} 
where the indices $i_{n}$ label the coordinates  $\xi^{i_n}$ on the D5-brane world volume.
Inserting this expansion into the Chern-Simons part of (\ref{eqn:effectiveaction}),
one finds up to second order 
\begin{eqnarray} \label{eqn:CS}
        S_{\text{CS}} &=& \mu_5\int \Big[\tfrac{\ell^2}{4}  c^\Sigma F\wedge F 
        -\tfrac{\ell}{2} d(\tilde{\rho}_{(2)}^\Sigma-
        \mathcal{C}_{(2)}\mathcal{B}^\Sigma)\wedge A 
        +\tfrac{i}{4}\, \mathcal{L}_{A\bar B}^\alpha\,
        d\tilde{c}^{(2)}_\alpha\wedge (d\zeta^A\bar\zeta^{\bar B}
        -d\bar{\zeta}^{\bar{B}}\zeta^A)\nonumber \\
        &\phantom{=}&\phantom{\mu_5} 
         -\ell^2 \tfrac{i}{2} \mathcal{C}^{I\bar J}\, 
          d\mathcal{C}_{(2)}\wedge (da_I\bar{a}_{\bar J}-d\bar{a}_{\bar J}a_I )
          - \tfrac{i}{4} \cL_{a b A\bar B}\,
        d(\cB^a \tilde{\rho}_{(2)}^b)\wedge(d\zeta^A\bar{\zeta}^{\bar B}- d\bar{\zeta}^{\bar B}\zeta^A) \nonumber \\
        &\phantom{=}&\phantom{\mu_5}
        +\tfrac12(\mathcal{N}_{\mc AK}\, A_{(3)}^K
        +\mathcal{N}_{\mc A}^K\,  \tilde A ^{(3)}_K)\wedge d\hat \zeta^{\mc A}
        -\tfrac\ell2 \hat \zeta^{\mc A} (\mathcal{N}_{\mc A\tilde K}\, dV^{\tilde K}
         + \mathcal{N}_{\mc A}^{\tilde K}\, dU_{\tilde K})\wedge F
        \Big]\ ,
        \label{eqn:eff-cs-action}
\end{eqnarray}
where $\cB^\Sigma,\cB^a$ are introduced in \eqref{def-Ba} and we similarly define
$\tilde{\rho}_{(2)}^\Sigma = \int_{\Sigma_-} C_4$ as well as $c^\Sigma=\int_{\Sigma_+}C_2$.  
In the action $S_{\text{CS}}$ we also used the abbreviations
\beq \label{eqn:n-tilde}
    \displaystyle \mathcal{N}_{\mc AK}=\int_{\Sigma_+} \hat s_{\mc A}
     \lrcorner\alpha_K\ , \qquad 
    \mathcal{N}_{\mc A}^K = \int_{\Sigma_+}\hat s_{\mc A}
    \lrcorner\beta^K\ ,\qquad \mathcal{N}_{\mc A\tilde K}=\int_{\Sigma_-} \hat s_{\mc A}
    \lrcorner\alpha_{\tilde K}\ , \qquad \mathcal{N}_{\mc A}^{\tilde K} =
     \int_{\Sigma_-} \hat s_{\mc A} \lrcorner\beta^{\tilde K}\ ,   
\eeq
where the forms and their orientifold parity can be found in Table
\ref{tab:cohomology-basis}. 
We also evaluated the coupling 
\beq  \label{eqn:symbols-cs-eff-action}
  \cL_{ab A\bar B}=-i \int_{\Sigma_+}s_A\lrcorner \bar s_{\bar
    B}\lrcorner(\omega_a \wedge \omega_b) = \cL^\alpha_{A \bar B} \cK_{\alpha
    a b}\ ,
\eeq
where $\cL^{\alpha}_{A \bar B}$ was introduced in \eqref{eqn:metrics} and $\cK_{\alpha
ab} = \int_Y \omega_\alpha\wedge \omega_a \wedge \omega_b$ are the only non-vanishing triple intersection numbers
involving the negative $(1,1)$-forms $\omega_a$ of Table \ref{tab:cohomology-basis}.
We note that in the action \eqref{eqn:CS} and the definitions \eqref{eqn:n-tilde} 
we have used the expansion $\hat \zeta = \hat \zeta^\cA \hat s_\cA$ into a real basis $\hat s_\cA$
given in \eqref{zeta_expansion}. Clearly, the expressions involving $ \hat \zeta^\cA$, $\hat
s_\cA$ are readily rewritten into complex coordinates
$\zeta^A,\bar \zeta^A$.
Let us also recall that in general both combination $\Sigma_+$ and $\Sigma_-$
occur in (\ref{eqn:n-tilde}) depending on whether the integrand
transforms with a positive or negative eigenvalue under the involution
$\sigma$. However, terms involving $\Sigma_-$ can by translated to $\Sigma_+$
by using the function $P_{-}(y)$ introduced after \eqref{gauge-expansion}. 

Let us now discuss the  interpretation of the different terms appearing in the
action (\ref{eqn:CS}). The first term in $S_{\text{CS}}$ corresponds to the theta-angle 
term of the gauge theory on the D5-brane and thus contains the imaginary
part of the gauge-kinetic function. The second term is a Green-Schwarz term
which indicates the gauging of the scalar fields dual to the two-forms 
$\tilde{\rho}^a_{(2)}$ and $\cC_{(2)}$ with the D5-brane vector field $A$.
In fact, we will show in section \ref{subsec:N=1data} that this term indeed induces a 
gauging of one chiral multiplet in the four-dimensional spectrum and that the
corresponding D-term is precisely the one encountered in the reduction of the
DBI action in section \ref{DBIreduction}.

The interpretation of the remaining terms in \eqref{eqn:CS} is of more
technical nature. The third, fourth and fifth terms are mix terms which will
contribute in the kinetic terms of the scalars $c^{\alpha}$, $h$ and $\rho_a$ dual to the
two-forms $\tilde c^{(2)}_\alpha$, $\cC_{(2)}$ and $\tilde \rho^a_{(2)}$. In section
\ref{subsec:N=1data} they will help us to identify the correct complex
coordinates which cast the kinetic term into the standard $\cN=1$ form.
The sixth term contains the four-dimensional three-forms $A_{(3)}^K$ and $\tilde A
^{(3)}_K$. We will show in the next section \ref{subsec:fluxpotential} that
these terms are crucial in the calculation of the scalar potential. 
Finally, the last term in $S_{\text{CS}}$ indicates a mixing of the field strength 
on the D5-brane with the $U(1)$ bulk vector fields $V_K,U^K$. The precise form of 
the redefined gauge-couplings will be discussed in appendix \ref{kinmix}.

\subsection{The scalar potential \label{subsec:fluxpotential}}

In this section we will compute the scalar potential of the four-dimensional
effective theory. The potential due to 
background R--R and NS--NS fluxes $F_3=\langle dC_2 \rangle$ and $H_3=\langle dB_2 \rangle$ has already 
been studied in ref.~\cite{Grimm:2004uq}. Here we will show that there are
additional contributions in the presence of the space-time filling D5-branes. 

 A first contribution to the scalar potential is induced by the couplings 
 of the three-forms $A_{(3)}^K$ and $\tilde{A}^{(3)}_K$
 in the Chern-Simons action \eqref{eqn:CS}. Here it is crucial to keep 
 these forms in the spectrum despite the fact that a 
 massless three-form has no propagating degree of freedom in  four dimensions. 
 Moreover, if this potential is  treated quantum mechanically, as described
 in ref.~\cite{Beasley:2002db},  one is able to also account for 
 the possible R--R three-form flux 
 \beq \label{R-Rflux}
     F_3 =m^K \alpha_K -  e_K \beta^K\ ,
 \eeq 
where the flux quanta $(m^K,e_K)$ are interpreted as labeling the discrete
excited states of the system and $(\alpha_K,\beta^K)$ is the real symplectic 
basis introduced in \eqref{def-alpha_beta}. 
This is in accord with the fact that the duality condition $G^{(3)} = (-1) *_{10} G^{(7)}$ given in \eqref{selfduality} 
relates the three-form containing $F_3$ to a seven-form containing $(dA_{(3)}^K,d\tilde{A}^{(3)}_K)$.
 
Let us collect the terms involving the non-dynamical three-forms $A_{(3)}^K$ and $\tilde{A}^{(3)}_K$. 
The first contribution arises form the effective bulk supergravity action containing 
the kinetic term $ \tfrac14 \int dC_6\wedge *dC_6$ for the R--R-form field $C_6$.\footnote{The factor $\tfrac14$
arises due to the fact that we can eliminate $dC_2$ contributions in this analysis by the 
duality condition \eqref{selfduality}.} 
Together with the contribution from the effective Chern-Simons action (\ref{eqn:eff-cs-action}) 
we obtain
\beq
      S_{A_{(3)}}=\int \big[ \tfrac14 e^{-4\phi} \cV^2  d\vec{A}_{(3)}\wedge *E\, d\vec{A}_{(3)} +\tfrac12\mu_5\,\vec{\mathcal{N}}^T d\vec{A}_{(3)} \big]\ ,
\label{eqn:cs-term-containing-a3}
\eeq
where the factor $e^{-4\phi} \cV^2$ arises due to the rescaling to the four-dimensional Einstein frame.
Note that we have introduced a vector notation to keep the following equations more transparent. 
More precisely, we define the matrix $E$, the vector-valued forms $\vec{A}_{(3)}$ and the vector $\vec{\mathcal{N}}$, cf.\ \eqref{eqn:n-tilde}, as
\begin{equation}
        E= \begin{pmatrix}
                \int \alpha_K\wedge *\alpha_L & \int \alpha_K\wedge *\beta^L\\
                \int\beta^K\wedge * \alpha_L & \int \beta^K\wedge * \beta^L
        \end{pmatrix}, \qquad \vec{A}_{(3)}=
        \begin{pmatrix}
                A_{(3)}^K\\
                \tilde{A}^{(3)}_K
        \end{pmatrix},\qquad 
        \vec{\mathcal{N}}=\hat \zeta^\cA 
              \begin{pmatrix}
                \mathcal{N}_{\cA K}\\
                \mathcal{N}^K_\cA
        \end{pmatrix}.
        \label{eqn:e-matrix}
\end{equation}
Our aim is to integrate out the forms $dA_{(3)}^K$ and $d\tilde{A}^{(3)}_K$ similar to \cite{Beasley:2002db} by 
also allowing for the discrete excited states labeled by the background fluxes $(e_K, m^K)$ in \eqref{R-Rflux}. 
In fact, we can treat this as dualizing the three-forms $A_{(3)}^K$ and $\tilde{A}^{(3)}_K$ into constants 
$(e_K, m^K)$ \cite{Louis:2002ny}. We thus add to $S_{A_{(3)}}$ the Lagrange multiplier term $\tfrac12(e_K dA^K_{(3)}+ m^Kd\tilde{A}^{(3)}_K) $
such that 
\beq
  S'_{A_{(3)}}=\int \big[\tfrac14e^{-4\phi} \cV^2  d\vec{A}_{(3)}\wedge *E\, d\vec{A}_{(3)} +\tfrac12(\mu_5\,\vec{\mathcal{N}}+\vec{m})^T d\vec{A}_{(3)} \big]\ ,
\eeq
where we abbreviated $\vec{m}=(e_K, m^K)^T$. Formally replacing  $\vec{F}_{(4)}=d\vec{A}_{(3)}$ 
with its equations of motion, one finds the scalar potential  
\begin{equation}
        V_{ A_{(3)}}=\frac{e^{4\phi}}{4\mathcal
        V^2}\left(\mu_5\vec{\mathcal{N}}+\vec{m}\right)^T E^{-1} \left(\mu_5\vec{\mathcal{N}}+\vec{m}\right).
        \label{eqn:potential-from-a4}
\end{equation}
Here we used the results of appendix \ref{app:derivationofpotential} to determine the inverse matrix given by 
\begin{equation}
        E^{-1}=\begin{pmatrix}
                \int \beta^K\wedge * \beta^L &  -\int\beta^K\wedge * \alpha_L\\
                -\int \alpha_K\wedge *\beta^L & \int \alpha_K\wedge *\alpha_L
        \end{pmatrix}.
        \label{eqn:matrix-d}
\end{equation}
It is convenient to rewrite the potential $V_{A_{(3)}}$ of (\ref{eqn:potential-from-a4}) in a more compact form 
\begin{equation}
        V_{ A_{(3)}}=\frac{e^{4\phi}}{4 \mathcal{V}^2}\int_Y \mathcal{G}\wedge *
        \mathcal{G},\qquad \quad  \mathcal{G}=F_{3}+\mu_5 \hat \zeta^\cA \left(\mathcal{N}^K_{\cA}\alpha_K- \mathcal{N}_{\cA K}\beta^K\right)\ ,
        \label{eqn:v-written-in-g}
\end{equation}
where we identified $F_3$ as given in \eqref{R-Rflux}. 
We note that the scalar potential contains the familiar contribution from the R--R fluxes $F_3$. In addition, 
there are terms linear and quadratic in the D5-brane deformations $\hat
\zeta$. In section \ref{subsec:suppot} we will show  that the
scalar potential \eqref{eqn:v-written-in-g} supplemented by the second term of $V_{\text{DBI}}$ in (\ref{DBIpotential}) can be obtained from a superpotential.

In order to prepare for the derivation of $V_{ A_{(3)}}$ from this superpotential, it is necessary 
to introduce the holomorphic and anti-holomorphic variables $\zeta^A$ and $\bar \zeta^A$.
Therefore we expand the three-form $\mathcal G$ in a complex basis 
$(\Omega,\chi_\kappa,\bar{\chi}_{\bar\kappa},\bar\Omega)$ of $H^3_+(Y)=H^{(3,0)}\oplus H^{(2,1)}_+\oplus H^{(1,2)}_+\oplus H^{(0,3)}$.
Explicitly, we find the expansion
\bea
    \cG =\big({\textstyle \int}\Omega\wedge\bar\Omega \big)^{-1}
        \big[I\, \bar{\Omega} +G^{\bar\kappa \kappa} \bar I_{\bar\kappa}\, \chi_\kappa 
        - G^{\bar\kappa \kappa}I_\kappa\, \bar{\chi}_{\bar\kappa}-\bar I \Omega \big]\ ,
        \label{eqn:expansion-g-dolbeault}
\eea
where the coefficient  functions are given by 
\beq
        \label{eqn:omega-wedge-g}
        I=\int_Y\Omega\wedge\mathcal G = \int_Y \Omega\wedge F_3- \mu_5 \int_{\Sigma_+} \zeta\lrcorner\Omega\ , \qquad \
        I_\kappa=\int_Y\chi_\kappa\wedge\mathcal G = \int_Y \chi_\kappa\wedge F_3-\mu_5 \int_{\Sigma_+}\zeta\lrcorner\chi_\kappa\ .
\eeq
Here we have used the relations (\ref{eqn:expansion-omega-a}) and (\ref{eqn:zeta-omega-bar-zero})
as well as the familiar metric $G_{\kappa \bar\kappa }$ on the space of complex structure deformations of $Y$. Using its explicit form   
\begin{equation}
 G_{\kappa\bar\kappa}=-\frac{\int\chi_\kappa\wedge\bar\chi_{\bar\kappa}}{\int\Omega\wedge\bar\Omega}\ ,
\end{equation}
the expansion \eqref{eqn:expansion-g-dolbeault} together with \eqref{eqn:omega-wedge-g} is readily checked.
Finally, we insert (\ref{eqn:expansion-g-dolbeault}) into (\ref{eqn:v-written-in-g}) and use
$\ast\Omega=-i\Omega,\ \ast\chi_\kappa=i\chi_\kappa$ to cast the complete potential $V_{\rm F}$ into the form
\beq  \label{eqn:v-final}
      V_{\rm F}= \frac{i e^{4\phi}}{2\mathcal V^2 \int\Omega\wedge\bar\Omega}\Big[G^{\kappa\bar\kappa} I_\kappa \bar I_{\bar \kappa} + |I|^2+2\mu_5\ \mathcal{G}^{A\bar B}e^{-\phi}\int_{\Sigma_+}s_A\lrcorner \chi_{\kappa}\int_{\Sigma_+}\bar s_{\bar B}\lrcorner \bar \chi_{\bar\kappa}\delta z^{\kappa}\delta \bar{z}^{\bar \kappa}\Big]\ .
\eeq
Here we have added the potential term in (\ref{DBIpotential}) originating from the reduction of the DBI-action.
Once we have computed the $\cN=1$ K\"ahler metric of the effective theory in appendix \ref{app:F-termscalarpot}, we will be able to derive $V_{\text{F}}$ 
from a superpotential depending on the complex structure and D5-brane
deformations. This will show that $V_{\rm F}$
is indeed an F-term potential as indicated by the notation.

Let us now turn to the remaining potential terms arising from the DBI action \eqref{eqn:DBI} and the NS--NS fluxes
$H_3$. For simplicity, we will only discuss electric NS--NS flux such that $H_3$ admits the expansion 
\beq \label{NSNSflux}
   H_3 = -\tilde e_K \beta^K\ .
\eeq
In ref.~\cite{Grimm:2004uq} it was shown that the electric fluxes $\tilde e_K$ result in a gauging of the
scalar $h$ dual to $\cC_{(2)}$ in \eqref{eqn:kk-expansion-form-fields}. The effect of magnetic fluxes
$\tilde m^K \alpha_K$ is more involved since
they directly gauge the two-from $\cC_{(2)}$ \cite{Grimm:2004uq}.  In order to be able
to work with the scalar $h$, we will not
allow for the additional complication and set $\tilde m^K=0$. 
Together with the last term in \eqref{eqn:DBI} we find the potential
\beq
   V_{\rm D} =\mu_5 \frac{e^{3\phi}}{\mathcal{V}^2}
   \frac{(\mathcal{B}^{\Sigma})^2}{16v^{\Sigma}} + \frac{e^{2\phi}}{4 \cV^2}
   \int_Y H_3 \wedge * H_3 \ ,\label{Dtermpot}
\eeq
which will turn out to be a D-term potential arising due to the gauging of two chiral multiplets. Recall that the first potential term in 
the DBI action \eqref{DBIpotential} is cancelled by the contribution \eqref{eqn:orieaction} of the O5-planes.

\section{The $\cN=1$ characteristic data} \label{subsec:N=1data}

In this section we bring the four-dimensional effective action for the brane 
and bulk fields into the standard $\cN=1$ supergravity form. More precisely,
we first determine the correct complex coordinates $M^I$ forming the bosonic part
of the $\cN=1$ chiral multiplets.  Their kinetic terms are expressed by a
K\"ahler potential $K(M,\bar M)$, while their F-term scalar
potential is encoded by a  holomorphic superpotential $W$. The $\cN=1$ vector multiplets
contribute kinetic terms and theta-angles that are expressed through
holomorphic gauge-kinetic coupling functions $f(M)$. We will also identify a
D-term potential arising through the gauging of the scalars $M^I$ of the chiral multiplets. The general
form of the bosonic $\cN=1$ action is given by \cite{Wess:1992cp,Gates:1983nr}
\beq\label{N=1action}
  S^{(4)} = -\int \tfrac{1}{2}R * 1 +
  K_{I \bar J} \cD M^I \wedge * \cD \bar M^{\bar J}  
  + \tfrac{1}{2}\text{Re}f_{\kappa \lambda}\ 
  F^{\kappa} \wedge * F^{\lambda}  
  + \tfrac{1}{2}\text{Im} f_{\kappa \lambda}\ 
  F^{\kappa} \wedge F^{\lambda} + V*1\ ,
\eeq
where
\beq\label{N=1pot}
V=
e^K \big( K^{I\bar J} D_I W {D_{\bar J} \bar W}-3|W|^2 \big)
+\tfrac{1}{2}\, 
(\text{Re}\; f)^{-1\ \kappa\lambda} D_{\kappa} D_{\lambda}
\ .
\eeq
Note that $K_{I \bar J}$ and $K^{I \bar J}$ are the K\"ahler metric 
and its inverse, where locally one has $K_{I\bar J} = \partial_I \bar\partial_{\bar J} K(M,\bar M)$.
The scalar potential is expressed in terms of the K\"ahler-covariant derivative $D_I W= \partial_I W + 
(\partial_I K) W$. To simplify our results, we have set the four-dimensional
gravitational coupling $\kappa_4=1$ in the following discussion.

\subsection{The K\"ahler potential and $\cN=1$ coordinates \label{subsec:Kpot}}

Now we are ready to read off the $\cN=1$ data from the effective action.
Let us first define the $\cN=1$ complex coordinates $M^I$ which are the bosonic 
components of the chiral multiplets. We note that the $M^I$ consist of the D5-brane 
deformations $\zeta^A$ and Wilson lines $a_I$ introduced in section \ref{4Dspectrum}.
In addition there are the complex structure deformations $z^\kappa$ as well as
the complex fields 
\bea \label{N=1coords}
   t^\alpha &=& e^{-\phi} v^\alpha - i c^\alpha +\tfrac12\mu_5\, \cL^\alpha_{A\bar B} \zeta^A
   \bar \zeta^{\bar B}\ ,\nonumber\\
   P_a &=& \Theta_{ab}\, \mathcal{B}^b + i \rho_a\ ,\\
   S &=&  e^{-\phi} \cV + i \tilde{h} -\tfrac14 (\R \Theta)^{ab} P_a (P+\bar
   P)_b +\mu_5\,\ell^2\,\mathcal{C}^{I\bar J}a_{I}\bar a_{\bar J}\ ,\nonumber
\eea
where $v^\alpha,b^a$, $c^\alpha,\rho_a$ as well as $\mc B^a$ are given in \eqref{exp_JB}, \eqref{eqn:kk-expansion-form-fields} and (\ref{def-Ba}) as well as $\tilde h=h-\tfrac12\rho_a\mathcal{B}^a$. 
The complex symmetric tensor appearing in \eqref{N=1coords} is given by
$\Theta_{ab}=\cK_{ab \alpha} t^\alpha$ and $(\R \Theta)^{ab}$ denotes the
inverse of $\R \Theta_{ab}$. The function $\cL^\alpha_{A \bar B}$
is defined in \eqref{eqn:metrics}. Note that we
recover the $\cN=1$ coordinates found in refs.~\cite{Grimm:2004uq,Grimm:2005fa} from an analysis of
the effective bulk action if we set $\zeta^A=a_I = 0$. The completion \eqref{N=1coords} is inferred from
the couplings in the D5-brane action \eqref{eqn:DBI} and \eqref{eqn:CS}.

The full $\cN=1$ K\"ahler potential is determined by integrating the kinetic
terms of the complex scalars $M^I=(S,t^\alpha,P_a,z^\kappa,\zeta^A,a_I)$. 
It takes the form 
\beq \label{eqn:kaehler-pot}
    K =-\ln\big[ -i\int\Omega\wedge\bar\Omega \Big]+K_q\ ,\qquad  K_q=-2\ln\big[
      \sqrt{2}e^{-2\phi}\mathcal V \Big]\ ,
\eeq
where $K_q$ has to be evaluated in terms of the coordinates
\eqref{N=1coords}. In contrast to general compactifications with O3/O7
orientifold planes, this can be done explicitly for O5-orientifolds yielding 
\beq \label{eqn:Kaehlerpart}
        K_q
       = -\ln \big[ \tfrac{1}{48} \cK_{\alpha \beta \gamma} \Xi^\alpha\,
      \Xi^\beta\, \Xi^\gamma \big] 
      - \ln \big[S +\bar S + \tfrac{1}{4} (\R \Theta)^{ab} (P +\bar P)_a (P
     +\bar P)_b - 2\mu_5\,\ell^2\,\mathcal{C}^{I\bar J}a_{I}\bar a_{\bar J}\big]\ ,
\eeq
where we write
\beq
  \Xi^\alpha = t^\alpha +\bar t^\alpha -
        \mu_5\,\cL^{\alpha}_{A\bar B} \zeta^A \bar \zeta^{\bar B} \ .
                \label{eqn:xi-def}
\eeq
Note that the expression \eqref{eqn:kaehler-pot} for $K$ can already be
inferred from general Weyl rescaling arguments, e.g.~from the factor $e^K$ in
front of the $\cN=1$ potential \eqref{N=1pot}. However, the explicit form
\eqref{eqn:Kaehlerpart} displaying the field dependence of $K$ has to be
derived by taking derivatives of $K$ and comparing the result with the bulk
and D5-brane effective action.
Let us also note that the expression \eqref{eqn:Kaehlerpart} reduces to the results found in \cite{Kors:2003wf,Lust:2004cx} in the orbifold limit.

\subsection{The superpotential \label{subsec:suppot}}

Having defined the right $\mathcal{N}=1$ chiral coordinates as well as 
the K\"ahler potential we have to compute the effective superpotential 
$W$ to complete the $\mathcal{N}=1$ data of the chiral multiplets. 
Using the general supergravity formula (\ref{N=1pot}) for the scalar 
potential in terms of $W$ we are able, as presented below, to deduce the superpotential $W$ 
entirely by comparison to the derived scalar potential $V_{\rm F}$ 
(\ref{eqn:v-final}) after dimensional reduction. Thus, it is indeed 
an F-term potential of the $\mathcal{N}=1$ effective theory as indicated by 
the notation.

The superpotential $W$ yielding the potential $V_{\rm F}$ consists of 
two parts, a truncation of the familiar Gukov-Vafa-Witten flux superpotential
for the closed string moduli \cite{Gukov:1999ya} and a contribution encoding 
the dependence on the open string moduli of the wrapped D5-brane,
\begin{equation} \label{effsuperpot} 
 W=\int_Y F_3\wedge \Omega+\mu_5\int_{\Sigma_+}\zeta\lrcorner\Omega\ ,
\end{equation}
where we introduced the field strength $F_3=dC_2$.
Now, it is a straight forward but lengthy calculation to obtain the 
F-term contribution of the scalar potential (\ref{N=1pot}). The detailed 
calculations of appendix \ref{app:F-termscalarpot} yield a positive 
definite F-term potential 
\begin{equation}
 V=\frac{ie^{4\phi}}{2\mathcal V^2 \int\Omega\wedge\bar\Omega}\left[ \left|
     W\right|^2 +D_{z^\kappa}WD_{\bar z^{\bar\kappa}}\bar W
     G^{\kappa\bar\kappa} 
     +\mu_5\ \mathcal{G}^{A\bar B}e^{-\phi}\int_{\Sigma_+}s_A\lrcorner 
     \Omega\int_{\Sigma_+}\bar s_{\bar B}\lrcorner \bar \Omega \right]
\label{F-termpotential}
\end{equation}
of no-scale type. 
Here the covariant derivatives with respect to the complex structure coordinates $z^i$ and the open string moduli $\zeta^A$ have to be inserted,
\begin{equation}
       D_{z^\kappa}W = \int F_3\wedge\chi_\kappa +\mu_5\int\zeta\lrcorner\chi_\kappa\ ,\qquad
        D_{\zeta^A}W = \mu_5\int s_A\lrcorner\Omega+\hat{K}_{\zeta^A}W\ .\label{d_zW}
\end{equation}
Finally, we have to use the first order expansion of $s_A\lrcorner \Omega$ discussed in \eqref{eqn:expansion-omega-a} to make sense of the integration over the two-cycle $\Sigma_+$,
\begin{equation}
        \int_{\Sigma_+}s_A\lrcorner \Omega=\int_{\Sigma_+}s_A\lrcorner\chi_{\kappa}\delta z^\kappa.
\end{equation}
Inserting this into (\ref{F-termpotential}), the F-term potential perfectly matches the scalar potential $V_{\rm F}$ of (\ref{eqn:v-final}) obtained by dimensional reduction of the D5-brane as well as the bulk supergravity action.

We conclude with a discussion of the derivation and special structure 
of the F-term potential. We first note that the potential (\ref{F-termpotential}) is
positive definite unlike the generic F-term potential of supergravity. 
This is due to the no-scale structure \cite{Cremmer:1983bf,Ellis:1983sf,Barbieri:1985wq} of the underlying
$\mathcal{N}=1$ data. Indeed, the superpotential \eqref{effsuperpot} only depends on $z$ and $\zeta$ and is
independent of the chiral fields $S$, $P$, $a$ and $t$. Consequently, the
$\mathcal{N}=1$ covariant derivative $D_{M^I}W$ of the superpotential simplifies to
$K_{M^I}W$ when applied with respect to the fields $M^I=(S, P, a,
t)$. The K\"ahler potential \eqref{eqn:kaehler-pot} for these
fields has the form
\begin{equation} \label{noscalekaehlerpot}
 K=-m\ \text{ln}(t+\bar
t+f(\zeta,\bar\zeta))-n\ \text{ln}(S+\bar{S}+g(P+\bar P,t+\bar t)+h(a,\bar a))\ 
\end{equation}
with $m=3$ and $n=1$. In order to clarify our exposition
we concentrate on the one-modulus case for each chiral multiplet. The
generalization to an arbitrary number of moduli is
straightforward, cf.~appendix \ref{app:F-termscalarpot}, where also the functions $f$, $g$ and $h$ can be found. 
The contributions of the fields $M^I=(S, P, a,
t, \zeta)$ to the scalar potential $V$ take a characteristic form given by
\begin{equation}\label{noscalescalarpot}
 K^{I\bar J}D_{M^I}W D_{\bar{M}^{\bar J}}\bar{W}=|\partial_\zeta W|^2K^{\zeta\bar\zeta}+(n+m)|W|^2 
\end{equation}
as familiar from the basic no-scale type models of supergravity.\footnote{This no-scale structure will be  clearified further, extending the example of \cite{Grimm:2004uq}, in appendix C using the dual description of $S+\bar S$ in terms of a linear multiplet $L$.} Consequently, 
this turns the negative term $-3|W|^2$ in \eqref{N=1pot} into the positive 
definite term $|W|^2$ of \eqref{F-termpotential} for the case $n=1$ and
$m=3$. A similar structure for the underlying $\mathcal{N}=1$ data has been 
found for D3- and D7-branes as shown in \cite{DeWolfe:2002nn,Camara:2003ku,
Grana:2003ek,Camara:2004jj,Jockers:2004yj,Jockers:2005zy}.\footnote{See ref.~\cite{Correia:2007sv} for a similar discussion in heterotic M-theory.}
In particular, this form for the scalar potential $V$ implies that a generic
vacuum for the complex structure and D-brane deformations is de Sitter, i.e.~has a positive cosmological constant, while in a supersymmetric vacuum $V$ and $W$ vanishes.
However, the potential depends on the K\"ahler moduli only through an overall factor of the volume and thus drives the internal space to decompactify.

\subsection{The gauge-kinetic function, gaugings and D-term potential \label{subsec:gauging}}

In the following we will discuss the terms of the four-dimensional effective
action arising due to the $U(1)$ vector multiplets in the spectrum. Firstly, 
there are the kinetic terms of the D5-brane vector $A$ and the 
vectors $V^{\tilde K}$ arising from the expansion
\eqref{eqn:kk-expansion-form-fields} of the R--R form $C_4$.
The gauge-kinetic function is determined from the actions \eqref{eqn:DBI} and \eqref{eqn:CS} 
and reads
\beq \label{gauge-couplingD5bulk}
    f_{\Sigma \Sigma}(t^\Sigma) = \tfrac12\mu_5 \ell^2\, t^\Sigma \ , \qquad  \quad f_{\tilde K
      \tilde L}(z^\kappa) = - \tfrac{i}{2} \bar \cM_{\tilde K \tilde L} = - \tfrac{i}{2} \cF_{\tilde K \tilde L}\big|_{z^{\tilde K}=0}\ .
\eeq
Here $ f_{\Sigma \Sigma}$ is the gauge-coupling function for the D5-brane
vector $A$ and $f_{\tilde K \tilde L}$ is the  gauge-coupling function for the 
bulk vectors $V^{\tilde K}$. Note that the latter can be expressed via
$ \cF_{\tilde K \tilde L} = \partial_{z^{\tilde K}} \partial_{z^{\tilde L}} \cF$ as the second derivative of the $\cN=2$ prepotential
$\cF$ with respect to the $\cN=2$ coordinates $z^{\tilde K}$ which are then set to zero 
in the orientifold set-up \cite{Grimm:2004uq}. This ensures that the 
gauge-coupling function is holomorphic in the coordinates $z^\kappa$ which would be not
the case for the full $\cN=2$ matrix $ \bar \cM_{K L} $ given in \eqref{eqn:matrix-a-b-m}.
The gauge-kinetic function encoding the mixing between the D5-brane vector and
the bulk vectors is discussed in appendix \ref{kinmix}. 
The quadratic dependence of $f_{\Sigma \Sigma}$ on the open string moduli
$\zeta$ through the coordinate $t^\Lambda$ in \eqref{N=1coords} is not visible on 
the level of the effective action. 
These corrections as well as further mixing with the open string moduli are
due to one-loop corrections of the sigma model and thus not covered by our
bulk supergravity approximation nor the DBI- or Chern-Simons actions of the D5-brane.

Let us now turn to the potential terms induced by the gauging of
global shift symmetries. There will be two sources for such gaugings. The first
gauging arises due to the source term proportional to 
$d(\tilde{\rho}^\Sigma -\mathcal{C}_{(2)}\mathcal{B}^\Sigma)\wedge A$ in (\ref{eqn:CS}).
It enforces a gauging of the scalars dual to the two-forms $\tilde{\rho}^\Sigma$ and $\mathcal{C}_{(2)}$. In fact, eliminating $d\tilde{\rho}^\Sigma$ and
$d\mathcal{C}_{(2)}$ by their equations of motion, the kinetic terms of the 
dual scalars $\rho_a$ and $h$ contain the covariant derivatives
\begin{equation} \label{gaugedfields}
        \cD \rho_a=d\rho_a+\mu_5\ell\delta_a^\Sigma A\ ,\qquad \quad \cD
        h=dh+\mu_5\ell\mathcal{B}^\Sigma A\ ,
\end{equation}
where $A$ is the $U(1)$ vector on the D5-brane. We note that the plus sign in
the covariant derivative of $h$ arises due to the minus sign in the duality
conditions (\ref{selfduality}) and ensures that the complex scalar $S$ defined
in \eqref{N=1coords} remains neutral under $A$. However, the 
gaugings \eqref{gaugedfields} imply  a charge for the chiral field $P_\Sigma$. It
is gauged by the D5-brane vector $A$. Its covariant derivative is given by
\begin{equation} \label{gaugingA}
        \cD P_\Sigma =dP_\Sigma +i\mu_5\ell A\ .
\end{equation}

The second gauging arises in the presence of electric NS--NS
three-form flux $\tilde e_K$ introduced in \eqref{NSNSflux}. 
It was shown in ref.~\cite{Grimm:2004uq}, that the scalar $h$ is 
gauged by the bulk $U(1)$ vectors $V^{\tilde K}$ arising in the 
expansion \eqref{eqn:kk-expansion-form-fields}
of $C_4$. This forces us to introduce the covariant derivative 
\beq \label{gaugingS}
  \cD S = d S - i\tilde e_{\tilde K} V^{\tilde K}\ .
\eeq
The introduction of magnetic NS--NS three-form flux is more involved and leads
to a gauged linear multiplet $(\phi , \cC_{(2)})$ as described in
ref.~\cite{Grimm:2004uq}. 

Having determined the covariant derivatives \eqref{gaugingA} and \eqref{gaugingS} it is
straightforward to evaluate the D-term potential. Recall
 the general formula
for the D-term \cite{Wess:1992cp}
\beq \label{D-term_gen}
K_{I\bar J} \bar X^{\bar J}_k = i \partial_I D_k \ ,
\eeq
where $X^{I}$ is the Killing vector of the $U(1)$ transformations
defined as $\delta M^I = \Lambda^k_0 X_k^J \partial_J M^I$. For the 
gaugings \eqref{gaugingA} and \eqref{gaugingS} we find the Killing vectors 
$X^{P_\Sigma}=i\mu_5\ell$ and $X^S_{\tilde K} = - i\tilde e_{\tilde K}$
which are both constant. Integrating \eqref{D-term_gen} one evaluates using
$K_{P_\Sigma}$ and $K_S$ given in \eqref{firstderivatives} of appendix \ref{app:F-termscalarpot} the D-terms
\beq
  D = - \tfrac14 \mu_5 \ell e^{\phi} \cB^\Sigma \cV^{-1}  \ ,\qquad \quad D_{\tilde K} = \tfrac{1}{2} \, \tilde e_{\tilde K}\,e^{\phi}\cV^{-1}\ .
\eeq  
Inserting these D-terms into the $\cN=1$ scalar potential \eqref{N=1pot} and using the
gauge-kinetic functions \eqref{gauge-couplingD5bulk}, we precisely recover the D-term potential
\eqref{Dtermpot} found by dimensional reduction.

\section{The structure of the $\cN=1$ open-closed field space}\label{sec:n1-special-geometry}

In the last section we derived the data of the four-dimensional $\mc N=1$ effective 
action for an D5-brane coupled to Type IIB supergravity by dimensional reduction in the 
large radius limit. Our analysis incorporated the D-brane moduli 
to linear order around a background configuration. We found that at this order
the D5-brane moduli correct the $\cN=1$ dilaton and K\"ahler coordinates and mix with the complex
structure deformations only in the scalar potential. In this section we
discuss whether one can extend this analysis to higher orders in the
deformations. 

Let us first recall the more familiar situation in $\cN=2$ compactifications.
In these theories the complex and K\"ahler structure moduli decouple at generic
points in the moduli space. In particular, it is possible in Type IIB compactifications 
to study the K\"ahler potential 
and the gauge kinetic function analytically over the entire complex 
structure moduli space. This is due to the underlying $\cN=2$ special 
K\"ahler geometry which relates them to the holomorphic prepotential whose 
dependence on the moduli is exactly calculable by period integrals. 
In absence of such strong non-renormalizations 
arguments in $\mc N=1$ theories the same problem is much more 
difficult to address. Let us discuss it for the K\"ahler potential, 
the superpotential and the gauge-kinetic function, which are 
introduced in section \ref{subsec:N=1data}. 

The $\cN=1$ K\"ahler potential is not protected by any non-renormalization 
theorems against corrections. This makes it hard to infer reliable information 
on its precise form beyond the approximations in sections
\ref{subsec:N=1data}. For example, one expects that the split $K_{\rm
  cs}(z,\bar z) + K_q$ in (\ref{eqn:kaehler-pot}), 
with $K_q$ being independent of the complex structure
moduli $z$, will no longer persist at higher orders in the D5-brane
deformations $\zeta$. 
This is due to the fact that the definitions of the $\zeta$ and $\bar
\zeta$ depend on the complex structure of the background as discussed in
section \ref{relations}. Both perturbative and non-perturbative corrections are 
expected to modify $K$ and it is beyond the scope of this work to examine
their form.  

The situation improves when considering the $\cN=1$ superpotential. 
In a compactification with background three-form fluxes the dependence of the 
superpotential on the closed string moduli is encoded by  a truncation of the familiar
Gukov-Vafa-Witten superpotential \cite{Gukov:1999ya}
\begin{equation} \label{eq:closed-superpotential}
 W_{\rm closed}=\int_Y F_3\wedge\Omega =\int_{\hat \Gamma} \Omega\ .
\end{equation}
Here $F_3$ is the R--R three-form \eqref{R-Rflux} which is an element of
$H^3(Y,\bbZ)$ since we henceforth restrict to the case that $h^3_-(Y)=0$.
Using Poincar\'e duality this three-form is related to a 
three-cycle $\hat \Gamma \in H_3(Y,\mathbb{Z})$. It admits 
the explicit expansion $\hat \Gamma = m^K A_K-e_K B^K$,
where $(A_K,B^K)$ is the symplectic basis introduced in \eqref{def-AB} and 
$(e_K,m^K)$ are the flux quanta in \eqref{R-Rflux}.  
The superpotential $W_{\rm closed}$ depends on the complex structure moduli through $\Omega(z)$ and with analytic 
continuation of the periods it can be studied everywhere 
in the moduli space~\cite{Curio:2000sc,Douglas:2006es,Blumenhagen:2006ci,Lust:2004ks}.

Perturbative string theory completes the closed string holomorphic
superpotential by an open string holomorphic superpotential. On the Type IIA 
side it is generated by disc instanton contributions and is calculable by topological 
string theory~\cite{Kachru:2000ih,Aganagic:2000gs,Aganagic:2001nx}.  
In our application, i.e.~on the Type IIB  side, these terms localize 
to a chain integral\footnote{In this section we set $\mu_5=1$.}~\cite{Witten:1997ep,Kachru:2000an,Aganagic:2000gs,Aganagic:2001nx}   
\begin{equation}
        W_{\rm open}=\int_\Gamma\Omega\ ,
        \label{eqn:open-superpotential}
\end{equation}
where $\Gamma$ is a three-chain whose boundary is given 
by curves $\Sigma-\Sigma_0$, where $\Sigma_0$ 
is a fixed reference curve in the same homology class as $\Sigma$. The 
dependence on the closed string parameters is through $\Omega$ and 
on the open string variables is through the deformation parameters 
of $\Sigma$. Using the general power series expansion of a functional about a 
reference function, we recover our result for the superpotential
\eqref{effsuperpot} to linear order.\footnote{The general
Taylor expansion is given by 
$F[g]=\sum_{k=0}^{\infty}\int\,dx_1\cdots dx_k\frac{1}{k!}\left.\frac{\delta^k
    F[g]}{\delta g(x_1)\cdots \delta g(x_k)}\right|_{g=\tilde g}\delta
g(x_1)\cdots \delta g(x_k)$. 
For $W$ as a functional of the embedding $\iota$ and $\delta \iota\equiv\zeta$
as well as $\tilde g=\iota$ we to first order derive the 
second term of (\ref{effsuperpot}).}
In section \ref{non-compactCY} we will review how  
(\ref{eqn:open-superpotential}) is directly calculated on non-compact 
Calabi-Yau manifolds. As will be explained in more detail in section 
\ref{rel_cohom}, it is natural to view combination of the 
cycle $\hat \Gamma$ and the chain $\Gamma$ as an element 
$\Gamma^\Sigma$ of the relative homology group $H_3(Y,\Sigma,\mathbb{Z})$ 
and write 
\beq \label{open-closed_W}
   W=W_{\rm closed}+W_{\rm open}=\int_{\Gamma^\Sigma}\Omega\ .
\eeq 
In the $\mc N=1$ theory there is no spacetime argument 
for the decoupling of the K\"ahler and the complex structure
moduli.
However, it was argued that the K\"ahler moduli dependence 
to the superpotential can only arise through D-instanton contributions 
due to its holomorphicity, see e.g.\ \cite{Witten:1996bn}. This implies that at large volume these 
corrections are strongly suppressed by powers of the instanton action. 
It is believed that the part of the Type IIB superpotential which is independent of the K\"ahler moduli 
is exactly given by \eqref{open-closed_W}. Thus it can be computed from the 
topological sector of the physical string. 
In this sense the integrals 
(\ref{eq:closed-superpotential}) and 
(\ref{eqn:open-superpotential})
contain the exact analytic dependence of $W$ on the open moduli 
and closed complex structure moduli in Type IIB Calabi-Yau 
compactifications with D5-branes. In the orientifold set-ups,
\eqref{open-closed_W} should still be valid for D5-branes which are 
sufficiently separated from the O5-planes.

One can also calculate the topological string 
contributions to the gauge kinetic terms \eqref{gauge-couplingD5bulk}. They 
are given by the annulus amplitude which can be 
evaluated in the non-compact Calabi-Yau manifolds at large radius by 
localization~\cite{Graber:2001dw,Mayr:2002zi}
or more effectively by large-$N$ techniques~\cite{Aganagic:2003db}. 
In particular, in the Type IIB models the term is given by 
the Bergman kernel~\cite{Eynard:2007kz,Marino:2006hs}, 
whose analytic dependence on the moduli is exactly known 
and whose expansions in flat coordinates at various points 
in the moduli space have been studied
in~\cite{Bouchard:2008gu}.

This section contains six parts. First we discuss the 
problem of computing the superpotential for non-compact 
toric Calabi-Yau in section \ref{non-compactCY}. Here the Type IIB geometry 
is governed by a Riemann surface and all essential 
ideas are realized in the simplest context.
In the next subsection \ref{closedHodge} we prepare our discussion of the 
open-closed moduli dependence of the superpotential by reviewing the complex structure 
dependence of the Gukov-Vafa-Witten superpotential through the closed periods.
Then, we introduce the appropriate geometric quantities, namely 
relative (co)homology theory to describe also the open moduli 
contribution to the superpotential. Next, we proceed by describing a new
method to circumvent the difficulties in handling the relative group of 
the curve $\Sigma$. We associate a canonically constructed divisor $\Div$ 
to the given curve which enables us to replace the relative group in 
two different ways.
One possibility is to replace it by the cohomology of forms with logarithmic 
singularities along $\Div$ and to study the moduli dependence using the 
so-called mixed Hodge structure. Another possibility is to embed the
open-closed moduli into the complex structure deformations of a canonically 
constructed K\"ahler manifold $\tilde Y$. Then, we can investigate the complex 
structure moduli space of $\tilde Y$ instead. Next, we present an application 
of this rather abstract discussion by giving recipes to obtain Picard-Fuchs 
equations whose solutions describe the open-closed  moduli dependence of the superpotential.
Finally, we apply the described blow-up procedure to the non-compact example of the total space of the canonical bundle over the del Pezzo surface $\mc B_3$ where the D5-brane is represented by a point in $\mc B_3$.

\subsection{Non-compact Calabi-Yau spaces} 
\label{non-compactCY} 

While field theory considerations restrict $\mc N=1$ supergravity 
much less than $\mc N=2$ supergravity, we expect additional 
structures, when the $\mc N=1$ theory arises as the effective action 
of a string theory. The stringy origin of the 
superpotential and the gauge  kinetic terms can be explored 
best in Type II string theory in the background of 
non-compact Calabi-Yau spaces. The main ideas and concept 
related to these quantities are realized in this context 
in a very simple way, which makes it worthwhile to 
introduce them here. Moreover, explicit calculations are 
feasible and the mirror symmetry picture between the Type IIB geometry and 
Type IIA geometry has been developed for 
local Calabi-Yau spaces and used for predictions as well as checks.

While our focus will be on the Type IIB geometries, let us 
briefly recall the Type IIA geometry first. In the non-compact case of
interest the internal manifold $X$ is  
typically a complex line bundle over a del Pezzo surface. 
Here one specifies charge vectors $Q^\alpha_i\in\mathbb{Z}$ which
describe toric group actions. We use the notation 
of~\cite{Bouchard:2007ys}. More precisely $X$ is given 
by the quotient
\begin{equation} 
    X=(\mathbb{C}^{k+3}-Z)/(\mathbb{C}^*)^k\ .
    \label{quotient}
\end{equation} 
Here $(\mathbb{C}^*)^k$ acts by $x_i\mapsto \lambda_\alpha^{Q_i^\alpha} x_i$,
$\alpha=1,\ldots,k$  on the complex coordinates $x_i$ of $\mathbb{C}^{k+3}$ with 
$\lambda_\alpha\in \mathbb{C}^*$ and $Z$ is the Stanley-Reisner ideal. 
The geometry has vanishing first Chern
class, iff the constraint $\sum_{i=1}^{k+3} Q_i^\alpha=0$ holds 
$\forall \alpha$. The mirror of D5-branes are D6-branes 
wrapping Harvey-Lawson special Lagrangians. Their superpotential 
arises form disks ending on $L$ and has 
been first calculated in these geometries in~\cite{Aganagic:2000gs,
Aganagic:2001nx}.

We are mainly interested in the mirror Type IIB geometry with Calabi-Yau space 
$Y$ and D5-branes
\cite{Katz:1996fh,Hori:2000kt,Aganagic:2000gs,Aganagic:2001nx}. 
$Y$ is a conic bundle 
\begin{equation} 
   uv= H(x,y;z)
  \label{eq:local-Y}
\end{equation}
branched over a Riemann surface ${\cal Y}$ given by 
$H(x,y;z)=0$~\cite{Hori:2000kt}. Here $u,v$ are in $\mathbb{C}$, 
$x,y$ are in  $\mathbb{C}^*$ and the variables $z$ parametrize 
the complex structure of $Y$.
The function $H$ is 
given by $H=\sum_{i=1}^{k+3} x_i$, where $x_i\in \mathbb{C}^*$ are homogeneous 
coordinates w.r.t.~an additional $\mathbb{C}^*$-action and subject to the 
constraints 
\begin{equation} 
(-1)^{Q_0^\alpha} \prod_{i=1}^{k+3}x_i^{Q_i^\alpha}=z_\alpha\ , \quad 
\forall \alpha \ . 
\label{mirrormap}
\end{equation} 
The $z_\alpha$ denote the complex structure moduli\footnote{They are dual to the
complexified K\"ahler parameters $t_\alpha$ of compact two-cycles in $X$, 
so $\alpha=1,\ldots, h_{comp}^{(1,1)}(X)$.} of
${\cal Y}$, while $x$ and $y$ in (\ref{eq:local-Y})
denote the independent variables that remain after solving the constraints 
(\ref{mirrormap}) and using  the additional $\mathbb{C}^*$-action on the coordinates $x_i$. 

The main simplification of the non-compact models is the  dimensional
reduction in the $B$-model geometry. The holomorphic three-form of $Y$ 
reduces to a meromorphic  differential~\cite{Katz:1996fh} 
\begin{equation} 
\lambda=\log(x)\frac{{\rm d}y}{y}
\end{equation} 
on the genus $g$ Riemann surface ${\cal Y}$. The three-cycles in
$H_3(Y,\mathbb{Z})$ reduce either to one-cycles $a_i,b^i$, 
$i=1,\ldots,g$ in $H_1({\cal Y},\mathbb{Z})$ or to one-cycles $c_k$ enclosing 
the poles of $\lambda$ at $p_i$. The flat closed string 
modulus, its mirror map and the closed string prepotential are 
encoded in periods of $\lambda$ over paths in the homology 
of ${\cal Y} \setminus\{p_i\}$.  The closed string potential reduces to 
$W_{\rm closed}=\int_{\hat \Gamma} \lambda$, where $\hat \Gamma={e'}^jc_j+ e^i a_i-m_k b^k$. 

The holomorphic cycle $\Sigma$ in $Y$, which is mirror to the special
Langrangian on $X$, reduces to a point $x$ on ${\cal Y}$, so that 
the triple $({\cal Y},\lambda,x)$ contains the non-trivial information 
of the Type IIB geometry with one non-compact D5-brane. It provides
the geometrical realization of the non-trivial superpotential. The latter is
obtained by reduction of \eqref{eqn:open-superpotential} to the Riemann surface
\begin{equation} 
        W_{\text{open}}(x,z,m)=\int_{\Gamma^x} \lambda(m,z)\ ,
\end{equation}
where the integral is over a path $\Gamma^x$ from an 
irrelevant reference point $x_0$ to $x$. After the 
mirror map, $W_{\text{open}}(x,z,m)$ has been identified 
with the disk instanton generating function~\cite{Aganagic:2001nx}.   
Beside the  open modulus $x$ dependence, whose domain is 
simply the Riemann surface ${\cal Y}$, the integral 
depends on the complex modulus $z$ of $\cal Y$ and 
potentially on constants $m_i$, which are the 
non-vanishing residua of $\lambda(m,z)$. The 
evaluation of the integrals 
\begin{equation} 
\int_{\hat \Gamma} 
\lambda+\int_{\Gamma^x} \lambda=\int_{\hat\Gamma^x} \lambda\ , 
\label{onedimensionalchain}
\end{equation} 
is a simple example of a  problem in relative homology. 
Here $\hat \Gamma$ is a one-cycle of ${\cal Y}$ and ${\hat\Gamma^x}$ a relative one-cycle, i.e. an element 
of the group $H_1({\cal Y},\{p_i\},\mathbb{Z})$
which contains the one-cycles of ${\cal Y}$ as well as one-chains which end on $p_i$.
On the Riemann surface (\ref{onedimensionalchain}) can be solved by evaluating 
the integrals explicitly~\cite{Aganagic:2001nx}. The specific 
elements  $H_1({\cal Y},\{p_i\},\mathbb{Z})$, that yield 
the closed string flat coordinates, the closed string mirror flat
coordinates and the superpotential have been described 
in~\cite{Aganagic:2001nx}.

Differential equations for ordinary periods are encoded  
in the variation of Hodge structure. They can be quite 
generically derived using the Griffith residua formulas for 
the periods~\cite{Griffiths:1968,Griffiths:1968a}. 
Differential equations for relative period integrals, i.e.\ the 
integrals over the elements of the relative homology 
$\int_{\hat\Gamma^x} \lambda$ are mathematically encoded in the
variation of the mixed Hodge structure. In certain
situations they can be derived from residua expressions
for the normal function~\cite{Griffiths:1979rt}. 
For the local models such differential equations 
have been described in~\cite{Lerche:2002ck,Lerche:2002yw}. 

On a Riemann surface ${\cal Y}$ the integral  
$W=\int_{\hat \Gamma^x}\lambda$ defines an 
Abel-Jacobi map, albeit with meromorphic 
1-forms instead of the holomorphic ones. 
Other canonical invariants of the pair 
$({\cal Y},\lambda(z,m))$ have been 
studied~\cite{Eynard:2007kz} and can be 
associated to analytic expressions for the  
topological string amplitudes on $Y$~\cite{Marino:2006hs,Bouchard:2007ys,Bouchard:2008gu}.
Most notably the Bergman kernel is identified 
with the annulus amplitude and gives a global 
definition of the gauge kinetic function.  

\subsection{Hodge structure for complex structure moduli \label{closedHodge}}

First, we describe the situation of closed strings only where we focus on the complex structure moduli.
Generally, infinitesimal deformations of the complex structure are described by elements of $H^1(Y,TY)$, cf.\ \cite{Kodaira1986}.
For K\"ahler manifolds the infinitesimal study of the complex structure moduli space can be carried out by the study of the variation of the Hodge structure on its cohomology groups.
For Calabi-Yau manifolds as discussed in section \ref{sec:closedStringSpectrum} the analysis simplifies since there is an unique non-vanishing holomorphic three-form $\Omega$.
On the one hand, this enables us to map the infinitesimal deformations in $H^1(Y,TY)$ simply to forms in $H^{(2,1)}(Y)$.
On the other hand, the fact $h^{(3,0)}=1$ allows us to study the variation of the Hodge structure explicitly, as will be discussed below.
Here, we review the concepts of complex structure deformations and the simplifications for K\"ahler threefolds with $h^{(3,0)}=1$, in particular Calabi-Yau manifolds, as will be relevant for our later discussion.

If we consider $H^3(Y)$ over every point of the complex structure moduli space $\mc M^{\rm{cs}}$, it forms a holomorphic vector bundle over $\mc M^{cs}$ which we will denote by $\mc H^3(Y)$.
We define a decreasing filtration on $H^3(Y)$, the Hodge filtration, which equips $H^3(Y)$ with a (pure) Hodge structure
\begin{eqnarray}
        \nn F^3H^3(Y) &=& H^{(3,0)}(Y)\ ,\\
        \nn F^2H^3(Y) &=& H^{(3,0)}(Y)\oplus H^{(2,1)}(Y)\ ,\\
        \nn F^1H^3(Y) &=& H^{(3,0)}(Y)\oplus H^{(2,1)}(Y)\oplus H^{(1,2)}(Y)\ ,\\
        F^0H^3(Y) &=& H^{(3,0)}(Y)\oplus H^{(2,1)}(Y)\oplus H^{(1,2)}(Y)\oplus H^{(0,3)}(Y)=H^3(Y)\ ,
        \label{eqn:closed-hodge-filtraion}
\end{eqnarray}
where we recover the familiar decomposition of the de Rham group $H^3(Y)$ into $(p,q)$-forms for K\"ahler manifolds.
This filtration is decreasing since $F^mH^3(Y)$ is contained in $F^{m-1}H^3(Y)$ for all $m$.
We study the filtration $F^mH^3(Y)$ instead of $H^{(p,q)}(Y)$ because the $F^mH^3(Y)$ form a holomorphic subbundle $\mc F^m_{\text{cs}}$ of $\mc H^3(Y)$, but $H^{(p,q)}(Y)$ do not.
The bundle $\mc H^3(Y)$ has a flat connection $\nabla_{\rm{cs}}$ which is called the Gau{\ss}-Manin connection.
It has the so-called Griffiths transversality property
\begin{equation}
        \nabla_{\rm{cs}}\mc F^m_{\rm{cs}}\subset \mc F^{m-1}_{\rm{cs}}\otimes \Omega^1_{\mc M^{\rm{cs}}}\ .
        \label{eqn:closed-griffigths-transversality}
\end{equation}
This together with $h^{(3,0)}=1$ is one of the main ingredients for the formulation of the $\mc N=2$ special geometry for Calabi-Yau manifolds.
We can study the variation of the complex structure by looking at how $\Omega$ changes under the complex structure deformations.
The form $\Omega$ and its derivatives $\nabla_{\text{cs}}^k\Omega$ span the complete space $H^3(Y)$, thus a derivative of any element of $H^3(Y)$ can be expressed as a linear combination of $\nabla_{\rm{cs}}^{k}\Omega$.
These linear combinations yield the Picard-Fuchs equations.

\subsection{Relative cohomology \label{rel_cohom}}

As discussed at the beginning of this section, the $\mc N=1$ superpotential is expressed as integrals of the holomorphic three-form over cycles and chains whose boundaries contain the curve $\Sigma$.
In order to give a unified description of integrals of these kinds
it is necessary to generalize the well-known homology theory for the manifold $Y$.
This is achieved by relative homology which, by definition, includes additionally to the closed three-cycles also three-chains with boundary containing the curve $\Sigma$ on which the D5-brane is supported.
Therefore, we review in the following its construction and essential properties and refer the reader to ref.~\cite{Spanier:1966} for a more detailed description.

First, we start dual to homology with the definition of the relative de Rham cohomology $H^k(Y,S)$ where $S$ denotes an arbitrary submanifold embedded into the ambient space $Y$ by $\iota:S\hookrightarrow Y$. This definition will guide us directly to the appropriate algebraic definition of relative homology exhibiting all the intuitive features mentioned above by simply applying Stokes theorem.
To construct the relative cohomology group $H^k(Y,S)$, we define relative forms by forming the direct sum of modules
\begin{equation}
        \Omega^k_\iota=\Omega^k(Y,S)=\Omega^k(Y)\oplus\Omega^{k-1}(S)\ .
        \label{eqn:relative-complex}
\end{equation}
Then, the relative differential $d$ on $\Omega_\iota^k$ is given by
\begin{equation}
        d(\Theta,\theta)=(d_Y\Theta,\iota^*\Theta-d_S\theta)\ ,
        \label{eqn:relative-differential}
\end{equation}
where $d_Y$, $d_S$ denote the de Rham differentials on $Y$ and $S$, respectively.
It is easily checked that $d^2=0$, thus, we obtain a complex of relative forms $(\Omega_{\iota}^\bullet,d)$. As usual the relative cohomology measures the difference between $d$-closed and $d$-exact relative forms. 
Hence, relative cohomology groups $H^k(Y,S)$ are constructed from the forms (\ref{eqn:relative-complex}) and the differential (\ref{eqn:relative-differential}) as quotients of closed relative $k$-forms by exact relative $k$-forms. In particular, an element in $H^k(Y,S)$ is represented by a pair of forms $(\Theta,\theta)$ obeying $d(\Theta,\theta)=0$ or equivalently
\begin{equation}
         d_Y\Theta=0\ ,\qquad \iota^\ast\Theta=d_S \theta\ .
\end{equation}
This implies that $\Theta$ is a non-trivial element in $H^k(Y)$ whose restriction $\iota^\ast\Theta$ to $S$ is trivial in $H^{k}(S)$. Furthermore, the equivalence relation in relative cohomology allows us to represent a class with representative $(\Theta,\iota^\ast \theta)$ seemingly very different, i.e.~
\begin{equation}
        (\Theta,\iota^\ast \theta)\sim (\Theta,\iota^\ast \theta)-d(\theta,0)=(\Theta-d_Y\theta,0)\ .
\end{equation}
This is particularly helpful to relate calculations with usual forms and chains to those with relative forms and cycles by carefully treating the de Rham exact form $d_Y\theta$ for the pullback forms $(0,\iota^\ast \theta)$ in relative cohomology.
Note that the relative cohomology covers also the de Rham cohomology as a special case obtained by setting $S$ to the empty set.

Parallel to the definition of relative cohomology, we define the relative homology group by introducing relative chains by 
\begin{equation}
 C_k^{\iota}=C_k(Y)\oplus C_{k-1}(S)\ .
\label{eqn:relative-chains}
\end{equation}
Next, we need an appropriate definition of a relative boundary operator. This is achieved by first introducing a natural pairing between relative forms and chains defined as
\begin{equation}
        \langle (\Theta,\theta),(A,a)\rangle=\int_A \Theta -\int_a\theta\ ,
\label{eqn:pairing}
\end{equation}
where we represent also relative chains by a pair $(A,a)$.
Then, the relative boundary operator $\partial$ on $C_k^\iota$ is introduced as the unique operator that is dual to the relative de Rham differential $d$ with respect to the pairing (\ref{eqn:pairing}).
By considering an exact relative form $d(\Omega,\omega)$ and application of Stokes theorem we obtain
\begin{equation}
        \partial(A,a)=(\partial_Y A-\iota_\ast a,-\partial_S a)\ ,
\end{equation}
where $\partial_Y$ and $\partial_S$ denote the boundary operators on $Y$ and $S$, respectively.
This squares also to zero and we define the relative homology groups $H_k(Y,S)$ as $\partial$-closed $k$-chains divided out by $k$-chains, that are $\partial$-boundaries of $(k+1)$-chains.
Then, it is easily checked that the pairing (\ref{eqn:pairing}) descends to a well-defined pairing between the (co-)homology groups as well. 
Again, every element in the relative group $H_k(Y,S)$ is represented by chains $(A,a)$ obeying $\partial(A,a)=0$ or
\begin{equation}
        \partial_S a=0\ , \qquad \partial_Y A=\iota_\ast (a)\,
\end{equation}
i.e.~$a\in H_{k-1}(S)$ is again trivial in $H_{k-1}(Y)$.
Therefore, these groups consist, as expected, of $k$-chains which are closed up to boundaries on $S$ and $k$-chains which have no boundaries, i.e.~are usual cycles. We note that there might be no additional $k$-chains in $H_{k}(Y,S)$ in case that there are no $(k-1)$-cycles $a$ which are trivial in the homology $H_{k-1}(Y)$. This happens, for example, when we consider $H_3(Y,S)$ for a non-trivial two-cycle $S$ in $Y$ like the curve $\Sigma$.

There is also a relative version of Poincar\'e duality which relates the relative (co-)homology groups due to the pairing (\ref{eqn:pairing}) in the usual fashion as
\begin{equation}
 H^k(Y,S)\cong H_{6-k}(Y,S).
\end{equation}

To gain a better understanding of the relative cohomology groups, one notes that there is the following short exact sequence of modules
\begin{equation}
        \xymatrix{
                0 \ar[r] & \Omega^{k-1}(S) \ar[r]^{\alpha} & \Omega^k(Y,S) \ar[r]^\beta & \Omega^k(Y) \ar[r] & 0
        },
        \label{eqn:relative-short-exact-sequence}
\end{equation}
which is just the definition \eqref{eqn:relative-complex} rewritten in an equivalent way.
More precisely, the map $\alpha$ is the natural embedding to the second summand of \eqref{eqn:relative-complex} and $\beta$ is the projection to the first summand.
From this sequence one obtains the long exact cohomology sequence
\begin{equation}
        \xymatrix @R=0.15in {
        \cdots \ar[r] & H^{k-1}(Y) \ar[r] & H^{k-1}(S) \ar[r] &
                H^k(Y,S)\ar[lld]\\ 
                &H^k(Y) \ar[r] & H^k(S) \ar[r] & H^{k+1}(Y,S)\ar[r]& \cdots }
        \label{eqn:relative-long-exact-sequence}
\end{equation}
The definition of an exact sequence gives the splitting of the relative cohomology group 
\begin{equation}
        H^k(Y,S) = \Kern \left( H^k(Y)\rightarrow H^k(S)
        \right)\oplus 
        \operatorname{Coker}\left( H^{k-1}(Y)\rightarrow H^{k-1}(S) \right)\ ,
        \label{eqn:decomposition-relative-group}
\end{equation}
where we observed the first summand already in the explicit construction presented above.
In the following, we denote the first and second summand by $H^k_v(Y)$ and $H^{k-1}_v(S)$ for convenience.

We now consider the case of $S=\Sigma-\Sigma_0$ where the two-cycle $\Sigma$ is wrapped by the D5-brane.
Since $\Sigma$ is complex one-dimensional, the first summand of $H^3(Y,\Sigma)$ only consists of $H^3(Y)$.\footnote{In order to work with the developed formalism of relative cohomology we have to consider $H^3(Y,\Sigma-\Sigma_0)$. However, we simplify our notation by just writing $H^3(Y,\Sigma)$ for the relative group.}
The second summand just consists of two-forms on $\Sigma$ which do not arise from the pull-back of non-trivial two-forms of $Y$.
As an example we note that the two-forms $s_A\lrcorner\chi_\kappa$ introduced in section \ref{relations} are elements of $H^2_v(\Sigma)$ when considered as forms on $\Sigma$.

As an application of the pairing \eqref{eqn:pairing} we rewrite the superpotential \eqref{open-closed_W} as
\begin{equation}
        W = \hat N_A\int_{\hat \Gamma_A}\Omega +
        N_a\int_{\Gamma_a}\Omega 
        = \hat N_A\hat \Pi^A+N_{a}\Pi^{a}
                \equiv \sum_iN_i\langle\Omega ,\Gamma_i^{\Sigma}\rangle\ ,
        \label{eqn:total-w-in-periods}
\end{equation}
where $\hat\Gamma_A\equiv (\hat\Gamma_A,0)$ and $\Gamma_a\equiv (\Gamma_a,\Sigma-\Sigma_0)$ denote a basis of three-cycles and three-chains in $H_3(Y,\Sigma)$ and $\Omega\equiv (\Omega, 0)$ the holomorphic three-form in $H^3(Y,\Sigma)$. As introduced before, $\hat N_A,N_a$ are the flux numbers and brane windings, respectively. On the right hand side of the equation $\Gamma_i^\Sigma$ form an integral basis of the relative homology group $H_3(Y,\Sigma)$.
Thus, we view the superpotential consisting of Gukov-Vafa-Witten potential and the chain integral as sum of relative periods.

We conclude with a remark about the expansion (\ref{eqn:total-w-in-periods}). 
In contrast to the chain integral (\ref{eqn:open-superpotential}) where we integrate over an arbitrary chain $\Gamma$ with $\partial \Gamma=\Sigma-\Sigma_0$, the chain integrals in the above expansion are performed with an integral basis $\Gamma^\Sigma_i$ of $H_3(Y,\Sigma)$.
As in the non-compact case of section \ref{non-compactCY} this integral basis of the relative group may consist of chains $\Gamma_a$ that have, in order to be integral, contributions of cycles $\hat \Gamma_A$ as well. Thus, the chain integrals in (\ref{eqn:total-w-in-periods}) may incorporate closed periods.
 
\subsection{From curves to divisors}
\label{sssec:curves-to-divisors}
Now let us turn to the open-closed moduli space $\mc M$.
As discussed in section \ref{openspectrum}, the infinitesimal open moduli are described by the holomorphic sections in the normal bundle of the curve which the D5-brane wraps.
Analogously to the consideration of $H^3(Y)$ for the closed string moduli, we use the elements of the relative group $H^3(Y,\Sigma)$ to probe the open-closed moduli space.
Mimicking as much of the familiar structure for complex structure moduli as possible, we proceed as follows.
We again obtain an absolute cohomology group by using the Lefschetz and Poincar\'e duality\footnote{In the following wee will use this isomorphism between relative and absolute group quite frequently without referring to it at every place.}
\begin{equation}
        H_3(Y,\Sigma)\cong H^3(Y,\Sigma)\cong H_3(Y-\Sigma)\cong H^3(Y-\Sigma)\ . 
        \label{eqn:lefschetz-poincare-for-curve}
\end{equation}
In order to infinitesimally analyze the moduli dependence of the objects in this group, we have to study the so-called mixed Hodge structure of $H^3(Y,\Sigma)$.
For completeness we have given the mixed Hodge structure of $H^3(Y,\Sigma)$ in appendix \ref{app:mixed-hodge}.
However, it will turn out, for practical as well as conceptual purposes, it is mathematically more adequate to consider codimension one objects, i.e.\ divisors, than higher codimensional objects.

The cohomology group \eqref{eqn:lefschetz-poincare-for-curve} as well as the mixed Hodge structure governing the moduli dependence only depend on the open manifold $U\equiv Y-\Sigma$.
Hence, we can replace $Y$ and $\Sigma$ by objects $\tilde Y$ and $\Div$ satisfying
\begin{equation}
        \tilde Y-\Div = U = Y-\Sigma\ .
        \label{eqn:open-manifold}
\end{equation}
The deformations of the pair $(Y,\Sigma)$ which we denote by $\operatorname{Def}(Y,\Sigma)$ are described more adequately by an auxiliary pair $(\tilde Y, \Div)$.
One canonical way to construct $\tilde Y$ and $\Div$ is to blow-up $Y$ along $\Sigma$ \cite{Griffiths:1978yf}.
We set $\Div$ to be the exceptional divisor of the blow-up procedure.
By construction, it is clear that $H^3(\tilde Y-\Div)\cong H^3(Y-\Sigma)$.
Furthermore, the deformation theory $\operatorname{Def}(Y,\Sigma)$ is equivalent to $\operatorname{Def}(\tilde Y,\Div)$ such that the variation of mixed Hodge structures of $H^3(Y,\Sigma)$ and $H^3(\tilde Y,\Div)$ over the moduli space are equivalent.

Before we proceed let us discuss the geometry of $\Div$ and $\tilde Y$ in more detail.
First, we turn to the exceptional divisor $\Div$.
It is the projectivization of the normal bundle of $\Sigma$ in $Y$, i.e.\ $\P (N_Y\Sigma)$ which is a $\P^1$-bundle over $\Sigma$.
On any projectivization of a complex vector bundle there exists a natural line bundle which is called tautological bundle $T$ which is the analogue of $\mc O_{\P^n}(-1)$ on $\P^n$.
The line bundle $T$ is also the normal bundle of $\Div$ in $\tilde Y$.
Since $T$ does not have any holomorphic section, $\Div$ is rigid and thus has no deformation moduli.
Furthermore, the cohomology ring of $\Div$ is generated by $\eta=c_1(T)$ as an $H^\bullet(\Sigma)$-algebra, i.e.
\begin{equation}
        H^\bullet(\Div)=H^\bullet(\Sigma)\langle\eta\rangle
        \label{eqn:cohomology-ring-exceptional-divisor}
\end{equation}
with the following relation
\begin{equation}
        \eta^2=c_1(N_Y\Sigma)\wedge\eta=-c_1(\Sigma)\wedge\eta\ .
        \label{eqn:relation-divisor}
\end{equation}
Thus, $H^\bullet(\Div)$ is generated by $c_1(T)=\eta$ with elements of $H^\bullet(\Sigma)$ as coefficients.
Consequently, the Hodge diamond looks as follows
\begin{equation}
        \begin{tabular}{ccccc}
                &&1&&\\
                &g&&g&\\
                0&&2&&0\\
                &g&&g&\\
                &&1&&
        \end{tabular}\ ,
        \label{tab:hodge-diamond-exceptional-divisor}
\end{equation}
where $g$ is the genus of $\Sigma$.
Here, the holomorphic one-forms are the Wilson lines $a_I$ of $\Sigma$, the $(2,1)$-forms are of the form $a_I\wedge\eta$ and the two $(1,1)$-forms are given by $\eta$ and $c_1(N_\Div\Sigma)$.
Using twice the adjunction formula, one time for $\Sigma$ as a divisor in $\Div$ and another time for $\Div$ as a divisor in $\tilde Y$, we obtain with (\ref{eqn:first-chern-of-blowup}):
\begin{equation}
c_1(N_\Div\Sigma)=-c_1(\Sigma)-2\eta\ .
\end{equation}

Now we describe the geometry of $\tilde Y$ in more detail.
We first observe that the blow-up $\tilde Y$ is again a compact K\"ahler manifold \cite{Voisin2002a} since the blow-up of a K\"ahler manifold along a complex submanifold is always K\"ahler, too.
Secondly, $\tilde Y$ can still be embedded into $\P^N$ for some $N$, i.e.\ it is projective, if $Y$ is projective.
In the case of $Y$ being a Calabi-Yau threefold this is always true.
This implies that we can always find algebraic equations defining $Y$ and $\tilde Y$.
However, $\tilde Y$ is not a Calabi-Yau manifold anymore.
Using the general formula for the first Chern class of a blow-up \cite{Griffiths:1978yf}
\begin{equation}
        c_1(\tilde Y) = \pi^*c_1(Y)-\eta\ ,
        \label{eqn:first-chern-of-blowup}
\end{equation}
we see that the first Chern class of $\tilde Y$ does not vanish.
Here, we used the usual notation $\pi^*$ for the pullback of forms from $Y$ to $\tilde Y$ induced by the projection $\pi:\tilde Y\rightarrow Y$. 
Furthermore, we also use that the Poincar\'e dual of the exceptional divisor is just the first Chern class of its normal bundle in $\tilde Y$.
Secondly, the cohomology ring of $\tilde Y$ has the form \cite{Griffiths:1978yf}
\begin{equation}
        H^\bullet (\tilde Y) = \pi^*H^\bullet(Y)\oplus H^\bullet(\Div)/\pi^*H^\bullet(\Sigma)\ .
        \label{eqn:cohomology-ring-x-tilde}
\end{equation}
Since $H^{3,0}(\Sigma)=H^{3,0}(\Div)=0$ for dimensional reasons, it follows that $H^{3,0}(\tilde Y)\cong \pi^*H^{3,0}(Y)$, i.e.\ it is still one-dimensional as for the original Calabi-Yau space $Y$.
However, the holomorphic three-form on $\tilde Y$ has $\Div$ as its zero locus as can be seen as follows.
The first Chern class of a holomorphic vector bundle $E$ describes the zero locus of a single section of the determinant line bundle $\det E$.
We can apply this for $E=T^*\tilde Y$ by reading \eqref{eqn:first-chern-of-blowup} in terms of its Poincar\'e dual $\Div$ and using $c_1(Y)=0$.

\subsection{Two ways towards Picard-Fuchs equations \label{subsec:PF}}

Now we are aiming at the description of the moduli dependence of $H^3(\tilde Y-\Div)$.
This dependence can be characterized by Picard-Fuchs equations.
In the following we will discuss two possible ways to derive these equations in principle.
The cases of most interest are those where $Y$ and $\tilde Y$ are described as 
complete intersections in (weighted) projective spaces where powerful methods 
like residue representation of cohomology, Griffiths-Dwork reduction method etc.\ are available.

The first way is to use the mixed Hodge structure\footnote{For more details, cf.\ appendix \ref{app:mixed-hodge}.} and its variations.
However, we will quickly specialize to the case of the divisor $\Div$.
In general, the mixed Hodge structure is a free abelian group $H_\Z$ with a decreasing Hodge filtration $F^mH_\C$ and an increasing weight filtration $W_kH_\C$ where $H_\C$ is the complexification $H_\Z\otimes_\Z \C$.
For a divisor $\Div$ this takes the following form.
First, we note the following isomorphism
\begin{equation}
        \xymatrix{\phi:H^3(\tilde Y-\Div)\ar[r]^(0.4)\sim & \displaystyle\bigoplus_{p+q=3}H^q(\tilde Y,\Omega^p_{\tilde Y}(\log \Div))}.
        \label{eqn:iso-normal-crossing-divisor}
\end{equation}
By $\Omega_{\tilde Y}^k(\log D)$ we mean holomorphic $k$-forms on $\tilde Y$ that are locally generated by e.g.\ $dz^1,dz^2$ and $d\log z_3=dz^3/z_3$ with holomorphic functions as coefficients for a divisor locally given by $z_3=0$.\footnote{Because of $d\log z_3$ these forms are denoted by $\Omega^1_{\tilde Y}(\log \Div)$. In general they have logarithmic singularities along $\Div$. As usual $\Omega_Y^k(\log \Div)$ is then given by the $k$-th exterior power of $\Omega^1_Y(\log \Div)$.}
One can use $\Omega_{\tilde Y}^k(\log \Div)$ to define the Hodge and weight filtrations for $H^3(\tilde Y-\Div)$.
Then the filtrations have the form
\begin{equation}
        F^mH^3 = \bigoplus_{p\geq m}H^{3-p}(\tilde Y,\Omega_{\tilde Y}^p(\log \Div))
        \label{eqn:hodge-filtration-log-detailed}
\end{equation}
and
\begin{equation}
        W_{-1}H^3 = 0\ , \quad
        W_{0}H^3 =H^3(\tilde Y)\ ,\quad
        W_{1}H^3 = H^3(\tilde Y-\Div)\ .
                \label{eqn:weight-filtration-log}
\end{equation}
Additionally, the mixed Hodge structure has graded weights $\operatorname{Gr}_k^WH^3=W_{-k+3}H^3/W_{-(k+1)+3}H^3$ that take the following form for the divisor $\Div$
\begin{equation}
        \operatorname{Gr}^W_{3}H^3 =W_0H^3/W_{-1}H^3\cong H^3(\tilde Y)\quad,\quad
        \operatorname{Gr}^W_{2}H^3 =W_1H^3/W_0H^3\cong H^{2}(\Div)\ .
        \label{eqn:graded-weights-log}
\end{equation}
The reason to consider these (graded) weights is the following:
The mixed Hodge structure is defined such that the Hodge filtration $F^m H^3$ induces a pure Hodge structure on each graded weight, i.e.\ on $\operatorname{Gr}_2^WH^3$ and on $\operatorname{Gr}_3^WH^3$.  
Thus, the following two induced filtrations on $\operatorname{Gr}_3^WH^3$
\begin{equation}
        H^3(\tilde Y)\cap F^3H^3 \subset  H^3(\tilde Y)\cap F^2H^3 \subset H^3(\tilde Y)\cap F^1H^3 \subset H^3(\tilde Y)\cap F^0 H^3 = H^3(\tilde Y)
        \label{eqn:pure-hodge-structure-weight-3}
\end{equation}
and on $\operatorname{Gr}_2^WH^3$
\begin{equation}
        H^2(\Div)\cap F^2H^3  \subset H^2(\Div)\cap F^1H^3 \subset H^2(\Div)\cap F^0H^3 =H^2_v(\Div)
        \label{eqn:pure-hodge-structure-weight-2}
\end{equation}
lead to pure Hodge structures on $H^3(Y)$ and $H^2_v(\Div)$, respectively.
Here, for example, $H^2(\Div)\cap F^2H^3$ should be understood as follows: The second summand $H^2_v(\Div)$ of \eqref{eqn:decomposition-relative-group} represents the part of $H^2(\Div)$ which is contained in the relative group $H^3(\tilde Y,D)$.
Thus, we use the isomorphism $\phi$ of \eqref{eqn:iso-normal-crossing-divisor} to obtain their logarithmic counterparts.
Then, we intersect $\phi(H^2(\Div))$ with $F^2H^3$.
Analogously to the case of closed string moduli, $H^3(\tilde Y-\Div)$ forms a bundle $\mc H^3$ over the open-closed moduli space $\mc M$ with the Gau{\ss}-Manin connection $\nabla$.
Each $F^mH^3$ forms a subbundle $\mc F^m$ of $\mc H^3$.
As already discussed, the Gau{\ss}-Manin connection has the following important transversality property
\begin{equation}
        \nabla\mc F^p \subset \mc F^{p-1}\otimes\Omega_{\mc M}^1\ .
        \label{eqn:transversality}
\end{equation}
Combining this with (\ref{eqn:pure-hodge-structure-weight-3}) and (\ref{eqn:pure-hodge-structure-weight-2}) and assuming that $\{\nabla_{z,u}\mc F^k\}$ span $\mc F^{k-1}$, we see that
\begin{equation} \label{eqn:chain-variations-closed-open-moduli}
        \xymatrix @C=0.4in {
        \mc H^3(\tilde Y)\cap \mc F^3 \ar[dr]^(0.6){\nabla_u}\ar[r]^{\nabla_z} & 
        \mc H^3(\tilde Y)\cap \mc F^2 \ar[dr]^(0.6){\nabla_u}\ar[r]^{\nabla_z} & 
        \mc H^3(\tilde Y)\cap\mc F^1 \ar[dr]^(0.6){\nabla_u}\ar[r]^{\nabla_z} & 
                \mc H^3(\tilde Y)\cap\mc F^0 \ar[d]^(0.4){\nabla_z,\nabla_u}\\
        &\mc H^2(\Div)\cap\mc F^2 \ar[r]_{\nabla_z,\nabla_u} &  
        \mc H^2(\Div)\cap\mc F^1  \ar[r]_{\nabla_z,\nabla_u} &
                \mc H^2(\Div)\cap \mc F^0
                }
\end{equation}
where $z$ denotes the closed string moduli and $u$ the open string moduli.
Here, again, we should understand the groups under the isomorphism $\phi$, i.e.\ all forms occurring in \eqref{eqn:chain-variations-closed-open-moduli} are logarithmic three-forms.
If we want to obtain a two-form representative of e.g.\ $\eta\in\mc H^2(\Div)\cap \mc F^2$, we consider $\phi^{-1}(\eta)$ which is an element of $H^2_v(\Div)$ and thus also an element of $H^2(\Div)$.
As we can see the variations of the mixed Hodge structure has two levels: The closed string sector and a sector which mixes the open and closed moduli.
As has been pointed out in \cite{Griffiths:1979rt}, there exist differential equations obeyed by the relative periods of $H^3(Y_t,D_t)$ where $D_t$ denotes a family of divisors in the family of manifolds $Y_t$.
In particular this covers our setting for the blow-up $\tilde Y$ by $\Div$.
The resulting equations for the relative periods of $H^3(\tilde Y,\Div)$ are the advertised Picard-Fuchs equations.
One possible way to obtain these Picard-Fuchs equations explicitly may be given by residue representations for the relative forms of $H^3(\tilde Y,\Div)\cong\bigoplus H^q(\tilde Y,\Omega_{\tilde Y}^p\log \Div)$ making explicit use of algebraic equations defining $\tilde Y$ and $\Div$ as complete intersections in the ambient space, cf.\ section \ref{sssec:curves-to-divisors}.
The main difficulty of this approach is to find explicit residue representation of $H^3(\tilde Y,\Div)$.

The second ansatz relies on the study of the complex structure moduli of the blow-up $\tilde Y$.
Since ${\rm Def}(\tilde Y,\Div)$ form a subset of 
deformations of $\tilde Y$, we can use the available techniques for ordinary
complex structure deformations to describe the relevant Picard-Fuchs equations.
Using the algebraic equations
for $\tilde Y$ as a complete intersection, it is possible to apply the
Griffiths-Dwork  reduction method for residue representation of the unique
holomorphic three-form $\tilde \Omega$ which is the proper transform of
$\Omega$ in $H^{(3,0)}(\tilde Y)$. 
This can be seen from $\iota^\ast(\tilde\Omega)\equiv0$ on the divisor $\Div$ as argued in section \ref{sssec:curves-to-divisors} implying that $\tilde\Omega$ is an element of the first summand $\Kern(H^3(\tilde Y)\rightarrow H^3(\Div))$ in (\ref{eqn:decomposition-relative-group}), i.e.~it can be represented as $(\tilde\Omega,0)$ in the relative cohomology on $\tilde Y$. Thus, \eqref{eqn:cohomology-ring-x-tilde} allows us to represent $\tilde\Omega$ as a pull-back form of $H^3(Y)$.  
In this way we obtain Picard-Fuchs operators $\mc L_i$ for $\tilde \Omega$ with
\begin{equation}
\mathcal{L}_i\tilde\Omega=d\alpha_i\ ,
\end{equation}
where $\alpha_i$ denote two-forms constructed by the Griffiths-Dwork method.
Furthermore, we expect that the full effective superpotential $W$ is a linear combination of the solutions to the corresponding Picard-Fuchs system with the inhomogeneous piece given by functions obtained by integrating $d\alpha_i$ over chains.
Indeed, we can replace all quantities occurring in the expansion of the
superpotential into relative periods (\ref{eqn:total-w-in-periods}) 
by corresponding relative periods on $\tilde Y$. First, we use the isomorphism (\ref{eqn:lefschetz-poincare-for-curve}) to replace 
\begin{equation}
        H_3(Y,\Sigma)\cong H_3(\tilde Y,\Div)
\end{equation}
as well as the corresponding integral basis $\Gamma^\Sigma_i$ and
$\Gamma^\Div_j$. Then, we replace the holomorphic three-form $\Omega$ on $Y$
by its proper transform $\tilde \Omega$ on $\tilde Y$. This leads to the following expression for the superpotential,
\begin{equation}
        W=\sum_j\tilde N_j\left<\tilde \Omega,\Gamma^\Div_j\right>\ ,
\end{equation}
where $\tilde N_j$ denote appropriately chosen integers.
Next, we observe that the superpotential is annihilated by the Picard-Fuchs operators $\mc L_i$ for $\tilde \Omega$ as it just consists of the integral of $\tilde \Omega$ over the relative cycles of $H_3(\tilde Y, \Div)$.
Due to the rigidness of the exceptional divisor $\Div$ in $\tilde Y$ all deformations are now complex structure deformations of $\tilde Y$.
Thus we can choose a topological integral basis of  $H_3(\tilde Y, \Div)$ which is 
not affected by the complex structure deformations on $\tilde Y$.    
This is in contrast to the original chains which depend 
on deformations of the boundary curves $\Sigma$ in $Y$. It is 
a main advantage of the prescribed blow-up procedure that all 
moduli dependence of the relative periods of $\tilde \Omega$ is 
captured by the dependence of $\tilde \Omega$ itself. 

The superpotential $W$ is a linear combination of the solutions to the
Picard-Fuchs system on $\tilde Y$.  In general there might be more 
complex structure deformations of $\tilde Y$ than  ${\rm Def}(\tilde Y,\Div)$, 
so that one has to identify the deformations, that correspond to the original 
deformation problem ${\rm   Def}(\tilde Y,\Div)$ and to restrict the 
dependence of the solutions to the Picard-Fuchs system accordingly. 

Comparison of the two methods reveals their advantages and drawbacks.
On the one hand, it is necessary for the starting point of the first approach
to find the residue representation of the logarithmic forms.
Then the remaining calculations should follow straight forwardly.
On the other hand, it is clear for the second approach how to start, i.e.\ the
residue representation of the holomorphic three-form of $\tilde Y$.
However, the identification of the right moduli for the pair $(\tilde Y, \Div)$ from the complex structure moduli $H^1(\tilde Y, T\tilde Y)$ is crucial to obtain the relevant moduli dependence.


\subsection{An explicit example of the blow-up}\label{blow-up-example}

In this section we construct an example for which the blow-up procedure can be 
carried out explicitly. We will start with a non-compact example and later 
comment on possible compact realizations. 
The non-compact Calabi-Yau space we will consider is a complex line bundle 
$Y \rightarrow \cB_n$ over a del Pezzo surface $\cB_n$.  The del Pezzo 
surface $\cB_n$ is a $\bbP^2$ for which 
$n$ generic points are blown-up to $\bbP^1$'s.
We also wrap a space-time filling D5-brane on $Y$ such that it sits at a  
point $x$ on $\cB$ and also extends along the non-compact complex fiber in 
$Y$. The D5-brane can move on the del Pezzo surface which corresponds to 
moving the point $x$. Let us first examine 
what is the minimal number of blow-ups $n$ in $\cB_n$ for which the point 
$x$ can be moved with respect to a fixed reference point $x_0$ in $\cB_n$ 
such that the movement cannot be compensated by 
a coordinate redefinition. We count eight coordinate redefinition symmetries 
of $\bbP^2$ which is the dimension of $PGL(3,\C)$ acting on the projective 
coordinates $(x_1,x_2,x_3)$. Hence, we have to mark
at least four points in $\bbP^2$, each specified by two coordinates, to 
fix the coordinate freedom on $\bbP^2$. The movement of the fifth point 
then cannot be compensated by a coordinate redefinition.
Thus, the fifth point gives rise to two complex open moduli describing its position in $\P^2$.
Hence, we are lead to minimally consider $\cB_3$ with one fixed reference 
point $x_0$ in order to have open moduli.\footnote{%
This should be compared to the non-compact examples of section \ref{non-compactCY}
where the D5-brane is a point on a Riemann surface $\mathcal{Y}$. If 
$\mathcal{Y}$ has genus $g=1$, one needs to fix the reference point $x_0$
to fix the freedom of coordinate choice.} 

The canonical class of $\cB_3$ is given by $K_{\cB_3}=-3\ell+e_1+e_2 + e_3$,
where $\ell$ is the hyperplane divisor and $e_i$ are the three exceptional 
$\bbP^1$ blow-up divisors. The Calabi-Yau 
manifold $Y$ is then given by 
\beq
   Y = \cO_{\cB_3}(K_{\cB_3}) \longrightarrow \cB_3\ 
\eeq
and can described torically as in (\ref{quotient}) by the four charge vectors 
\begin{equation} 
\begin{array}{rl} 
Q^{1}&=(-1,          -1,  \phantom{-}1,\phantom{-}0, \phantom{-}0, \phantom{-}0, \phantom{-}1)\ ,\\
Q^{2}&=(-1,\phantom{-}1,  \phantom{-}0,\phantom{-}0, \phantom{-}0, \phantom{-}1,           -1)\ ,\\
Q^{3}&=(-1,\phantom{-}0,  \phantom{-}1,          -1, \phantom{-}1, \phantom{-}0, \phantom{-}0)\ ,\\
Q^{4}&=(-1,\phantom{-}1,            -1,\phantom{-}1, \phantom{-}0, \phantom{-}0, \phantom{-}0)\ .
\end{array}
\end{equation} 
The latter can be viewed as coefficients of linear relations among the vectors
$(1,0,0)$, $(1,1,0)$, $(1,1,1)$, $(1,0,1)$, $(1,-1,0)$, $(1,-1,-1)$ and  $(1,0,-1)$ which
span the non-compact toric fan for $Y$ from the origin in $\mathbb{R}^3$. In
the plane $(1,x,y)$, $(x,y)\in \mathbb{R}^2$ the fan contains the hexagonal 
toric polyhedron for $\cB_3$, see Figure 1.
\begin{figure}[!ht]
\begin{center}
\includegraphics[height=3cm]{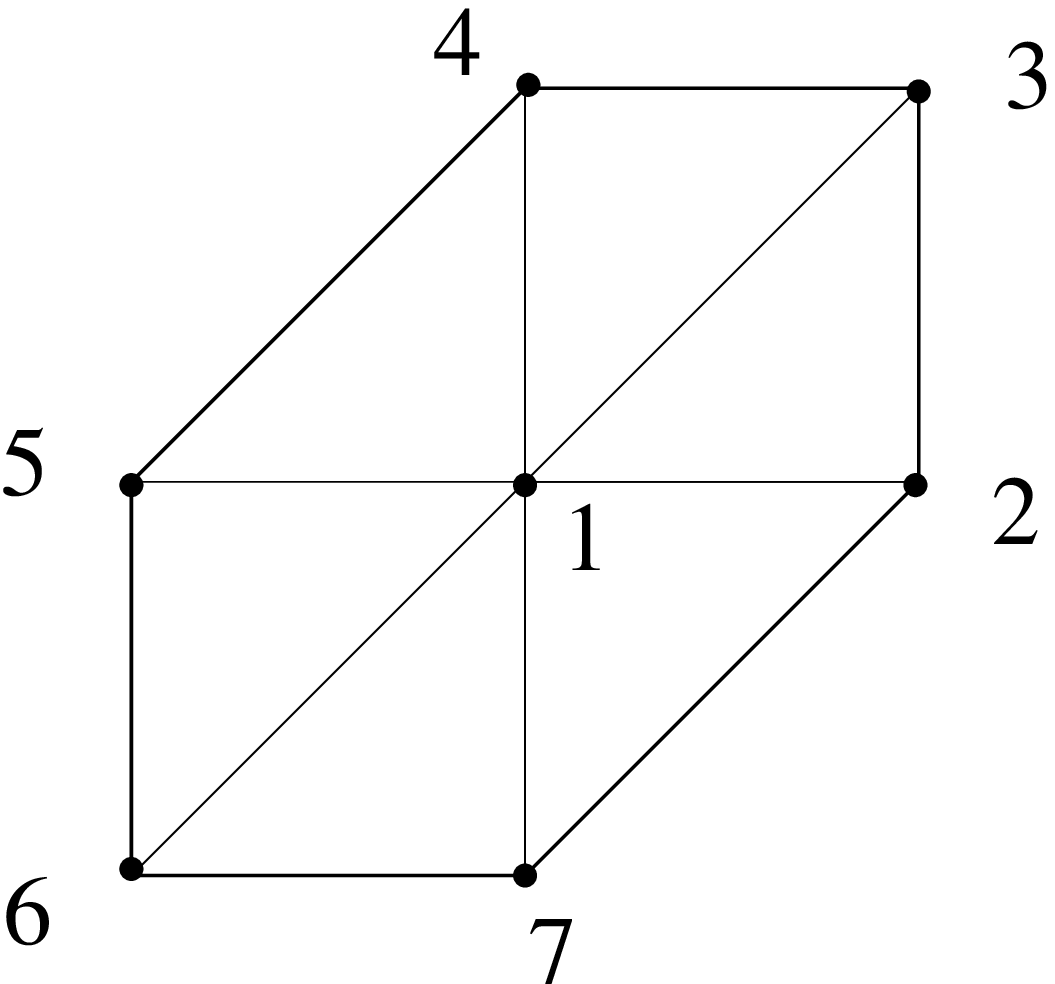}
\caption{\small  \label{poly} Polyhedron for $\cB_3$.}
\end{center}
\end{figure}
Each point in the Figure 1 is associated to a coordinate $x_i\in \mathbb{C}$
and the Stanley-Reisner ideal $Z$ is generated  by all sets $\{x_{i_1}=\ldots=x_{i_r}=0\}$,
where $\{i_1,\ldots,i_r\}$ are not indices of a common triangle in the figure.      

Since $Y$ is toric, it has no complex structure moduli. 
However, once we include the D5-brane on the fiber at $x$ (and fix the reference line at $x_0$) one finds two complex
open moduli $\zeta_1,\zeta_2$ which correspond to the two complex 
dimensions in which $x$ can move on $\cB_3$. 

Next we want to use the insights of section 
\ref{sssec:curves-to-divisors} and blow up the line 
$\Sigma$ wrapped by the D5-brane and a reference line $\Sigma_0$ into a divisor.
We note that $\Sigma$ intersects $\cB_3$ in the point $x$ while a 
reference line $\Sigma_0$ intersects $\cB_3$ in the rigid point $x_0$. We recall that the 
blow-up divisors are the projectivizations of the normal bundles 
$\bbP(N_Y \Sigma)$ and $\bbP(N_Y \Sigma_0)$.
However, for $x$ and $x_0$ not on the exceptional $\bbP^1$'s in $\cB_3$ we 
can simply identify the blow-up divisors as the blow-ups of $x$ and $x_0$ 
into two new $\bbP^1$'s. Therefore,
the new base of $\tilde Y$ is the del Pezzo surface $\cB_5$. 
We can construct $\tilde Y$ as the line bundle
\beq
  \tilde Y = \cO_{\cB_5}(K_{\cB_3}) \longrightarrow \cB_5\ ,
\eeq
where $K_{\cB_3}=-3\ell+e_1+e_2 + e_3$ only includes $e_1,e_2,e_3$ as in $Y$.
Now, however, the first Chern class does not vanish
\beq
   c_1 (\tilde Y ) = - \nu^*(e_4) - \nu^*(e_5)\ ,
\eeq
where $\nu: \tilde Y \rightarrow \cB_5$ is the projection to the base. 
This is in accord with the general formula 
\eqref{eqn:first-chern-of-blowup} and matches our 
expectation that $\tilde Y$ is not Calabi-Yau.

We can also investigate what happened to the open moduli of the D5-brane 
in this set-up. Clearly, after blowing-up the
exceptional $\bbP^1$'s cannot be moved within $\cB_5$. This corresponds to 
the general fact the blow-up divisors are rigid. Thus, the two deformations 
$\zeta_1,\zeta_2$ of $\Sigma$ have disappeared,
but the del Pezzo surface $\cB_5$ has now two complex structure 
deformations $z_1,z_2$. These complex structure 
deformations can be canonically identified with $\zeta_1,\zeta_2$, and, by 
studying the periods depending on $z_1,z_2$, we implicitly solve the 
original deformation problem for the curve $\Sigma$. Hence the complex
structure moduli space of $\tilde Y$ captures the deformation space 
of the brane moduli on $Y$. 

Even for this non-compact Calabi-Yau threefolds, we have to ensure tadpole cancellation.
Since all directions normal to the D5-brane are compact, O5-planes with negative D5-brane charge have to be included in order to obtain a vanishing net R--R charge.
Therefore we consider the following involution on the del Pezzo base whose action on the basis $(\ell,e_1,e_2,e_3)$ of the cohomology lattice is given by \cite{BBGW}
\begin{equation}
        \sigma=
        \begin{pmatrix}
                2 & 1 & 1 & 1\\
                -1 & 0 & -1 & -1\\
                -1 & -1 & 0 & -1\\
                -1 & -1 & -1 & -0
        \end{pmatrix}.
        \label{eqn:orientifold-involution-delpezzo}
\end{equation}
This involution has four fixpoints on the del Pezzo surface.
We extend this involution to $Y$ by demanding it to act trivially on the fiber such that the O5-planes extend along the fiber and intersect $\mc B_3$ in four points.
Therefore a consistent configuration requires eight D5-branes in the covering space.
We conclude the example by noting that this non-compact situation can be generalized to compact examples.
We replace the fibration of $Y$ with an elliptic fibration giving rise to a well-known elliptically fibered Calabi-Yau.
The methods discussed in section \ref{subsec:PF} should be directly applicable to these examples and the open mirror symmetry can be studied in detail.

\section{Conclusions} 
 
In this work we have analyzed the four-dimensional $\mc N=1$ effective action for
a D5-brane wrapping a two-cycle in a Calabi-Yau orientifold.
We have performed the dimensional reduction 
of the six-dimensional Dirac-Born-Infeld and 
Chern-Simons actions coupled to the ten-dimensional bulk supergravity
action. We were able to derive the $\mathcal{N}=1$ characteristic 
data encoding the kinetic terms for the chiral and vector multiplets
including the gaugings. Of particular interest was the derivation of the 
$\cN=1$ potential which was shown to consist of both F- and D-term contributions.
 
Before performing the actual dimensional reduction 
we discussed that it is important to consider 
the interrelations of the complex structure deformations of the Calabi-Yau orientifold 
$Y/\mathcal{O}$ and the moduli of the D5-brane. This was captured by the infinitesimal analysis of section 
\ref{relations} where concrete relations on the open-closed moduli space 
were derived. We found an explicit form for the deformations $\delta
(\iota^\ast g)$ of the induced metric on the two-cycle $\Sigma$ due to 
complex structure deformations of the ambient space. These variations 
led to an essential contribution to the F-term potential. 
In order to complete the calculation of the F-term potential, we also 
had to consider the couplings of four-dimensional 
non-dynamical three-form fields in the D5-brane action.
After performing a formal dualization procedure 
for these fields we were able to derive the complete
scalar potential in the presence of a D5-brane and background 
R--R three-form flux $F_3$. In fact, the correct interpretation of the 
flux quantum numbers of $F_3$ was given in \cite{Beasley:2002db} as labeling 
quantum mechanical states of the system. Together with 
the knowledge of the $\cN=1$ K\"ahler potential we then 
determined the complete effective superpotential 
\eqref{open-closed_W} entirely by dimensional reduction of the 
bosonic fields. 

After the discussion of this F-term potential we identified the remaining 
terms in the scalar potential as D-terms.
One D-term arose due to the NS--NS-tadpole and needed to be cancelled 
by the tension of the O5-planes in order for the set-up to be stable. 
The other terms were induced by gaugings of chiral 
fields by the brane vector and the bulk vectors. We 
showed that if the D5-brane and its orientifold image are in different homology 
classes, a D-term enforces the NS--NS B-field moduli to be identical to the D5-brane 
gauge flux. The second D-term was induced by non-trivial NS--NS three-form flux.
Studying the dimensional reduction of the complete action we also succeeded 
in giving a complete list of the $\mc N=1$ coordinates incorporating the 
corrections due to open string moduli. Besides the effective superpotential, 
we read off the effective $\mc N=1$ K\"ahler potential and gauge kinetic function.

The derived effective action describing a generic compactification 
allows for various phenomenological applications. Let us mention three examples here.
Firstly, it can be used to study mechanisms of D-brane inflation using 
e.g.\ D5-branes on the vanishing $S^2$ of the conifold \cite{Becker:2007ui} or
D-brane Wilson line moduli \cite{Avgoustidis:2006zp}.
Secondly, our results can be used to study dynamical supersymmetry breaking 
in the presence of D5-branes. 
In particular, \cite{Aganagic:2007py} used geometric transitions to construct 
stringy scenarios of dynamical supersymmetry breaking with dynamical D5-branes on 
vanishing two-cycles. These scenarios were constructed in non-compact Calabi-Yau 
geometries where many of the bulk and D5-brane fields are non-dynamical. 
To study the compact embedding of these models the derived effective action 
of the full supergravity with D5-branes will be of importance. This also
applies to explicit GUT model constructions in Type IIB compactifications 
on non-trivial Calabi-Yau orientifolds. It would be interesting to 
find explicit models with intersecting D5-branes 
using similar techniques as developed for intersecting D7-branes 
in refs.~\cite{BBGW,Grimm:2008ed}. 

Since the main focus of our work concentrated on the derivation of the 
effective action for a generic compactification, we have not addressed  
the question of moduli stabilization so far. However, there are some immediate
conclusions which can be drawn from our analysis. Most importantly, one notes
that the dilaton multiplet $S$ does not appear in the superpotential which is 
induced by three-form fluxes or the presence of the D5-brane. Similar to the heterotic
string the flux, which allows to tunebly stabilize the dilaton in a 
compactification with O3- and O7-planes, is projected out in the O5-orientifolds.
However, the R--R flux does induce an additional D-term potential with a
different dilaton power. Unfortunately, this is not sufficient to stabilize
the dilaton since both the D-terms as well as the F-terms contribute positive
definite terms to the scalar potential. The latter fact can be traced back to 
the presence of the no-scale structure with a positive definite 
scalar potential \eqref{F-termpotential}.
Clearly, this no-scale structure can be broken due to perturbative and non-perturbative 
corrections. It would be interesting to investigate whether these corrections
can stabilize the dilaton and compare the situation with the 
well-known heterotic string results.
Furthermore, it is of equal relevance to study the backreaction of the
included fluxes on the geometry. In a fully backreacted set-up the background might 
no longer be a Calabi-Yau manifold or may be strongly warped.

In the second part of this work we discussed the geometric structure
underlying the effective $\mc N=1$ theory.
Our analysis was concentrated on the effective flux and D5-brane superpotential.
This superpotential can be expressed in terms of relative periods
which encode the closed string flux as well as the brane windings.
To investigate the moduli dependence of the superpotential
we developed a canonical procedure to study the deformations of 
the complex structure of $Y$ and the deformations of the curve $\Sigma$ on 
an equal footing.
We associated to $\Sigma$ a divisor $\Div$ by means of the blow-up 
along $\Sigma$ of $Y$ to $\tilde Y$.
We gave two possible ways to describe the deformations of the pair 
$(\tilde Y, \Div)$.
For the first one, we used $H^3(\tilde Y, \Div)$ to replace $H^3(Y,\Sigma)$ 
and the fact $\operatorname{Def}(Y,\Sigma)=\operatorname{Def}(\tilde Y, \Div)$.
Then we employed the representation of $H^3(\tilde Y,\Div)$ by cohomologies of
the forms with logarithmic singularities along $\Div$ to define a mixed Hodge
structure and its variations.
This enabled us to recover as many methods as possible familiar from the
closed string moduli.
In particular, we can use the flat Gau{\ss}-Manin connection to obtain
Picard-Fuchs equations obeyed by relative periods of $H^3(\tilde Y, \Div)$.
For the second approach, we embed the deformations of the pair $(\tilde
Y,\Div)$ into the complex structure deformations of $\tilde Y$.
This way the derivation of the Picard-Fuchs equations reduces to the
Griffiths-Dwork method for $\tilde Y$ and the identification of moduli.

For future works it would be interesting to work out explicit examples 
in more detail using one or both of the two presented methods.
This would involve the determination of the embedding equations for $\tilde Y$
and/or $\Div$, the residue representations of logarithmic forms and the
analysis of the mapping of $\operatorname{Def}(\tilde Y,\Div)$ into $\operatorname{Def}(\tilde Y)$.
A successful computation would allow us to compare with the results of
\cite{Walcher:2006rs,Morrison:2007bm,Krefl:2008sj,Knapp:2008uw,Jockers:2008pe}
and investigate the question of unobstructed open moduli.
It would also be interesting to study the connection of the exceptional divisor $\Div$ with the divisor given in \cite{Lerche:2002ck,Lerche:2002yw,Jockers:2008pe}.

\vskip 1cm 
 {\noindent  {\Large \bf Acknowledgements}} 
 \vskip 0.5cm  
We gratefully acknowledge helpful discussions with Babak Haghighat, Olaf Hohm, Daniel Huybrechts, Hans
 Jockers, Sheldon Katz, Peter Mayr, Marco Rauch, Emanuel Scheidegger, Stephan Stieberger, Johannes
 Walcher and Eric Zaslow. 
This work was partially supported by the 
European Union 6th framework program MRTN-CT-2004-503069 
``Quest for unification", MRTN-CT-2004-005104 ``ForcesUniverse", MRTN-CT-2006-035863  
``UniverseNet" and SFB-Transregio 33 ``The Dark Universe" by the DFG. 
The work of T.-W.H.\ and D.K.\ is partly supported by the German Excellence Initiative via the graduate school ``Bonn Cologne Graduate School". The work of D.K.\ is partly supported by the ``Deutsche Telekom Stiftung".

\newpage

\noindent {\bf \LARGE Appendices}

\begin{appendix}

\section{The $\cN=2$ gauge-kinetic coupling function} \label{app:derivationofpotential}

In this appendix we collect some useful formulas applied in the derivation of
the $\cN=1$ scalar potential and the $\cN=1$ gauge-kinetic function for the
bulk vectors.  Both quantities depend on the complex structure deformations of 
the internal Calabi-Yau manifold $Y$. In the underlying $\cN=2$ theory the complex
structure deformations are in vector multiplets together with vectors
$V^K$ in the expansion $C_4 = V^K \alpha_K + \ldots$, where we abuse the
notation and use the same indices as for the $\cN=1$ case. However, note that here
$K=0,\ldots,h^{(2,1)}$ and $V^0$ is actually the graviphothon in the gravity
multiplet. The four-dimensional $\cN=2$ action for the vectors $V^K$ is of the form
\beq
   S_{V^K} = \int \big[ \tfrac14 \I \cM_{KL} dV^K \wedge * dV^L + \tfrac14 \R \cM_{KL} dV^K \wedge
   dV^L\big]\ .
\eeq
The complex matrix $\cM_{KL}$ can be expressed in terms of the periods
$(X^K,\cF_K)$ in the expansion $\Omega = X^K \alpha_K - \cF_K \beta^K$ as 
\begin{equation}
         \mathcal{M}_{KL}=\bar{\mathcal{F}}_{KL}+2i\frac{(\Im\mathcal{F})_{KL}X^M(\Im\mathcal{F})_{LN}X^N}{X^N(\Im\mathcal{F})_{NM}X^M}\ ,
        \label{eqn:matrix-a-b-m}
\end{equation}
where $\cF_{KL}= \partial_{X^K} \cF_L$.
To derive this expression one uses the natural scalar product on the 
cohomology $H^3(Y)$. This can be encoded in the following matrix \cite{Craps:1997gp}
\begin{equation}
        E= \begin{pmatrix}
                \int \alpha_K\wedge *\alpha_L & \int \alpha_K\wedge *\beta^L\\
                \int\beta^K\wedge * \alpha_L & \int \beta^K\wedge * \beta^L
        \end{pmatrix}
       =\begin{pmatrix}
                -(A+BA^{-1}B) & -BA^{-1}\\
                -A^{-1}B & -A^{-1}
        \end{pmatrix}\ ,
\end{equation}
where $A=\Im\mathcal{M}$ and $B=\Re\mathcal{M}$. 
A matrix of this form can be easily inverted where the inverse matrix reads
\begin{equation}
        E^{-1}=\begin{pmatrix}
                -A^{-1} & A^{-1}B\\
                BA^{-1} & -(A+BA^{-1}B)
        \end{pmatrix}=\begin{pmatrix}
                \int \beta^K\wedge * \beta^L &  -\int\beta^K\wedge * \alpha_L\\
                -\int \alpha_K\wedge *\beta^L & \int \alpha_K\wedge *\alpha_L
        \end{pmatrix}\ .
        \label{eqn:matrix-d-2}
\end{equation}
These matrices will be used in the derivation of the $\cN=1$
scalar potential in section \ref{subsec:fluxpotential}, where the indices $K=0,\ldots,
h^{(2,1)}_+$ are in the positive eigenspace $H^3_+(Y)$. 
The complex matrix $\cM_{KL}$ will also appear in the
$\cN=1$ gauge-kinetic coupling function in section \ref{subsec:gauging} where now the indices
$K=1,\ldots,h^{(2,1)}_-$ are in the negative eigenspace $H^{3}_-(Y)$.

\section{Kinetic mixing of bulk and brane gauge groups \label{kinmix}}

The reduction of the Chern-Simons action to the effective Lagrangian
(\ref{eqn:CS}) contains also mixing terms between the bulk vector 
fields $V^{\tilde K}$, $U_{\tilde L}$ and the D5-brane U(1)-field $F$. 
Since the vectors $U_{\tilde L}$ are the magnetic duals to the vector 
$V^{\tilde K}$, a dualization procedure has to be performed in order 
to reveal the effective action of the propagating fields, only. Here, 
we will present this dualization in detail and how it affects the 
kinetic term of the D5-brane vector $F$ such that a further 
intertwining between open and closed moduli appears.

First, we have to collect all terms of the effective action that are relevant 
for the dualization procedure. These are the kinetic terms of the bulk vectors 
$V^{\tilde K}$, $U_{\tilde L}$ of the bulk supergravity action, the 
kinetic as well as instanton term of the D5-brane vector $F$ given in the 
DBI-action (\ref{eqn:DBI}) and the Chern-Simons action (\ref{eqn:CS}),
respectively, and mixing terms between bulk and brane vectors of
(\ref{eqn:CS}). Thus, the starting point of the dualization is the action
\begin{equation}
 S_{\rm vec}= - \int\big[ \tfrac18 d\vec{V}^T\wedge\ast E\, d\vec{V}+ 
\tfrac{1}{2}\mu_5\ell^2 \big(v^{\Sigma} e^{-\phi}F\wedge\ast F-c^\Sigma
F\wedge F\big) +\tfrac12\mu_5\ell\vec{\hat{\mathcal{N}}}^T\, d\vec{V}\wedge F
        \big] \ , \label{Lbeforedualization}
\end{equation}
where we again used the matrix $E$ introduced in (\ref{eqn:e-matrix}) and the convenient shorthand notation
\begin{equation}
 \vec{V}=\begin{pmatrix}
                V^{\tilde K}\\
                U_{\tilde K}
         \end{pmatrix},\qquad \qquad
\vec{\hat{\mathcal{N}}}=\hat\zeta^{\mathcal{A}}\begin{pmatrix}
                \mathcal{N}_{\mathcal{A}\tilde K}\\
                \mathcal{N}_{\mathcal{A}}^{\tilde K}
                \end{pmatrix}
                =
                \begin{pmatrix}
                \mathcal{N}_{\tilde K}\\
                \mathcal{N}^{\tilde K}
         \end{pmatrix}.
\end{equation}
Next we have to add the Lagrange multiplier term $\tfrac14 dV^{\tilde K}\wedge F_{\tilde K}$ to the Lagrangian (\ref{Lbeforedualization}) in order to integrate out the magnetic field strength $F_{\tilde K}=dU_{\tilde K}$. However, the equations of motion for the vectors $V^{\tilde K}$ and their duals $U_{\tilde{L}}$ are not compatible with each other after the naive addition of this term. In order to restore consistency of the equations of motion, we have to shift the field strengths $dV^{\tilde K}$, $dU_{\tilde K}$ in the kinetic terms appropriately by
\begin{equation}
    dV^{\tilde K}\;\rightarrow\; \tilde F^{\tilde{K}}=dV^{\tilde K}-2\mu_5\ell \cN^{\tilde K}F\ ,\qquad \qquad 
    dU_{\tilde L}\;\rightarrow\; \tilde F_{\tilde{L}}=dU_{\tilde L}-2\mu_5\ell \cN_{\tilde L}F.
\end{equation}
Now, we can integrate out the magnetic dual $\tilde F_{\tilde{L}}$ consistently and obtain 
\begin{eqnarray}
S_{\rm vec}&=& \int \big[\tfrac14\text{Im}\mathcal M_{\tilde{K}\tilde{L}}  F^{\tilde{K}}\wedge\ast  F^{\tilde{L}} +\tfrac14\text{Re}\mathcal M_{\tilde{K}\tilde{L}}  F^{\tilde{K}} \wedge F^{\tilde{L}}\nonumber \\ &-&\tfrac12 \mu_5\ell^2\big((v^{\Sigma} e^{-\phi}+2\mu_5\text{Im}\mathcal{M}_{\tilde {K}\tilde{L}}(N^{\tilde K}+\bar{N}^{\tilde K})(N^{\tilde L}+\bar{N}^{\tilde L}))F\wedge\ast F\nonumber \\
\nn &+&(c^\Sigma+i\mu_5\text{Im}\mathcal{M}_{\tilde{K}\tilde{L}}(N^{\tilde K}N^{\tilde L}-\bar{N}^{\tilde K}\bar{N}^{\tilde L})) F\wedge F\big)\\
&+& \mu_5\ell\big(\text{Im}\mathcal M_{\tilde{K}\tilde{L}} \ast F+\text{Re}\mathcal M_{\tilde{K}\tilde{L}}F \big)\wedge F^{\tilde K}\big(N^{\tilde L}+\bar{N}^{\tilde L}\big)\big] . \label{dualizedL}
\end{eqnarray}
Here we introduced the complex fields 
\begin{equation}
        N^{\tilde K}=\int_{\Sigma_-}\zeta\lrcorner \beta^{\tilde K},\qquad \bar{N}^{\tilde K}=\int_{\Sigma_-}\bar{\zeta}\lrcorner\beta^{\tilde K}.
\end{equation}
The crucial point of this dualization is the change of the gauge-kinetic term in (\ref{dualizedL}) compared to the form in (\ref{Lbeforedualization}) before dualization.

\section{Derivation of the F-term scalar potential}\label{app:F-termscalarpot}

The calculation of the F-term contribution of the scalar potential (\ref{N=1pot})
using the superpotential (\ref{effsuperpot}) and K\"ahler potential
\eqref{eqn:kaehler-pot} is straightforward but tedious. To simplify this 
computation it is convenient to exploit one of the
shift symmetries of the K\"ahler potential $S \rightarrow S + i \Lambda$ and
dualize the chiral multiplet with bosonic scalar $S$ into a 
linear multiplet with bosonic components $(L,C_2)$. Here $L$ is a real scalar associated to $\R S$ while
$C_2$ is a two-form dual to $\I S$. In the context of O5 orientifolds without
D5-brane moduli this dualization has been carried out in refs.~\cite{Grimm:2004uq,Grimm:2005fa}, and
we refer the reader to these references for more details on the linear
multiplet formalism and references. Here we will be mainly interested in the scalar potential 
in the new scalar variables $L$ and $M^I=(P_a,a_I,t_\alpha, \zeta^A)$. First 
we express the K\"ahler potential \eqref{eqn:kaehler-pot} in terms of the 
new variables $L=-K_S = \frac12 e^\phi\mathcal{V}^{-1}$ and $M^I$ such that 
\beq \label{KL_pot}
  K = -\ln\big[ -i\int\Omega\wedge\bar\Omega \Big] -\ln \big[ \tfrac{1}{48} \cK_{\alpha \beta \gamma} \Xi^\alpha\,
      \Xi^\beta\, \Xi^\gamma \big] + \ln [L]\ ,
\eeq
where $\Xi^\alpha$ is given in \eqref{eqn:xi-def}. The kinetic terms in the 
effective action with a linear multiplet are now obtained as derivatives 
of the kinetic potential 
\beq \label{tildeK_pot}
  \tilde K(L,M^I,\bar M^I) = K + (S + \bar S) L \ , 
\eeq
where $S+\bar S$ is now a function of $(L,M^I)$. In fact, we have performed a 
Legendre transformation starting with $S+\bar S, K$ to obtain $L,\tilde
K$. In terms of these data
the scalar potential takes the general form 
\beq \label{V_lin}
  V = e^{K}(\tilde K^{IJ} D_I W D_{\bar J} \bar W - (3-L K_L) |W|^2)\ ,
\eeq
where $D_I W = \partial_I W + K_I W$ and $K_L = \partial_L K$. Note that in
front of $|W|^2$ as well as in $D_I W$ only the derivatives of the 
K\"ahler potential \eqref{KL_pot} appear.

With this formalism at hand we evaluate the scalar potential. We first take
derivatives of \eqref{KL_pot} and \eqref{tildeK_pot} such that
\begin{eqnarray}        
        K_{t_\alpha}= -\frac{e^{\phi}}{4\mathcal V}\mathcal K _\alpha\; , 
        \qquad K_{P_a}=0\; ,\qquad
        K_{a_I}= 0\; ,\qquad
        K_{\zeta^A}= \tfrac12\mu_5 e^{\phi} \mathcal{G}_{A\bar B}\bar \zeta^{\bar B}\; ,
\label{firstderivatives}
\end{eqnarray}
as well as 
\begin{eqnarray}        
        \tilde{K}_{t_\alpha}= \frac{e^{\phi}}{4\mathcal V}(\mathcal K _{\alpha ab}\mathcal{B}^a\mathcal{B}^b-\mathcal K _\alpha)\; , 
        \quad \tilde{K}_{P_a}=-\frac{e^{\phi}}{2\mathcal V} \mathcal B ^a\; ,\quad
        \tilde{K}_{a_I}= \frac{\mu_5\ell^2 e^{\phi}}{\mathcal V} \mathcal C ^{I\bar J}\bar a _{\bar J}\; ,\quad
        \tilde{K}_{\zeta^A}= K_{\zeta^A}\; .
\label{firstderivatives_2}
\end{eqnarray}
From this we can easily determine the metric $\tilde{K}_{I\bar J}$ for the remaining fields which is block-diagonal with one block $\tilde{K}_{a_I\bar a_{\bar J}}=\mu_5\ell^2e^\phi\mathcal{V}^{-1} C^{I\bar J}$ for the Wilson lines and another block of the following type
\begin{eqnarray}
 \tilde{K}_{I\bar J}=
\begin{pmatrix}
        A+B^\dagger GB &-B^\dagger G & 0\\
-GB& G+D^\dagger C D&D^\dagger C\\
0&C D& C
\end{pmatrix}
\label{tildemetric}
\end{eqnarray}
for the moduli $(\zeta,t,P)$.
Its inverse $\tilde{K}^{I\bar J}$ is then given by
\begin{eqnarray}
 \tilde{K}^{I\bar J}=
\begin{pmatrix}
        A^{-1} &A^{-1}B^\dagger  & -A^{-1}B^\dagger D^\dagger\\
BA^{-1}& G^{-1}+BA^{-1}B^\dagger&-(G^{-1}+BA^{-1}B^\dagger) D^\dagger\\
-DBA^{-1}&-D(G^{-1}+BA^{-1}B^\dagger)& C^{-1}+D(G^{-1}+BA^{-1}B^\dagger) D^\dagger\\
\end{pmatrix}.
\label{tildemetricinverse}
\end{eqnarray}
Here, we abbreviated the various matrices as follows,
\begin{eqnarray}
        A=\tfrac12e^\phi\mu_5\mathcal{G}_{A\bar B}\,,\quad G=e^{2\phi}(G_{\text{ks}})_{\alpha\beta}\,,\quad B=\mu_5 \mathcal{L}^\alpha_{A\bar B}\bar{\zeta}^{\bar B}\,,\quad C=-\frac{e^\phi}{2\mathcal{V}}(\Re \Theta)_{ab}\,,\quad D=\tfrac12\mathcal{K}_{ab\alpha}\mathcal{B}^b,
\end{eqnarray}
where the matrix $\mc G$ is defined in (\ref{eqn:metrics}) and we introduced the K\"ahler metric on the K\"ahler moduli space
\begin{equation}
        (G_{\text{ks}})_{\alpha\beta}=\frac{1}{4\mathcal{V}}\left(\frac{\mathcal{K}_{\alpha}\mc K_{\beta}}{4\mc V}-\mc K_{\alpha\beta}\right).
\end{equation}
Now we use this to compute the F-term scalar potential.
First we note the no-scale structure of $K$ and $W$. 
The superpotential does not depend on the moduli $(S,a,P)$ 
as well as on $t^\alpha$ such that the covariant derivative 
$D_I=\partial_I+K_I$ reduces just to $K_I$.
Moreover, for the dual linear multiplet to $S$  we find a contribution $1\cdot |W|^2$ to the 
scalar potential $V$ which is an immediate consequence of $K_L L=1$ in \eqref{V_lin}.
The block matrix for the Wilson lines $a$ does not contribute to $V$ since $K_{a_I}=0$. 
However, the block for the moduli $(\zeta,t,P)$ yields a contribution of the form
\begin{equation}
        D_{(\zeta,t,P)}WD_{(\bar \zeta,\bar t,\bar P)}\bar W \tilde K^{(\zeta, t,P)(\bar \zeta,\bar t,\bar P)}=\frac{\mathcal K_\alpha (G_{\text{ks}})^{\alpha\beta}\mathcal K_\beta}{(4\mc V)^2} |W|^2+2\mu_5 \left(\int_{\Sigma_+}s_A\lrcorner \Omega\int_{\Sigma_+}\bar s_{\bar B}\lrcorner \bar \Omega\right) e^{-\phi}\mc G^{A\bar B}.
        \label{eqn:str-derivative-kaehler-metric-2}
\end{equation}
Using the various intersection matrices $v^\alpha=\int J\wedge \tilde \omega^\alpha$, $\mathcal{K}_{\alpha}$, $\mathcal{K}_{\alpha\beta}$ and its formal inverse $\mathcal{K}^{\alpha\beta}$ as well as the inverse metric
\begin{equation}
        G_{\text{ks}}^{\alpha\beta}=2v^\alpha v^\beta-4\mc V\mathcal{K}^{\alpha\beta}\ ,
\end{equation}
we deduce the useful relation 
\begin{equation}
        \mathcal K_\alpha (G_{\text{ks}})^{\alpha\beta}\mathcal K_\beta = (8\mc V)^2v^\alpha(G_{\text{ks}})_{\alpha\beta}v^\beta=3(4\mc V)^2. 
        \label{eqn:kaehler-moduli-identity}
\end{equation}
Finally, we obtain the F-term contribution to the scalar potential $V$ of the form
\begin{eqnarray}
        V 
&=&\frac{ie^{4\phi}}{2\mathcal V^2 \int\Omega\wedge\bar\Omega}\left[ \left| W\right|^2+D_{z^\kappa}WD_{\bar z^{\bar\kappa}}\bar W G^{\kappa\bar\kappa} +2\mu_5 e^{-\phi}\mc G^{A\bar B}\int_{\Sigma_+}s_A\lrcorner \Omega\int_{\Sigma_+}\bar s_{\bar B}\lrcorner \bar \Omega \right]\ .
        \label{eqn:other-form-of-v}
\end{eqnarray}

\section{Detailed discussion of mixed Hodge structure}\label{app:mixed-hodge}
In this appendix we give a detailed description of the mixed Hodge structures for the relative groups $H^3(Y,\Sigma)$ and $H^3(\tilde Y, \Div)$.
Our main references are \cite{Green:1993qv,Voisin2002a}.

First we discuss $H^3(Y,\Sigma)$.
Let $\iota:\Sigma \hookrightarrow Y$ be an embedding of $\Sigma$ into $Y$ and $\Omega_Y^k$ the sheaf of local holomorphic sections in $\bigwedge^kT^*Y$.
Let us consider the following complex of sheaves
\begin{equation}
        \Omega_\iota^\bullet = \left\{ \Omega_Y^\bullet\oplus\Omega_\Sigma^{\bullet-1},\partial \right\}
         \label{eqn:sheaves-complex}
\end{equation}
with the differential $\partial(\alpha,\beta)=(\partial\alpha, f^*\alpha-\partial\beta)$.
We also have a complex of cochains
\begin{equation}
        C^\bullet(\iota,G)=C^\bullet(Y,G)\oplus C^{\bullet-1}(\Sigma,G)
        \label{eqn:cochains-complex}
\end{equation}
with $\delta(\alpha,\beta)=(\delta\alpha,\iota^*\alpha-\delta\beta)$ and $G$ denoting the coefficient, e.g.\ $\C$, $\Z$.
Furthermore, we define the following double complex
\begin{equation}
        C^{p,q}_\iota:=C^p(\Omega_\iota^q)=\{C^p(Y,\Omega_Y^q)\oplus C^{p}(\Sigma,\Omega_\Sigma^{q-1});\delta,\partial\}
\end{equation}
from which we construct the hypercohomology groups\footnote{For hypercohomology and spectral sequences see for example \cite{Griffiths:1978yf}.} $\mathbb H^k(\Omega_\iota^\bullet)$.
We define $H^k(\iota,G):=H^k(C^\bullet(\iota,G))$.
Then we have $H^k(\iota,\C)=H^k(Y,\Sigma,\C)\cong \mathbb H^k(\Omega_\iota^\bullet)$.
The spectral sequence computing $\mathbb H^k(\Omega^\bullet_\iota)$ has $E_1^{p,q}(\Omega_\iota^\bullet)=H^q_\delta(\Omega^p_\iota)$ and degenerates at the $E_2$-term which has the form $H^p_\partial(H_\delta^q(\Omega_\iota^\bullet))$.
Thus, $\mathbb H^k(\Omega_\iota^\bullet) = \bigoplus_{p+q=k}E_2^{p,q}(\Omega_\iota^\bullet)$.
The Hodge filtration on $H^k(Y,\Sigma)$ is given as follows\footnote{Also the familiar Hodge filtration on $H^k(Y)$ \eqref{eqn:closed-hodge-filtraion} can be shown to be $F^mH^k(Y)=\operatorname{Im}(\mathbb H^k(\Omega_Y^{\geq m})$.}
\begin{equation}
        F^m\mathbb H^k(\Omega_\iota^\bullet)=\Image(\mathbb H^k(\Omega^{\geq m}_\iota))\ ,
        \label{eqn:hodge-filtration-pair}
\end{equation}
where $\Image(\cdot)$ denotes the image of the induced map on the cohomology from the embedding of $\Omega^{\geq m}_\iota$ into $\Omega^\bullet_\iota$.
Now, we want to describe $F^m\mathbb H^k(\Omega_\iota^\bullet)$ in easier terms.
We obtain for $E_2^{p,q}(\Omega_\iota^{\geq m})$
\begin{equation}
        E_2^{p,q}(\Omega_\iota^{\geq m}) =
        \begin{cases}
                E_2^{p,q}(\Omega_\iota^\bullet) & \text{for }p > m\ ,\\
                \Kern \left(\xymatrix{H^q_\delta(\Omega^p)\ar[r] & H^q_\delta(\Omega^{p+1}})\right) & \text{for }p=m\ ,\\
                0 & \text{otherwise .}
        \end{cases}
        \label{eqn:e2pq-m}
\end{equation}
If we consider the image of $E_2^{m,q}(\Omega_\iota^{\geq m})$ in $\mathbb H^k(\Omega_\iota^\bullet)$, it is obvious that it equals $E_2^{p,q}(\Omega_\iota^\bullet)$.
Thus
\begin{equation}
        F^m\mathbb H^k(\Omega_\iota^\bullet)=\Image \left( \mathbb H^k(\Omega_\iota^{\geq m}) \right) = \bigoplus_{p\geq k} E_2^{p,k-p}(\Omega_\iota^\bullet)\ .
        \label{eqn:filtration-easy}
\end{equation}
Furthermore, the weight filtration for $H^k(Y,\Sigma)$ is defined as follows
\begin{align}
        \nonumber W_k\mathbb H^k(\Omega_\iota^\bullet)&=\mathbb H^k(\Omega_\iota^\bullet)\ ,\\
        W_{k-1}\mathbb H^k(\Omega_\iota^\bullet) &= \Image\left( \xymatrix{\mathbb H^k(\Omega_\Sigma^{\bullet-1})\ar[r]& \mathbb H^k(\Omega_\iota^\bullet)} \right)\ ,\\
        \nonumber W_{k-2}\mathbb H^k(\Omega_\iota^\bullet) &= 0\ .
        \label{eqn:weight-filtration}
\end{align}
For convenience we write $W_m$ for $W_m\mathbb H^k(\Omega_\iota^\bullet)$.
We want to show the following
\begin{equation}
        W_{k-1}\cong \operatorname{Coker} \left( \iota^*:\xymatrix{H^{k-1}(Y,\C)\ar[r] & H^{k-1}(\Sigma,\C)} \right)= H^{k-1}_v(\Sigma,\C)\ .
        \label{eqn:isom-weight-2}
\end{equation}
Using the fact $E_1^{p,q}(\Omega_\iota^\bullet)\cong H^q(\Omega_Y^p)\oplus H^q(\Omega_\Sigma^{p-1})$ and $E_2^{p,q}(\Omega_\iota^\bullet)=H^p_{\partial}(H^q_\delta(\Omega_\iota^\bullet))$,
we see that $E_1^{p,q}(\Omega_\Sigma^{\bullet-1})$ gets mapped to classes of $E_2^{p,q}(\Omega_\iota^\bullet)$ of the $H^q(\Omega_\Sigma^{p-1})$-part which are closed\footnote{This means that if we would ignore the modding out by $\partial(E_1^{p-1,q}(\Omega_\iota^\bullet))$, then the image of $E_1^{p,q}(\Omega_\Sigma^{\bullet-1})$ would be just itself since $E_1^{p,q}(\Omega_\Sigma^\bullet)=E_\infty^{p,q}(\Omega_\Sigma^\bullet)$.} under $\partial$ without involving classes of $H^q(\Omega_Y^p)$.
Additionally, we mod out classes of the form $\partial(\alpha, 0)=(0, \iota^*\alpha)$ which are the images under $\iota^*$.
Since the spectral sequence computing $\mathbb H^k(\Omega_\Sigma^{\bullet -1})$ degenerates at the $E_1$-term, we see $W_{k-1}$ corresponds exactly to $\operatorname{Coker}\iota^*$ which consists of classes of $(k-1)$-forms on $\Sigma$ which do not contain pull-back of $(k-1)$-forms on $Y$.
Thus, we obtain the isomorphism \eqref{eqn:isom-weight-2}.
We now define the graded weights as follows
\begin{align}
        \operatorname{Gr}_m^W\mathbb H^k(\Omega_\iota^\bullet) = W_m/W_{m-1}\ .
        \label{eqn:graded-weights}
\end{align}
Using the decomposition (\ref{eqn:decomposition-relative-group}) and \eqref{eqn:isom-weight-2}, we can write
\begin{equation}
        \operatorname{Gr}_k^W\mathbb H^k(\Omega_\iota^\bullet)\cong H^k_v(Y,\C)\ ,\quad
        \operatorname{Gr}_{k-1}^W\mathbb H^k(\Omega_\iota^\bullet)\cong H^{k-1}_v(\Sigma,\C)\ .
        \label{eqn:graded-weights-detailed}
\end{equation}

Now, we give a detailed description for the mixed Hodge structure of $H^3(\tilde Y, \Div)$.
Let $D$ be a smooth divisor of $\tilde Y$, i.e.\ $D$ can be locally written as $\left\{ z_n=0 \right\}$ where $n$ is the (complex) dimension of $\tilde Y$.
For $\tilde Y$ and $D$ we have the isomorphisms $H^\bullet(\tilde Y,D,\C)\cong H^\bullet(\tilde Y-D,\C)\cong \mathbb H^\bullet(\Omega^\bullet_{\tilde Y}(\log D))$.
For the hypercohomology of the $\log$-complex there exists Hodge- and weight-filtration which gives rise to a mixed Hodge structure.
The filtrations has the following form
\begin{equation}
        F^pH^k = \Image\left(\mathbb H^k(\Omega_{\tilde Y}^{\geq p}(\log D)) \right)\ ,\quad
        W_qH^k = \Image\left(\mathbb H^k(W_{q-k}\Omega_{\tilde Y}^\bullet(\log D)) \right)\ ,
        \label{eqn:hodge-weight-filtrations-log-complex}
\end{equation}
where
\begin{equation}
        W_q\Omega_{\tilde Y}^p(\log D)= 
        \begin{cases}
                0 &\text{for }q<0\ ,\\
                \Omega_{\tilde Y}^p(\log D)&\text{for }q\geq p\ ,\\
                \Omega_{\tilde Y}^{p-q}\wedge\Omega_{\tilde Y}^q(\log D)&\text{for }0\leq q \leq p\ .
        \end{cases}
        \label{eqn:wm-omega-log}
\end{equation}
On $H^k(\tilde Y-D)$, $F^\bullet H^k$ and $W_{\bullet}H^k$ gives
a mixed Hodge structure.
Since the hypercohomology computing $\mathbb H^\bullet (\Omega_{\tilde Y}^\bullet(\log D))$ degenerates at the first term, we obtain $F^mH^k = \bigoplus_{p\geq m}E_1^{p,k-p}(\Omega_{\tilde Y}^\bullet(\log D))$.
The weight filtration can then be described as follows
\begin{equation}
        \label{eqn:weight-filtration-log-detailed}
        W_{-1}H^k=0\ , \quad W_{0}H^k = H^k(\tilde Y,\C)\ ,\quad W_{1}H^k = H^k(\tilde Y-D,\C)\ .
\end{equation}
Defining the graded weights to be $\operatorname{Gr}_m^WH^k=W_{-m+k}H^k/W_{-(m+1)+k}H^k$, we obtain 
\begin{equation}
        \label{eqn:graded-weights-log-app}
\operatorname{Gr}^W_{k}H^k \cong H^k(\tilde Y,\C)\ ,\quad
        \operatorname{Gr}^W_{k-1}H^k \cong H^{k-1}(D,\C)\ .
\end{equation}

\end{appendix}      

\newpage

\providecommand{\href}[2]{#2}\begingroup\raggedright\endgroup


\begin{thebibliography}{10}

\bibitem{Douglas:2006es}
M.~R. Douglas and S.~Kachru, ``{Flux compactification},''
  \href{http://dx.doi.org/10.1103/RevModPhys.79.733}{{\em Rev. Mod. Phys.} {\bf
  79} (2007)  733--796},
\href{http://arxiv.org/abs/hep-th/0610102}{{\tt arXiv:hep-th/0610102}}.

\bibitem{Blumenhagen:2006ci}
R.~Blumenhagen, B.~Kors, D.~Lust, and S.~Stieberger, ``{Four-dimensional String
  Compactifications with D-Branes, Orientifolds and Fluxes},''
  \href{http://dx.doi.org/10.1016/j.physrep.2007.04.003}{{\em Phys. Rept.} {\bf
  445} (2007)  1--193},
\href{http://arxiv.org/abs/hep-th/0610327}{{\tt arXiv:hep-th/0610327}}.

\bibitem{Denef:2008wq}
F.~Denef, ``{Les Houches Lectures on Constructing String Vacua},''
\href{http://arxiv.org/abs/0803.1194}{{\tt arXiv:0803.1194 [hep-th]}}.

\bibitem{Lust:2004ks}
D.~Lust, ``{Intersecting brane worlds: A path to the standard model?},'' {\em
  Class. Quant. Grav.} {\bf 21} (2004)  S1399--1424,
\href{http://arxiv.org/abs/hep-th/0401156}{{\tt arXiv:hep-th/0401156}}.

\bibitem{Blumenhagen:2005mu}
R.~Blumenhagen, M.~Cvetic, P.~Langacker, and G.~Shiu, ``{Toward realistic
  intersecting D-brane models},'' {\em Ann. Rev. Nucl. Part. Sci.} {\bf 55}
  (2005)  71--139,
\href{http://arxiv.org/abs/hep-th/0502005}{{\tt arXiv:hep-th/0502005}}.

\bibitem{Camara:2003ku}
P.~G. Camara, L.~E. Ibanez, and A.~M. Uranga, ``{Flux-induced SUSY-breaking
  soft terms},'' \href{http://dx.doi.org/10.1016/j.nuclphysb.2004.04.013}{{\em
  Nucl. Phys.} {\bf B689} (2004)  195--242},
\href{http://arxiv.org/abs/hep-th/0311241}{{\tt arXiv:hep-th/0311241}}.

\bibitem{Grana:2003ek}
M.~Grana, T.~W. Grimm, H.~Jockers, and J.~Louis, ``{Soft supersymmetry breaking
  in Calabi-Yau orientifolds with D-branes and fluxes},''
  \href{http://dx.doi.org/10.1016/j.nuclphysb.2004.04.021}{{\em Nucl. Phys.}
  {\bf B690} (2004)  21--61},
\href{http://arxiv.org/abs/hep-th/0312232}{{\tt arXiv:hep-th/0312232}}.

\bibitem{Camara:2004jj}
P.~G. Camara, L.~E. Ibanez, and A.~M. Uranga, ``{Flux-induced SUSY-breaking
  soft terms on D7-D3 brane systems},''
  \href{http://dx.doi.org/10.1016/j.nuclphysb.2004.11.035}{{\em Nucl. Phys.}
  {\bf B708} (2005)  268--316},
\href{http://arxiv.org/abs/hep-th/0408036}{{\tt arXiv:hep-th/0408036}}.

\bibitem{Jockers:2004yj}
H.~Jockers and J.~Louis, ``{The effective action of D7-branes in N = 1
  Calabi-Yau orientifolds},''
  \href{http://dx.doi.org/10.1016/j.nuclphysb.2004.11.009}{{\em Nucl. Phys.}
  {\bf B705} (2005)  167--211},
\href{http://arxiv.org/abs/hep-th/0409098}{{\tt arXiv:hep-th/0409098}}.

\bibitem{Jockers:2005zy}
H.~Jockers and J.~Louis, ``{D-terms and F-terms from D7-brane fluxes},''
  \href{http://dx.doi.org/10.1016/j.nuclphysb.2005.04.011}{{\em Nucl. Phys.}
  {\bf B718} (2005)  203--246},
\href{http://arxiv.org/abs/hep-th/0502059}{{\tt arXiv:hep-th/0502059}}.

\bibitem{Kors:2003wf}
B.~Kors and P.~Nath, ``{Effective action and soft supersymmetry breaking for
  intersecting D-brane models},''
  \href{http://dx.doi.org/10.1016/j.nuclphysb.2004.01.002}{{\em Nucl. Phys.}
  {\bf B681} (2004)  77--119},
\href{http://arxiv.org/abs/hep-th/0309167}{{\tt arXiv:hep-th/0309167}}.

\bibitem{Lust:2004cx}
D.~Lust, P.~Mayr, R.~Richter, and S.~Stieberger, ``{Scattering of gauge,
  matter, and moduli fields from intersecting branes},''
  \href{http://dx.doi.org/10.1016/j.nuclphysb.2004.06.052}{{\em Nucl. Phys.}
  {\bf B696} (2004)  205--250},
\href{http://arxiv.org/abs/hep-th/0404134}{{\tt arXiv:hep-th/0404134}}.

\bibitem{Grimm:2004ua}
T.~W. Grimm and J.~Louis, ``{The effective action of type IIA Calabi-Yau
  orientifolds},''
  \href{http://dx.doi.org/10.1016/j.nuclphysb.2005.04.007}{{\em Nucl. Phys.}
  {\bf B718} (2005)  153--202},
\href{http://arxiv.org/abs/hep-th/0412277}{{\tt arXiv:hep-th/0412277}}.

\bibitem{Becker:1995kb}
K.~Becker, M.~Becker, and A.~Strominger, ``{Five-branes, membranes and
  nonperturbative string theory},''
  \href{http://dx.doi.org/10.1016/0550-3213(95)00487-1}{{\em Nucl. Phys.} {\bf
  B456} (1995)  130--152},
\href{http://arxiv.org/abs/hep-th/9507158}{{\tt arXiv:hep-th/9507158}}.

\bibitem{Witten:1997ep}
E.~Witten, ``{Branes and the dynamics of {QCD}},''
  \href{http://dx.doi.org/10.1016/S0550-3213(97)00648-2}{{\em Nucl. Phys.} {\bf
  B507} (1997)  658--690},
\href{http://arxiv.org/abs/hep-th/9706109}{{\tt arXiv:hep-th/9706109}}.

\bibitem{Brunner:1999jq}
I.~Brunner, M.~R. Douglas, A.~E. Lawrence, and C.~Romelsberger, ``{D-branes on
  the quintic},'' {\em JHEP} {\bf 08} (2000)  015,
\href{http://arxiv.org/abs/hep-th/9906200}{{\tt arXiv:hep-th/9906200}}.

\bibitem{Kachru:2000ih}
S.~Kachru, S.~H. Katz, A.~E. Lawrence, and J.~McGreevy, ``{Open string
  instantons and superpotentials},''
  \href{http://dx.doi.org/10.1103/PhysRevD.62.026001}{{\em Phys. Rev.} {\bf
  D62} (2000)  026001},
\href{http://arxiv.org/abs/hep-th/9912151}{{\tt arXiv:hep-th/9912151}}.

\bibitem{Kachru:2000an}
S.~Kachru, S.~H. Katz, A.~E. Lawrence, and J.~McGreevy, ``{Mirror symmetry for
  open strings},'' \href{http://dx.doi.org/10.1103/PhysRevD.62.126005}{{\em
  Phys. Rev.} {\bf D62} (2000)  126005},
\href{http://arxiv.org/abs/hep-th/0006047}{{\tt arXiv:hep-th/0006047}}.

\bibitem{Aganagic:2000gs}
M.~Aganagic and C.~Vafa, ``{Mirror symmetry, D-branes and counting holomorphic
  discs},''
\href{http://arxiv.org/abs/hep-th/0012041}{{\tt arXiv:hep-th/0012041}}.

\bibitem{Aganagic:2001nx}
M.~Aganagic, A.~Klemm, and C.~Vafa, ``{Disk instantons, mirror symmetry and the
  duality web},'' {\em Z. Naturforsch.} {\bf A57} (2002)  1--28,
\href{http://arxiv.org/abs/hep-th/0105045}{{\tt arXiv:hep-th/0105045}}.

\bibitem{Marino:2004uf}
M.~Marino, ``{Chern-Simons theory and topological strings},''
  \href{http://dx.doi.org/10.1103/RevModPhys.77.675}{{\em Rev. Mod. Phys.} {\bf
  77} (2005)  675--720},
\href{http://arxiv.org/abs/hep-th/0406005}{{\tt arXiv:hep-th/0406005}}.

\bibitem{Baumgartl:2007an}
M.~Baumgartl, I.~Brunner, and M.~R. Gaberdiel, ``{D-brane superpotentials and
  RG flows on the quintic},''
  \href{http://dx.doi.org/10.1088/1126-6708/2007/07/061}{{\em JHEP} {\bf 07}
  (2007)  061},
\href{http://arxiv.org/abs/0704.2666}{{\tt arXiv:0704.2666 [hep-th]}}.

\bibitem{Lerche:2002ck}
W.~Lerche, P.~Mayr, and N.~Warner, ``{Holomorphic N = 1 special geometry of
  open-closed type II strings},''
\href{http://arxiv.org/abs/hep-th/0207259}{{\tt arXiv:hep-th/0207259}}.

\bibitem{Lerche:2002yw}
W.~Lerche, P.~Mayr, and N.~Warner, ``{N = 1 special geometry, mixed Hodge
  variations and toric geometry},''
\href{http://arxiv.org/abs/hep-th/0208039}{{\tt arXiv:hep-th/0208039}}.

\bibitem{Aganagic:2003db}
M.~Aganagic, A.~Klemm, M.~Marino, and C.~Vafa, ``{The topological vertex},''
  \href{http://dx.doi.org/10.1007/s00220-004-1162-z}{{\em Commun. Math. Phys.}
  {\bf 254} (2005)  425--478},
\href{http://arxiv.org/abs/hep-th/0305132}{{\tt arXiv:hep-th/0305132}}.

\bibitem{Graber:2001dw}
T.~Graber and E.~Zaslow, ``{Open string Gromov-Witten invariants: Calculations
  and a mirror 'theorem'},''
\href{http://arxiv.org/abs/hep-th/0109075}{{\tt arXiv:hep-th/0109075}}.

\bibitem{Walcher:2006rs}
J.~Walcher, ``{Opening mirror symmetry on the quintic},''
  \href{http://dx.doi.org/10.1007/s00220-007-0354-8}{{\em Commun. Math. Phys.}
  {\bf 276} (2007)  671--689},
\href{http://arxiv.org/abs/hep-th/0605162}{{\tt arXiv:hep-th/0605162}}.

\bibitem{Jockers:2008pe}
H.~Jockers and M.~Soroush, ``{Effective superpotentials for compact D5-brane
  Calabi-Yau geometries},''
\href{http://arxiv.org/abs/0808.0761}{{\tt arXiv:0808.0761 [hep-th]}}.

\bibitem{Griffiths:1979rt}
P.~A. Griffiths, ``A Theorem Concerning the Differential Equations Satisfied by
  Normal Functions Associated to Algebraic Cycles,'' {\em American Jounal of
  Mathematics} {\bf 101} (1979) no.~1, 94--131.

\bibitem{Brunner:2003zm}
I.~Brunner and K.~Hori, ``{Orientifolds and mirror symmetry},''
  \href{http://dx.doi.org/10.1088/1126-6708/2004/11/005}{{\em JHEP} {\bf 11}
  (2004)  005},
\href{http://arxiv.org/abs/hep-th/0303135}{{\tt arXiv:hep-th/0303135}}.

\bibitem{Bergshoeff:2001pv}
E.~Bergshoeff, R.~Kallosh, T.~Ortin, D.~Roest, and A.~Van~Proeyen, ``{New
  formulations of D = 10 supersymmetry and D8 - O8 domain walls},''
  \href{http://dx.doi.org/10.1088/0264-9381/18/17/303}{{\em Class. Quant.
  Grav.} {\bf 18} (2001)  3359--3382},
\href{http://arxiv.org/abs/hep-th/0103233}{{\tt arXiv:hep-th/0103233}}.

\bibitem{Martucci:2006ij}
L.~Martucci, ``{D-branes on general N = 1 backgrounds: Superpotentials and
  D-terms},'' {\em JHEP} {\bf 06} (2006)  033,
\href{http://arxiv.org/abs/hep-th/0602129}{{\tt arXiv:hep-th/0602129}}.

\bibitem{Grimm:2004uq}
T.~W. Grimm and J.~Louis, ``{The effective action of N = 1 Calabi-Yau
  orientifolds},''
  \href{http://dx.doi.org/10.1016/j.nuclphysb.2004.08.005}{{\em Nucl. Phys.}
  {\bf B699} (2004)  387--426},
\href{http://arxiv.org/abs/hep-th/0403067}{{\tt arXiv:hep-th/0403067}}.

\bibitem{Marino:1999af}
M.~Marino, R.~Minasian, G.~W. Moore, and A.~Strominger, ``{Nonlinear instantons
  from supersymmetric p-branes},'' {\em JHEP} {\bf 01} (2000)  005,
\href{http://arxiv.org/abs/hep-th/9911206}{{\tt arXiv:hep-th/9911206}}.

\bibitem{Douglas:2000ah}
M.~R. Douglas, B.~Fiol, and C.~Romelsberger, ``{Stability and BPS branes},''
  {\em JHEP} {\bf 09} (2005)  006,
\href{http://arxiv.org/abs/hep-th/0002037}{{\tt arXiv:hep-th/0002037}}.

\bibitem{Beasley:2002db}
C.~Beasley and E.~Witten, ``{A note on fluxes and superpotentials in M-theory
  compactifications on manifolds of G(2) holonomy},'' {\em JHEP} {\bf 07}
  (2002)  046,
\href{http://arxiv.org/abs/hep-th/0203061}{{\tt arXiv:hep-th/0203061}}.

\bibitem{Louis:2002ny}
J.~Louis and A.~Micu, ``{Type II theories compactified on Calabi-Yau threefolds
  in the presence of background fluxes},''
  \href{http://dx.doi.org/10.1016/S0550-3213(02)00338-3}{{\em Nucl. Phys.} {\bf
  B635} (2002)  395--431},
\href{http://arxiv.org/abs/hep-th/0202168}{{\tt arXiv:hep-th/0202168}}.

\bibitem{Wess:1992cp}
J.~Wess and J.~Bagger, ``{Supersymmetry and supergravity},''. Princeton, USA:
  Univ. Pr. (1992) 259 p.

\bibitem{Gates:1983nr}
S.~J. Gates, M.~T. Grisaru, M.~Rocek, and W.~Siegel, ``{Superspace, or one
  thousand and one lessons in supersymmetry},'' {\em Front. Phys.} {\bf 58}
  (1983)  1--548,
\href{http://arxiv.org/abs/hep-th/0108200}{{\tt arXiv:hep-th/0108200}}.

\bibitem{Grimm:2005fa}
T.~W. Grimm, ``{The effective action of type II Calabi-Yau orientifolds},''
  {\em Fortsch. Phys.} {\bf 53} (2005)  1179--1271,
\href{http://arxiv.org/abs/hep-th/0507153}{{\tt arXiv:hep-th/0507153}}.

\bibitem{Gukov:1999ya}
S.~Gukov, C.~Vafa, and E.~Witten, ``{CFT's from Calabi-Yau four-folds},''
  \href{http://dx.doi.org/10.1016/S0550-3213(00)00373-4}{{\em Nucl. Phys.} {\bf
  B584} (2000)  69--108},
\href{http://arxiv.org/abs/hep-th/9906070}{{\tt arXiv:hep-th/9906070}}.

\bibitem{Cremmer:1983bf}
E.~Cremmer, S.~Ferrara, C.~Kounnas, and D.~V. Nanopoulos, ``{Naturally
  Vanishing Cosmological Constant in N=1 Supergravity},''
\href{http://dx.doi.org/10.1016/0370-2693(83)90106-5}{{\em Phys. Lett.} {\bf
  B133} (1983)  61}.

\bibitem{Ellis:1983sf}
J.~R. Ellis, A.~B. Lahanas, D.~V. Nanopoulos, and K.~Tamvakis, ``{No-Scale
  Supersymmetric Standard Model},''
\href{http://dx.doi.org/10.1016/0370-2693(84)91378-9}{{\em Phys. Lett.} {\bf
  B134} (1984)  429}.

\bibitem{Barbieri:1985wq}
R.~Barbieri, E.~Cremmer, and S.~Ferrara, ``{FLAT AND POSITIVE POTENTIALS IN N=1
  SUPERGRAVITY},''
\href{http://dx.doi.org/10.1016/0370-2693(85)90209-6}{{\em Phys. Lett.} {\bf
  B163} (1985)  143}.

\bibitem{DeWolfe:2002nn}
O.~DeWolfe and S.~B. Giddings, ``{Scales and hierarchies in warped
  compactifications and brane worlds},''
  \href{http://dx.doi.org/10.1103/PhysRevD.67.066008}{{\em Phys. Rev.} {\bf
  D67} (2003)  066008},
\href{http://arxiv.org/abs/hep-th/0208123}{{\tt arXiv:hep-th/0208123}}.

\bibitem{Correia:2007sv}
F.~Paccetti~Correia and M.~G. Schmidt, ``{Moduli stabilization in heterotic
  M-theory},'' \href{http://dx.doi.org/10.1016/j.nuclphysb.2008.01.005}{{\em
  Nucl. Phys.} {\bf B797} (2008)  243--267},
\href{http://arxiv.org/abs/0708.3805}{{\tt arXiv:0708.3805 [hep-th]}}.

\bibitem{Curio:2000sc}
G.~Curio, A.~Klemm, D.~Lust, and S.~Theisen, ``{On the vacuum structure of type
  II string compactifications on Calabi-Yau spaces with H-fluxes},''
  \href{http://dx.doi.org/10.1016/S0550-3213(01)00285-1}{{\em Nucl. Phys.} {\bf
  B609} (2001)  3--45},
\href{http://arxiv.org/abs/hep-th/0012213}{{\tt arXiv:hep-th/0012213}}.

\bibitem{Witten:1996bn}
E.~Witten, ``{Non-Perturbative Superpotentials In String Theory},''
  \href{http://dx.doi.org/10.1016/0550-3213(96)00283-0}{{\em Nucl. Phys.} {\bf
  B474} (1996)  343--360},
\href{http://arxiv.org/abs/hep-th/9604030}{{\tt arXiv:hep-th/9604030}}.

\bibitem{Mayr:2002zi}
P.~Mayr, ``{Summing up open string instantons and N = 1 string amplitudes},''
\href{http://arxiv.org/abs/hep-th/0203237}{{\tt arXiv:hep-th/0203237}}.

\bibitem{Eynard:2007kz}
B.~Eynard and N.~Orantin, ``{Invariants of algebraic curves and topological
  expansion},''
\href{http://arxiv.org/abs/math-ph/0702045}{{\tt arXiv:math-ph/0702045}}.

\bibitem{Marino:2006hs}
M.~Marino, ``{Open string amplitudes and large order behavior in topological
  string theory},'' \href{http://dx.doi.org/10.1088/1126-6708/2008/03/060}{{\em
  JHEP} {\bf 03} (2008)  060},
\href{http://arxiv.org/abs/hep-th/0612127}{{\tt arXiv:hep-th/0612127}}.

\bibitem{Bouchard:2008gu}
V.~Bouchard, A.~Klemm, M.~Marino, and S.~Pasquetti, ``{Topological open strings
  on orbifolds},''
\href{http://arxiv.org/abs/0807.0597}{{\tt arXiv:0807.0597 [hep-th]}}.

\bibitem{Bouchard:2007ys}
V.~Bouchard, A.~Klemm, M.~Marino, and S.~Pasquetti, ``{Remodeling the
  B-model},''
\href{http://arxiv.org/abs/0709.1453}{{\tt arXiv:0709.1453 [hep-th]}}.

\bibitem{Katz:1996fh}
S.~H. Katz, A.~Klemm, and C.~Vafa, ``{Geometric engineering of quantum field
  theories},'' \href{http://dx.doi.org/10.1016/S0550-3213(97)00282-4}{{\em
  Nucl. Phys.} {\bf B497} (1997)  173--195},
\href{http://arxiv.org/abs/hep-th/9609239}{{\tt arXiv:hep-th/9609239}}.

\bibitem{Hori:2000kt}
K.~Hori and C.~Vafa, ``{Mirror symmetry},''
\href{http://arxiv.org/abs/hep-th/0002222}{{\tt arXiv:hep-th/0002222}}.

\bibitem{Griffiths:1968}
P.~A. Griffiths, ``Periods of Integrals on Algebraic Manifolds, I,'' {\em
  American Jounal of Mathematics} {\bf 90} (1968) no.~3, 568--626.

\bibitem{Griffiths:1968a}
P.~A. Griffiths, ``Periods of Integrals on Algebraic Manifolds, II,'' {\em
  American Jounal of Mathematics} {\bf 90} (1968) no.~3, 805--865.

\bibitem{Kodaira1986}
K.~Kodaira, {\em Complex Manifolds and Deformations of Complex Structures}.
\newblock Springer-Verlag, 1986.

\bibitem{Spanier:1966}
E.~H. Spanier, {\em Algebraic Topology}.
\newblock Springer-Verlag, 1966.

\bibitem{Griffiths:1978yf}
P.~Griffiths and J.~Harris, {\em Principles of Algebraic Geometry}.
\newblock John Wiley and Sons, Inc., 1978.

\bibitem{Voisin2002a}
C.~Voisin, {\em Hodge Theory and Complex Algebraic Geometry, I}.
\newblock Cambridge University Press, 2002.

\bibitem{BBGW}
R.~Blumenhagen, V.~Braun, T.~W. Grimm, and T.~Weigand, ``{GUTs in Type IIB
  Orientifold Compactifications},''
\href{http://arxiv.org/abs/0811.2936}{{\tt arXiv:0811.2936 [hep-th]}}.

\bibitem{Becker:2007ui}
M.~Becker, L.~Leblond, and S.~E. Shandera, ``{Inflation from Wrapped Branes},''
  \href{http://dx.doi.org/10.1103/PhysRevD.76.123516}{{\em Phys. Rev.} {\bf
  D76} (2007)  123516},
\href{http://arxiv.org/abs/0709.1170}{{\tt arXiv:0709.1170 [hep-th]}}.

\bibitem{Avgoustidis:2006zp}
A.~Avgoustidis, D.~Cremades, and F.~Quevedo, ``{Wilson line inflation},''
  \href{http://dx.doi.org/10.1007/s10714-007-0454-y}{{\em Gen. Rel. Grav.} {\bf
  39} (2007)  1203--1234},
\href{http://arxiv.org/abs/hep-th/0606031}{{\tt arXiv:hep-th/0606031}}.

\bibitem{Aganagic:2007py}
M.~Aganagic, C.~Beem, and S.~Kachru, ``{Geometric Transitions and Dynamical
  SUSY Breaking},''
  \href{http://dx.doi.org/10.1016/j.nuclphysb.2007.11.032}{{\em Nucl. Phys.}
  {\bf B796} (2008)  1--24},
\href{http://arxiv.org/abs/0709.4277}{{\tt arXiv:0709.4277 [hep-th]}}.

\bibitem{Grimm:2008ed}
T.~W. Grimm and A.~Klemm, ``{U(1) Mediation of Flux Supersymmetry Breaking},''
  \href{http://dx.doi.org/10.1088/1126-6708/2008/10/077}{{\em JHEP} {\bf 10}
  (2008)  077},
\href{http://arxiv.org/abs/0805.3361}{{\tt arXiv:0805.3361 [hep-th]}}.

\bibitem{Morrison:2007bm}
D.~R. Morrison and J.~Walcher, ``{D-branes and Normal Functions},''
\href{http://arxiv.org/abs/0709.4028}{{\tt arXiv:0709.4028 [hep-th]}}.

\bibitem{Krefl:2008sj}
D.~Krefl and J.~Walcher, ``{Real Mirror Symmetry for One-parameter
  Hypersurfaces},''
\href{http://arxiv.org/abs/0805.0792}{{\tt arXiv:0805.0792 [hep-th]}}.

\bibitem{Knapp:2008uw}
J.~Knapp and E.~Scheidegger, ``{Towards Open String Mirror Symmetry for
  One-Parameter Calabi-Yau Hypersurfaces},''
\href{http://arxiv.org/abs/0805.1013}{{\tt arXiv:0805.1013 [hep-th]}}.

\bibitem{Craps:1997gp}
B.~Craps, F.~Roose, W.~Troost, and A.~Van~Proeyen, ``{What is special Kaehler
  geometry?},'' \href{http://dx.doi.org/10.1016/S0550-3213(97)00408-2}{{\em
  Nucl. Phys.} {\bf B503} (1997)  565--613},
\href{http://arxiv.org/abs/hep-th/9703082}{{\tt arXiv:hep-th/9703082}}.

\bibitem{Green:1993qv}
M.~L. Green, {\em Algebraic Cycles and Hodge Theory}, ch.~Infinitesimal Methods
  in Hodge Theory.
\newblock Springer-Verlag, 1993.

\end{thebibliography}
\end{document}